\documentclass[aps,pra,twocolumn,a4paper,floatfix,
superscriptaddress, nofootinbib]{revtex4-2}
  \usepackage[T1]{fontenc}
  \usepackage{amsmath}
  \usepackage{amsthm}
  \usepackage{amsbsy}
  \usepackage{amssymb}
  \usepackage{enumitem}
\usepackage{lipsum}
\usepackage[colorlinks=true,urlcolor=blue,citecolor=blue,linkcolor=blue]{hyperref}
\usepackage{graphicx, color}
\usepackage[dvipsnames,table,xcdraw]{xcolor}
\usepackage{orcidlink}

\DeclareMathOperator*{\argmax}{arg\,max}

\newcommand{\ket}[1]{\left|#1\right\rangle}
\newcommand{\bra}[1]{\left\langle#1\right|}

\DeclareMathOperator{\Ext}{Ext}
\DeclareMathOperator{\Tr}{Tr}
\DeclareMathOperator{\Conv}{Conv}
\DeclareMathOperator{\Var}{Var}

\newtheorem*{statement*}{Statement}

\newtheorem{theorem}{Theorem}

 \usepackage[normalem]{ulem}

\begin{document}
\title{Nonclassical photocounting statistics with a single on-off detector}

\author{V. S. Kovtoniuk \orcidlink{0009-0008-8363-0898}}
\affiliation{Quantum Optics and Quantum Information Group, Bogolyubov Institute for Theoretical Physics of the National Academy of Sciences of Ukraine, Vulytsia Metrolohichna 14b, 03143 Kyiv, Ukraine}

\author{M. Bohmann \orcidlink{0000-0003-3857-4555}}
% \affiliation{Institute for Quantum Optics and Quantum Information (IQOQI),
% 	Austrian Academy of Sciences, Boltzmanngasse 3, 1090 Vienna, Austria}
% \affiliation{Vienna Center for Quantum Science and Technology (VCQ), Faculty of Physics,
% 	University of Vienna, Boltzmanngasse 5, 1090 Vienna, Austria}
\affiliation{Quantum Technology Laboratories GmbH, Clemens-Holzmeister-Straße 6/6, 1100 Vienna, Austria}

\author{A. A. Semenov \orcidlink{0000-0001-5104-6445}}
\affiliation{Quantum Optics and Quantum Information Group, Bogolyubov Institute for Theoretical Physics of the National Academy of Sciences of Ukraine, Vulytsia Metrolohichna 14b, 03143 Kyiv, Ukraine}
\affiliation{Department of Theoretical and Mathematical Physics, Kyiv Academic University, Boulevard Vernadskogo  36, 03142  Kyiv, Ukraine}
\affiliation{Department of Mathematics, Kyiv School of Economics, Vulytsia Mykoly Shpaka 3, 03113  Kyiv, Ukraine}

\begin{abstract}
Any single on-off photocounter, which can only detect the presence or absence of photons without discriminating their number, is not capable of identifying the nonclassical nature of light.
This limitation arises because any photocounting statistics obtained with such a detector can be easily reproduced with coherent states of a light mode.
We show that a simple modification of an on-off detector---introducing controlled attenuation as a tunable setting---enables such detectors to reveal nonclassical properties of radiation fields.
\end{abstract}

\maketitle

\section{Introduction}

% Nonclassicality of quantum states

The concept of nonclassical states in quantum optics \cite{titulaer65,mandel86,mandel_book,vogel_book,agarwal_book,Schnabel2017,sperling2018a,sperling2018b,sperling2020} derives from the observation that all measurements on coherent states and their statistical mixtures can be explained within the framework of classical electrodynamics.
Technically, it is related to the Glauber-Sudarshan $P$ representation \cite{glauber63c,sudarshan63}.
It establishes that the density operator $\hat{\rho}$ of an electromagnetic field mode can be expanded by projectors of coherent states $\ket{\alpha}$,
	\begin{align}\label{Eq:P-func}
	\hat{\rho}=\int_{\mathbb{C}} d^2\alpha P(\alpha)\ket{\alpha}\!\bra{\alpha},
	\end{align}
where $P(\alpha)$ is the Glauber-Sudarshan $P$ function.
If $P(\alpha)\geq 0$, i.e. it can be interpreted as a probability distribution, then the corresponding quantum state is classical.
All states that violate this condition are considered in quantum optics as nonclassical. 
Optical nonclassicality encompasses a number of nonclassical phenomena such as photon-number \cite{mandel79} and quadrature \cite{Stoler1970, Stoler1971,Wu1986,Wu1987, Vahlbruch} squeezing and their further generalizations to higher-order nonclassical effects \cite{agarwal92}.
Moreover, a number of techniques have been developed to detect and quantify optical nonclassicality; see, e.g., Refs. \cite{reid1986,Hillery1987,Lee1991,agarwal93,klyshko1996,vogel00,richter02,Asboth2005,kiesel08,rivas2009,kiesel10,kiesel11b,sperling12c,bartley13,sperling13b,Agudelo2013,park2015a,park2015b,luis15,Miranowicz2015a,Miranowicz2015b,Yadin2019,Luo2019,Perina2020,bohmann2020,bohmann2020b,Semenov2021,Innocenti2022,Innocenti2023,Fiurasek2024,Kovtoniuk2024,Manikandan2025}.

% Nonclassicality as a resource

Optical nonclassicality is often considered in the framework of quantum resource theories \cite{Horodecki2013,Chitambar2019}.
Indeed, nonclassicality can be interpreted as a generalization of quantum coherence \cite{Streltsov2017,sperling2018a}.
In particular, nonclassicality is a resource in quantum metrology \cite{Ge2020a,Ge2020b}, quantum linear networks such as boson-sampling schemes \cite{rahimi-keshari16}, quantum kernel methods \cite{Chabaud2024}, and so on.
Importantly, optical nonclassicality is the resource for generating quantum entanglement with passive linear optical elements \cite{Asboth2005,Miranowicz2015b}.

% Nonclassicality of photocounting statistics

Experimentally accessible information about a given quantum state of light is encoded in measurement statistics.
In typical scenarios, this information is not complete to reconstruct the density operator of the studied quantum state.
Counting of photons represent a typical example of such informationally-incomplete measurements.
In this case, the photocounting probability distribution $\mathcal{P}(n)$ is given by 
    \begin{align}\label{Eq:PhotocountEq}
		\mathcal P(n)=\Tr\left[\hat{\rho}\, \hat{\Pi}(n)\right],
	\end{align}
where $\hat{\Pi}(n)$ is the positive operator-valued measure (POVM).
For ideal photon-number-resolving detection the POVM is known from the photodetection theory \cite{mandel_book, kelley64}.
In more realistic scenarios, it may differ significantly from this ideal situation \cite{sperling12a,Semenov2024,Uzunova2022,Stolyarov2023}. 

By substituting Eq.~(\ref{Eq:P-func}) into Eq.~(\ref{Eq:PhotocountEq}), we can rewrite the latter in the $P$ representation \cite{cahill69,cahill69a},
 \begin{align}\label{Eq:PhotocountEq-P}
			\mathcal P(n)=\int_{\mathbb{C}}d^2\alpha P(\alpha)\Pi(n|\alpha),
	\end{align}
where $\Pi(n|\alpha)=\bra{\alpha} \hat{\Pi}(n)\ket{\alpha}$ is the photocounting probability distribution for the coherent state $\ket{\alpha}$.
If the state $\hat{\rho}$ is classical, i.e., if $P(\alpha)\geq0$, then $\mathcal{P}(n)$ can be sampled from the statistical mixture of coherent states.
However, if we treat Eq.~(\ref{Eq:PhotocountEq-P}) as an integral equation in $P(\alpha)$ for a given $\mathcal{P}(n)$ and $\Pi(n|\alpha)$, we observe that it may yield multiple solutions due to informational incompleteness.
If at least one of these solutions is non-negative, then $\mathcal{P}(n)$ can be sampled from a statistical mixture of coherent states even if the state $\hat{\rho}$ is nonclassical~\cite{Semenov2021,Innocenti2022,Innocenti2023,Kovtoniuk2024,Fiurasek2024}.

%on-off detectors

This situation is always inherent to on-off detectors, which can only detect the presence or absence of electromagnetic radiation without distinguishing between the number of photons.
The POVM of these detectors is easily obtained from the POVM of ideal photon-number-resolving detectors \cite{mandel_book, kelley64}.
The corresponding photon-number distribution for the coherent state $\ket{\alpha}$ is given by
    \begin{align}
		&\Pi(0|\eta;\alpha)=\exp\left(-\eta|\alpha|^2\right),\label{Eq:POVM-onoff-0}\\
		&\Pi(1|\eta;\alpha)=1-\Pi(0|\eta;\alpha).\label{Eq:POVM-onoff-1}
	\end{align}
Here $n=0$ and $n=1$ correspond to the case of no click and click, respectively, and $\eta\in[0,1]$ is the detection efficiency.
Obviously, any photocounting statistics obtained from such a detector can always be reproduced by a coherent state whose amplitude satisfies $|\alpha|^2=-\ln\mathcal{P}(0)/\eta$.
This is the reason why a single on-off detector  is not considered capable of detecting nonclassicality of quantum states.

We note, however, that there are different ways to identify nonclassical features by exploiting several on-off detectors \cite{sperling12a,sperling12c,bartley13,Spring2013} and that there are also methods that are formulated in terms of the no-click probabilities \cite{Moreva2017,Obsil2018,Bohmann2019,Lachman2024, Fiurasek2024}.

% Scope of the paper

\begin{figure}[ht!]
        \centering
        \includegraphics[width=0.95\linewidth]{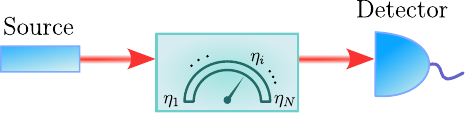}
        \caption{\label{Fig:Scheme} Scheme of the experiment for detecting nonclassicality of photocounting statistics with an on-off detector.
        The detector is preceded by an amplitude modulator, whose transmittance may take the values $\eta_i$, $i=1\ldots N$.}
    \end{figure}

In this paper, we show that a simple modification of on-off detectors makes it possible to detect nonclassicality of photocounting statistics.
Moreover, our technique works even for counterintuitive examples where the photocounting statistics obtained with ideal photon-number-resolving detectors are super-Poissonian.
The result can be achieved by treating the detection efficiency $\eta$ as a device setting that takes on a discrete, finite set of $N$ values $\eta_i$, $i=1\ldots N$, such that $\eta_1<\eta_2<\ldots<\eta_N$.
In experiments, this is easily implemented by placing an amplitude modulator in front of the detector, see Fig.~\ref{Fig:Scheme}.
In this case, the detection of nonclassicality is reformulated as a test for the impossibility of reproducing the set of no-click probabilities
    \begin{align}\label{Eq:SetOfProbabilities}
        \mathcal{P}(0|\eta_i)=\Tr\left[\hat{\rho}\, \hat{\Pi}(0|\eta_i)\right],
    \end{align}
for the state $\hat{\rho}$ with a statistical mixture of coherent states.
This formulation does not include the probabilities $\mathcal{P}(1|\eta_i)$, since they depend on $\mathcal{P}(0|\eta_i)$ according to Eq.~(\ref{Eq:POVM-onoff-1}). 

A similar task for general measurements has been considered in Refs.~\cite{Semenov2021,Kovtoniuk2024}, see also Refs.~\cite{rivas2009,Innocenti2022,Innocenti2023,Fiurasek2024} for important examples.
In particular, as discussed in Ref.~\cite{Kovtoniuk2024}, there are inequalities that tightly bound a convex set of classical probability distributions.
Violation of these inequalities is a necessary and sufficient condition for optical nonclassicality of measurement statistics and a sufficient condition for optical nonclassicality of quantum states.
In this paper, we apply this technique to reveal nonclassicality of photocounting statistics obtained with a single on-off detector having switchable detection efficiency.

The rest of the paper is organized as follows.
In Sec.~\ref{Sec:TightInequal}, we introduce the general framework for constructing tight inequalities that reveal nonclassicality with the device under consideration.
Intuitive cases with two and three settings are considered in Sec.~\ref{Sec:2and3}.
Complete sets of tight inequalities for an arbitrary number of settings are derived in Sec.~\ref{Sec:N}.
A special case involving uniformly distributed efficiencies, $\eta_i=i/N$, is considered in Sec.~\ref{Sec:Moments}.
In this section, we derive an alternative approach that leads to inequalities nonlinear in the probabilities $\mathcal{P}(0|\eta_i)$, as well as establish a relation between the no-click statistics and the photocounting statistics obtained from multiplexing schemes using several on–off detectors.
Applications of our approach to the counterintuitive example of phase-squeezed coherent states are presented in Sec.~\ref{Sec:Examples}, where we also account for statistical noise in the determination of transmittances.
A conclusion is given in Sec.~\ref{Sec:Conclusion}.

\section{Tight inequalities revealing nonclassicality}
\label{Sec:TightInequal}

In this section we recall tight inequalities for nonclassicality of measurement statistics introduced in Ref.~\cite{Kovtoniuk2024} (see also Ref.~\cite{Fiurasek2024}) and discuss the background of tailoring this method to the case considered in this paper.
It is particularly useful to use the geometric picture of this technique from the very beginning of our consideration.
Observe that all coherent-state probability distributions $\Pi(0|\eta;\alpha)$ depend on the magnitude of the coherent amplitude $\alpha$ and not on its phase.
This implies that we can employ a useful reparametrization by introducing a new parameter
	\begin{align}
		t=\exp\left(-\eta_1|\alpha|^2\right),
	\end{align}
where $\eta_1$ is the lowest efficiency among all $\eta_i$.
Then Eq.~(\ref{Eq:POVM-onoff-0}) is reduced to the form 
	\begin{align}\label{Eq:POVM_t}
		\Pi(0|\eta;t)=t^{\eta/\eta_1},
	\end{align}
where $t\in[0,1]$.
Considering $N$ settings $\eta=\eta_i$, where $i=1\ldots N$, we introduce the new vector 
    \begin{align}
        \boldsymbol{\Pi}(t)=\begin{pmatrix}
                \Pi_1(t)&\Pi_2(t)&\ldots&\Pi_N(t)
            \end{pmatrix}^\mathrm{T},
    \end{align}
composed of the components 
    \begin{align}
        \Pi_i(t)=\Pi(0|\eta_i;t)
    \end{align}
and parametrized by the parameter $t$. 
Similarly, we define the vector 
    \begin{align}
        \boldsymbol{\mathcal{P}}=\begin{pmatrix}
                \mathcal{P}_1&\mathcal{P}_2&\ldots&\mathcal{P}_N
            \end{pmatrix}^\mathrm{T},
    \end{align}
whose components 
    \begin{align}
        \mathcal{P}_i\equiv\mathcal{P}(0|\eta_i)
    \end{align}
are the no-click  probabilities given by Eq.~(\ref{Eq:SetOfProbabilities}).

The vector $\boldsymbol{\mathcal{P}}$ (the set of no-click probabilities) can be reproduced with classical electromagnetic fields if there exists $\varrho(t)\geq0$ such that
    \begin{align}\label{Eq:ConvexComb}
        \boldsymbol{\mathcal{P}}=\int_0^1 d t \varrho(t)\boldsymbol{\Pi}(t).
    \end{align}
For classical states, $\varrho(t)$ can be directly derived from a non-negative $P$ function of the state.
However, it can also exist for nonclassical states, due to informational incompleteness of the measurement procedure.

The set $\mathcal{C}=\{\boldsymbol{\Pi}(t)|t{\in}[0,1]\}$ defines a closed curve in the $N$-dimensional Euclidean space of the vectors $\boldsymbol{\mathcal{P}}$.
Then Eq.~(\ref{Eq:ConvexComb}) states that the vector $\boldsymbol{\mathcal{P}}$ can be reproduced by classical radiation fields if and only if it belongs to the convex combination of the set $\mathcal{C}$, $\mathcal{H}=\Conv \mathcal{C}$. 
Applying to this statement the hyperplane supporting theorem \cite{boyd_book} and the methods of Refs.~\cite{Semenov2021,Kovtoniuk2022,Kovtoniuk2024}, we conclude that the vector $\boldsymbol{\mathcal{P}}$ is classically treatable if and only if for any vector $\boldsymbol{\lambda}$ the inequality
    \begin{align}\label{Eq:InequalityGen}
        \boldsymbol{\lambda}\cdot\boldsymbol{\mathcal{P}}\leq
        \sup_{t\in[0,1]}\boldsymbol{\lambda}\cdot\boldsymbol{\Pi}(t)
	\end{align}
is satisfied.   
If we find at least a single instance of $\boldsymbol{\lambda}$ such that this inequality is violated, then the vector $\boldsymbol{\mathcal{P}}$ (the set of all no-click probabilities) cannot be reproduced with the statistical mixtures of coherent states.   

A straightforward way to deal with this problem may be to apply an optimization procedure to find vectors $\boldsymbol{\lambda}$ that violate inequality (\ref{Eq:InequalityGen}).
However, we can significantly reduce computational resources, by optimizing the set of $\boldsymbol{\lambda}$.
This optimal set $\Lambda$ corresponds to the set of tight inequalities \cite{Kovtoniuk2024}, which (i) can be used to derive any other inequality outside this set and (ii) are independent, i.e. no tight inequality can be derived from other tight inequalities.  
As discussed in Ref.~\cite{Kovtoniuk2024}, the vectors $\boldsymbol{\lambda}$ corresponding to tight inequalities are normal vectors to the supporting hyperplanes of the manyfold $\mathcal{H}$. 

Each supporting hyperplane should include at least one point of the curve $\mathcal{C}$.
For each value $t_1$ of the parameter $t$, we can define a set of vectors $\boldsymbol{\lambda}(t_1)$ that form closed convex cones,
    \begin{align}\label{Eq:Cond1}
        L(t_1)=\Big\{\boldsymbol{\lambda}(t_1)\Big| \argmax_{t\in[0,1]} \boldsymbol{\lambda}(t_1)\cdot\boldsymbol{\Pi}(t)=t_1\Big\}.
    \end{align}
In the considered scenario it is possible for a single value of $t_1$ to be associated with multiple values of $\boldsymbol{\lambda}(t_1)$.
Independent vectors $\boldsymbol{\lambda}(t_1)$ correspond to extreme rays $\Ext \left[L(t_1)\right]$.
Finally, the set 
    \begin{align}\label{Eq:Cond2}
        \Lambda=\bigcup_{t_1 \in [0,1]} \Ext \left[L(t_1)\right]
    \end{align}
defines all vectors $\boldsymbol{\lambda}(t_1)$ that generate tight inequalities.
The application of this method to the photocounting problem, where the vector $\boldsymbol{\mathcal{P}}$ represents a photocounting statistic, has been derived in Ref.~\cite{Kovtoniuk2024}.
The same procedure is straightforwardly generalized to the scenario where the vector $\boldsymbol{\mathcal{P}}$ represents the no-click probabilities considered in this paper.

\section{Two and three settings}
\label{Sec:2and3}

We start our analysis with the intuitively accessible cases of $N=2$ and $N=3$.
In these scenarios, the curve $\mathcal{C}$ and its convex hull $\mathcal{H}$ can be easily visualized, providing clear geometric insight into the structure of the problem.
These simple examples serve as a useful foundation for understanding the general approach developed in the following sections.
Here, we focus on an intuitive understanding of the results, while deferring a more rigorous  derivation, which is applicable to arbitrary $N$, to the next section.

\subsection{Two settings}

\begin{figure}[ht!]
        \centering
        \includegraphics[width=0.95\linewidth]{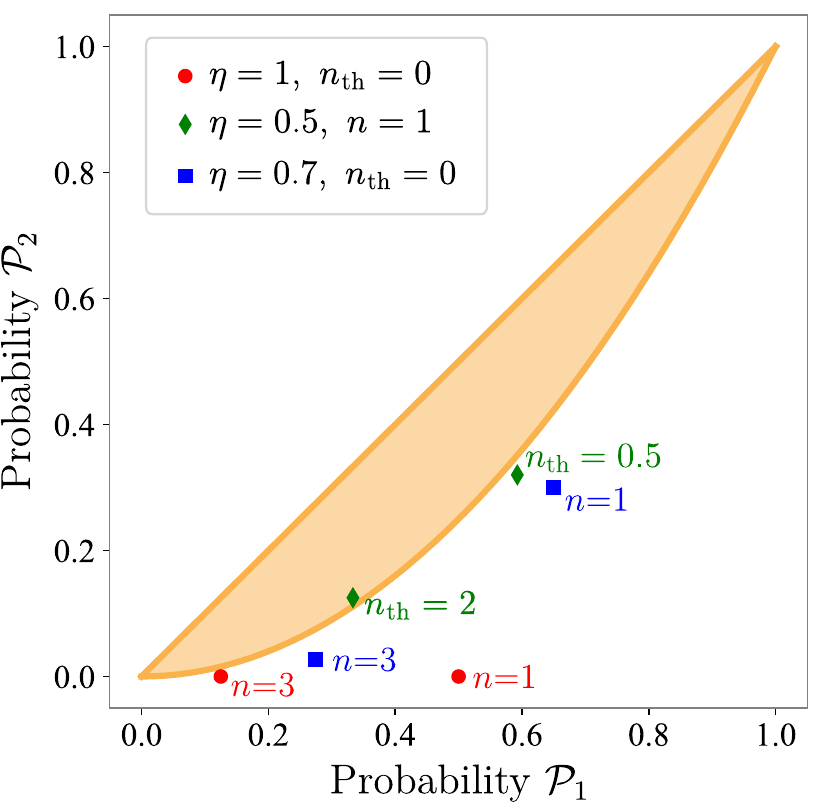}
        \caption{\label{Fig:Fock_N2} Classical region for $N=2$ settings (shaded), its boundary (solid line), and points related to the $n$-photon-added thermal states, cf. Eq.~(\ref{Eq:nPATS}),} attenuated with the efficiency $\eta_{\mathrm{c}}$.
        In this example, $\eta_1=1/2$ and $\eta_2=1$.
    \end{figure}

In the case of $N=2$, Eq.~(\ref{Eq:POVM_t}) reduces to 
	\begin{align}
		&\Pi_1(t)=t,\\
		&\Pi_2(t)=t^{\nu_2},
	\end{align}
where $\nu_2 = \eta_2/\eta_1$.
Therefore, the curve $\mathcal{C}$ is described by the equation 
    \begin{align}\label{Eq:Curve2D-1}
        \mathcal{P}_2=\mathcal{P}_1^{\nu_2},
    \end{align}
see Fig.~\ref{Fig:Fock_N2}.
Since $\nu_2>1$, its convex hull $\mathcal{H}$ is clearly the region bounded by this curve and the line $\mathcal{P}_2=\mathcal{P}_1$.

We now proceed to examine this construction in more technical detail.
According to Eq.~(\ref{Eq:Cond1}), at each point $t_1\in(0,1)$, the vectors $\boldsymbol{\lambda}(t_1)$ defining tight inequalities satisfy the equation
    \begin{align}\label{Eq:Ma
    xt1}
       \boldsymbol{\lambda}(t_1)\cdot\dot{\boldsymbol{\Pi}}(t_1)=0,
    \end{align}
where the overdot denotes differentiation.
Clearly, its solution yields the maximum of $\boldsymbol{\lambda}(t_1)\cdot\boldsymbol{\Pi}(t)$ with respect to $t$ at the point $t=t_1$.
It is given by
    \begin{align}\label{Eq:Vec2D}
		\boldsymbol{\lambda}(t_1)=\begin{pmatrix}\dot{\Pi}_2(t_1) \\-\dot{\Pi}_1(t_1)\end{pmatrix}%\nonumber\\
        =\begin{pmatrix}\nu_2 t_1^{\nu_2} \\ -1\end{pmatrix}.
	\end{align}
The vectors $\boldsymbol{\lambda}(t_1)$ are normal to the tangent line of the curve (\ref{Eq:Curve2D-1}), which are supporting hyperplanes of $\mathcal{H}$. 
Similar to the scenario described in Ref.~\cite{Kovtoniuk2024}, at the endpoints of this curve, $t=\tau\in\{0,1\}$, condition (\ref{Eq:Cond1}) defines the conic combinations of the vectors $\boldsymbol{\lambda}(\tau)$ and $\boldsymbol{\lambda}_{\nwarrow}=\begin{pmatrix} -1 & 1 \end{pmatrix}^{\mathrm{T}}$.
Therefore, according to Eq.~(\ref{Eq:Cond2}) the complete set of tight inequalities is given by the vectors $\boldsymbol{\lambda}(t)\in[0,1]$ and $\boldsymbol{\lambda}_{\nwarrow}$.

The set of linear tight inequalities can be combined into two nonlinear inequalities,
    \begin{align}
        &\mathcal{P}_1^{\nu_2}\leq\mathcal{P}_2,\label{Eq:NonlIneq2D-1}\\
        &\mathcal{P}_2\leq\mathcal{P}_1\label{Eq:NonlIneq2D-2}
    \end{align}
related to $\boldsymbol{\lambda}(t)\in[0,1]$ and $\boldsymbol{\lambda}_{\nwarrow}$, respectively. 
In fact, they define the convex set $\mathcal{H}$ of the curve $\mathcal{C}$ given by Eq.~(\ref{Eq:Curve2D-1}), cf. Fig.~\ref{Fig:Fock_N2}.
In addition, the inequality (\ref{Eq:NonlIneq2D-2}) represents the trivial fact that the no-click probability with efficiency $\eta_1$ is higher than the same probability with efficiency $\eta_2$ for $\eta_1<\eta_2$.

This elementary case of $N=2$ is already suitable to reveal nonclassicality of the Fock states $\ket{n}$ and even of the more general case of an $n$-photon-added thermal state,
    \begin{align}\label{Eq:nPATS}
        \hat{\rho}=\frac{1}{n!\left(1+n_{\mathrm{th}}\right)^{n}}\,\hat{a}^{\dag n}\,\hat{\rho}_{\mathrm{th}}\,\hat{a}^{n}.
    \end{align}
Here
    \begin{align}
        \hat{\rho}_{\mathrm{th}}=\frac{1}{1+n_{\mathrm{th}}}\left(\frac{n_{\mathrm{th}}}{1+n_{\mathrm{th}}}\right)^{\hat{n}}
    \end{align}
is a thermal state with mean photon number $n_{\mathrm{th}}$, $\hat{a}$ and $\hat{a}^{\dag}$ are the annihilation and creation operators, respectively, and $\hat{n}=\hat{a}^{\dag}\hat{a}$ is the photon-number operator.
For $n_{\mathrm{th}}=0$, the state~(\ref{Eq:nPATS}) reduces to the Fock state $\ket{n}$.
We further assume that this state undergoes attenuation with efficiency $\eta_{\mathrm{c}}$.

The no-click probability for this state is given by
    \begin{align}
        \mathcal{P}_i=\frac{\left(1-\eta_i\eta_{\mathrm{c}}\right)^{n}}{\left(1+\eta_i\eta_{\mathrm{c}}n_{\mathrm{th}}\right)^{n+1}}.
    \end{align}
Inequality~(\ref{Eq:NonlIneq2D-1}) is violated provided that $n_{\mathrm{th}}$ is below a threshold value determined by solving the corresponding equation, which is in general transcendental.
There are two cases in which inequality~(\ref{Eq:NonlIneq2D-1}) is always violated.
First, for $n_{\mathrm{th}}=0$, it reduces to $(1-\eta_1\eta_{\mathrm{c}})^{n \nu_2} \leq (1-\eta_2\eta_{\mathrm{c}})^n$, which is violated for all parameter values.
Second, for $\eta_{\mathrm{c}}=1$, the inequality~(\ref{Eq:NonlIneq2D-1}) is violated when $\eta_2=1$.
As shown in Fig.~\ref{Fig:Fock_N2}, all points related to the attenuated Fock states are placed outside the set $\mathcal{H}$, whereas for $n_{\mathrm{th}}$ exceeding the threshold value, the points lie inside it. 

\subsection{Three settings}

In the case of $N=3$, Eq.~(\ref{Eq:POVM_t}) is given by
	\begin{align}
		&\Pi_1=t,\label{Eq:Pi-3-1}\\
		&\Pi_2=t^{\nu_2},\label{Eq:Pi-3-2}\\
		&\Pi_3=t^{\nu_3},\label{Eq:Pi-3-3}
	\end{align}
where $\nu_i=\eta_i/\eta_1$ such that $1<\nu_2<\nu_3$.    
These equations parametrically define the curve $\mathcal{C}$ in a three-dimensional space.
Our task is to construct the convex hull $\mathcal{H}$ of this curve.
Appropriate mathematical methods to solve this problem are described in Refs.~\cite{Karlin1966,Krein1977,Pont2023}.
We use the idea of Ref.~\cite{Kovtoniuk2024} to obtain this result in an intuitively clear way.

\begin{figure}[ht!]
        \centering
        \includegraphics[width=0.95\linewidth]{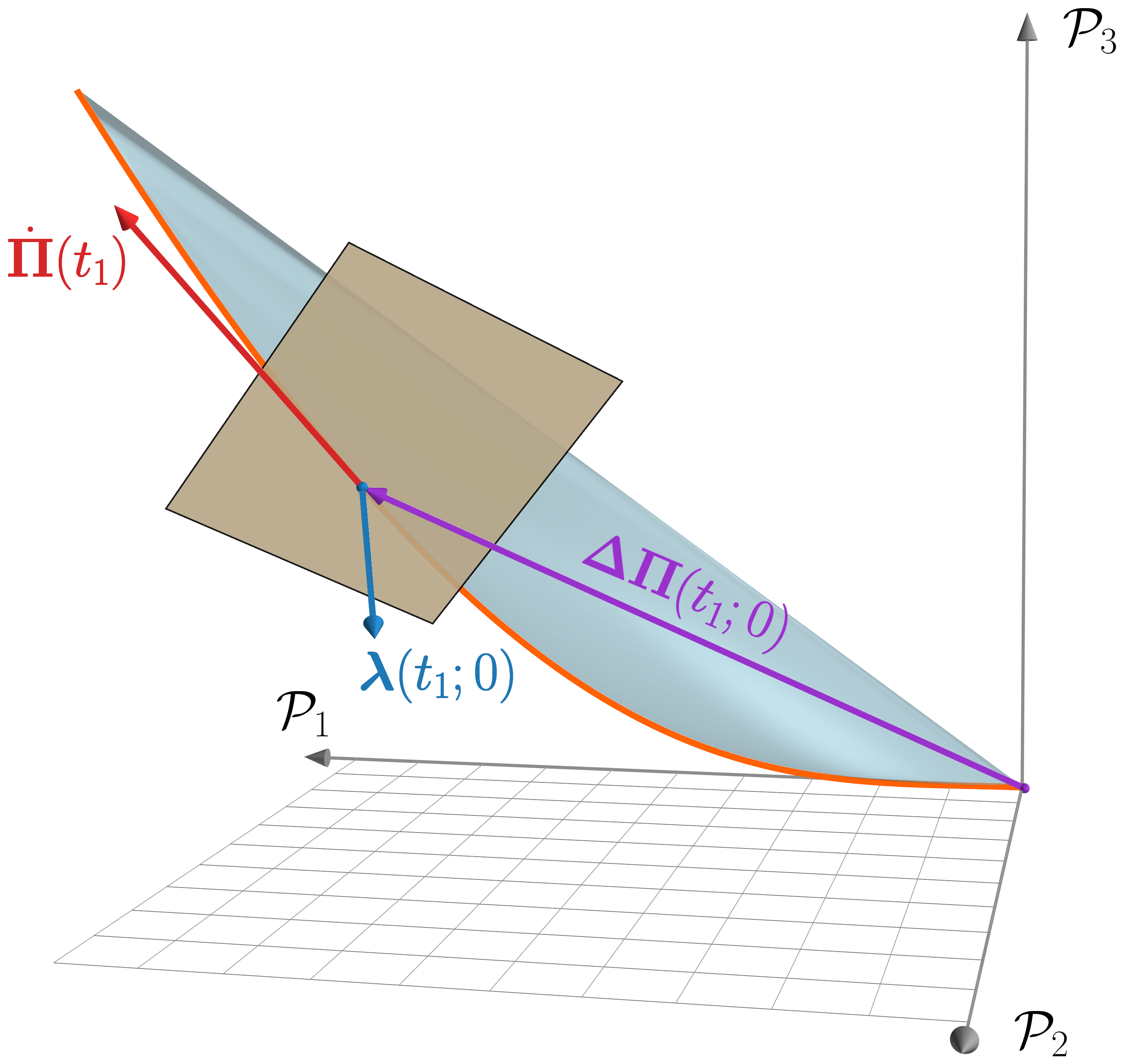}
        \caption{\label{Fig:3D} The curve $\mathcal{C}$ and its convex hull $\mathcal{H}$ for the case of $N=3$.
        Also the vectors $\boldsymbol{\lambda}(t_1;\tau)$ and the vectors $\Delta \boldsymbol{\Pi}(t_1,\tau)$ and $\dot{\boldsymbol{\Pi}}(t_1)$ defining supporting planes of $\mathcal{H}$ are shown for $\tau=0$.}
    \end{figure}

Let $t_1$ be the value of the parameter $t$ at which $\boldsymbol{\lambda}{\cdot}\boldsymbol{\Pi}(t)$ attains its maximum with respect to $t$. 
Clearly, a supporting hyperplane of $\mathcal{H}$ that contains the point $\boldsymbol{\Pi}(t_1)$ is parallel to the vector $\dot{\boldsymbol{\Pi}}(t_1)$, which is tangent to the curve $\mathcal{C}$ at $t=t_1$.
We also assume that each point $t_1$ corresponds to two supporting planes, each passing through one of the endpoints $\boldsymbol{\Pi}(\tau)$, where $\tau=0$ or $\tau=1$.
Hence, the vectors $\boldsymbol{\lambda}$ also depend on the parameter $\tau$.
These vectors, normal to the supporting plane and directed outward from $\mathcal{H}$, are given by
    \begin{align}\label{Eq:TightLambda3D}
		\boldsymbol{\lambda}(t_1;\tau) = \mathcal{N} (-1)^{1+\tau} \dot{\boldsymbol{\Pi}}(t_1) \times \Delta \boldsymbol{\Pi}(\tau, t_1),
    \end{align}
where 
    \begin{align}\label{Eq:DeltaPi}
        \Delta \boldsymbol{\Pi}(t^\prime,t^{\prime\prime})=\boldsymbol{\Pi}(t^\prime) - \boldsymbol{\Pi}(t^{\prime\prime}),
    \end{align}
see Fig.~\ref{Fig:3D}.
   
The explicit form of the vectors $\boldsymbol{\lambda}(t_1;\tau)$,
    \begin{align}
         & \boldsymbol{\lambda}(t_1;0) = \mathcal{N}
         \begin{pmatrix}
            -\left(\nu_{3} - \nu_{2}\right)t_{1}^{\nu_{2} + \nu_{3} - 1}  \\
            \left(\nu_{3}-1\right)t_{1}^{\nu_{3}} \\
           -\left(\nu_{2} - 1\right)t_{1}^{\nu_{2}} 
        \end{pmatrix}, \quad \label{Eq:TightLambda3D-Explct-1}\\
        & \boldsymbol{\lambda}(t_1;1)= \mathcal{N} \nonumber \\
        &  \times\begin{pmatrix}
            (\nu_3 - \nu_2) t_1^{\nu_3 + \nu_2 - 1} - \nu_3 t_1^{\nu_3 - 1} + \nu_2 t_1^{\nu_2 - 1} \\
           -(\nu_3 - 1) t_1^{\nu_3} + \nu_3 t_1^{\nu_3 - 1} - 1 \\
           (\nu_2 - 1) t_1^{\nu_2} - \nu_2 t_1^{\nu_2 - 1} + 1
        \end{pmatrix},\label{Eq:TightLambda3D-Explct-2}
    \end{align}
can be obtained by direct substitution of Eqs.~(\ref{Eq:Pi-3-1}), (\ref{Eq:Pi-3-2}) and (\ref{Eq:Pi-3-3}) into Eq.~(\ref{Eq:TightLambda3D}).
Here $\mathcal{N}$ is the norm of the vector $\boldsymbol{\lambda}$.
This factor becomes relevant only in the limiting cases $t_1\rightarrow 0$ and  $t_1\rightarrow 1$ for Eqs.~(\ref{Eq:TightLambda3D-Explct-1}) and (\ref{Eq:TightLambda3D-Explct-2}), respectively.
In these cases, the vectors are given by
    \begin{align}\label{Eq:TightLambda3D-Explct-1-Limit}
        \boldsymbol{\lambda}(0;0) = -
        \begin{pmatrix}
            0 \\
            0 \\
            1
        \end{pmatrix},
    \end{align}
and
        \begin{align}\label{Eq:TightLambda3D-Explct-2-Limit}
        \boldsymbol{\lambda}(1;1) = \mathcal{N}
        \begin{pmatrix}
            \nu_2 \nu_3 (\nu_3 - \nu_2) \\
            -\nu_3 (\nu_3 - 1) \\
            \nu_2 (\nu_2 - 1)
        \end{pmatrix},
    \end{align}
respectively.
These expressions are obtained by applying l'H\^{o}pital's rule.
The normalization factor $\mathcal{N}$ in Eq.~(\ref{Eq:TightLambda3D-Explct-2-Limit}) differs from that in Eq.~(\ref{Eq:TightLambda3D-Explct-2}).
In this context, it serves a purely technical purpose and is included for consistency with the examples later.

Using Eqs.~(\ref{Eq:TightLambda3D-Explct-1}) and (\ref{Eq:TightLambda3D-Explct-2}) in the general inequality (\ref{Eq:InequalityGen}) yields a set of tight inequalities,
    \begin{align}\label{Eq:IneqN3-1}
         -(\nu_3 - \nu_2) t_1^{\nu_3 + \nu_2 - 1} \mathcal{P}_1 &+ (\nu_3 - 1) t_1^{\nu_3} \mathcal{P}_2 \nonumber \\
        & - (\nu_2 - 1) t_1^{\nu_2} \mathcal{P}_3 \leq 0
    \end{align}
and
    \begin{align}\label{Eq:IneqN3-2}
        & (\nu_2 - \nu_3) t_1^{\nu_2 + \nu_3 - 1} \left( 1 - \mathcal{P}_1 \right) + \left(\nu_2 t_1^{\nu_2 - 1}-\nu_3 t_1^{\nu_3 - 1} \right) \mathcal{P}_1 \nonumber \\
        & + (\nu_3 - 1) t_1^{\nu_3} \left( 1 - \mathcal{P}_2 \right)  + \left( \nu_3 t_1^{\nu_3 - 1} - 1 \right) \mathcal{P}_2 \nonumber \\
        & - (\nu_2 - 1) t_1^{\nu_2} \left( 1 - \mathcal{P}_3 \right) - \left( \nu_2 t_1^{\nu_2 - 1} - 1 \right) \mathcal{P}_3  \leq 0.
    \end{align}
for $\tau= 0$ and $\tau = 1$, respectively.
Equation~(\ref{Eq:TightLambda3D-Explct-1-Limit}) yields the trivial inequality $\mathcal{P}_3\geq0$.
In contrast, Eq.~(\ref{Eq:TightLambda3D-Explct-2-Limit}) results in the nontrivial inequality
    \begin{align}\label{Eq:IneqN3-2-Lim}
        \nu_2 \nu_3 (\nu_3 - \nu_2) \mathcal{P}_1-\nu_3 (\nu_3 - 1)\mathcal{P}_2+\nu_2 (\nu_2 - 1)\mathcal{P}_3\nonumber \\
        \leq \nu_2 \nu_3 (\nu_3 - \nu_2)-\nu_3 (\nu_3 - 1)+\nu_2 (\nu_2 - 1),
    \end{align}
which provides a meaningful constraint on the classical set $\mathcal{H}$.

Linear inequalities~(\ref{Eq:IneqN3-1}) for all $t_1$ can be compactly combined into a single nonlinear inequality for the probabilities $\mathcal{P}_i$.
This is achieved by maximizing the left-hand side of this inequality over the interval $t_1\in[0,1]$ and requiring that the maximum remains nonpositive. 
This procedure leads to the inequality
    \begin{align}\label{Eq:IneqN3-Nonlin-1-Gen}
        \mathcal{P}_2^{\nu_3 - 1} \leq \mathcal{P}_1^{\nu_3 - \nu_2} \mathcal{P}_3^{\nu_3 - \nu_2}.
     \end{align}
For the inequalities~(\ref{Eq:IneqN3-2}), however, a similarly simple analytical reduction has not been identified. 

It is useful to consider the special case with $\eta_1=1/3$,  $\eta_2=2/3$, and  $\eta_3=1$, which corresponds to $\nu_2=2$ and $\nu_3=3$.
In this scenario, inequalities (\ref{Eq:IneqN3-1}) and (\ref{Eq:IneqN3-2}) can be interpreted as conditions for nonpositivity of quadratic trinomials of the variable $t_1$.
These inequalities hold whenever the corresponding trinomials have at least one real root or no real roots.
This results in two nonlinear inequalities
    \begin{align}\label{Eq:IneqN3-Nonlin-1}
        \mathcal{P}^2_2 \leq \mathcal{P}_1 \mathcal{P}_3
     \end{align}
and    
    \begin{align}\label{Eq:IneqN3-Nonlin-2}
         \big[\mathcal{P}_1 - \mathcal{P}_2\big]^2 \leq \big[1 - \mathcal{P}_1\big] \big[\mathcal{P}_2 - \mathcal{P}_3\big]
    \end{align}
for $\tau= 0$ and $\tau = 1$, respectively.
In this case, inequality (\ref{Eq:IneqN3-2-Lim}) reduces to 
\begin{align}
    3\mathcal{P}_1-3\mathcal{P}_2+\mathcal{P}_3\leq1,
\end{align}
which is automatically satisfied whenever inequality (\ref{Eq:IneqN3-Nonlin-2}) holds.

\section{Higher number of settings}
\label{Sec:N}

The technique described in the preceding section can be generalized to an arbitrary number of settings, $N$.
In this case, Eq.~(\ref{Eq:POVM_t}) takes the form
	\begin{align}\label{Eq:Moments}
		\Pi_i(t) = t^{\nu_i},
	\end{align}
where
	\begin{align}
		\nu_i = \frac{\eta_i}{\eta_1}.
	\end{align}
Here, $\nu_1 = 1$ and $1 < \nu_2 < \ldots < \nu_N$.
Of particular interest is the case where the discrete values $\eta_i$ are uniformly distributed over the interval $[0,1]$, i.e., $\eta_i = 1/N$.
Nevertheless, we also consider a more general situation.
A similar analysis was carried out in Ref.~\cite{Kovtoniuk2024} for the photocounting distribution with an odd-dimensional vector $\boldsymbol{\mathcal{P}}$.
In the present scenario, it is likewise convenient to distinguish between odd and even $N$, although a fully general treatment is also possible but considerably more cumbersome.

\begin{figure}
	\centering
	\includegraphics[width=1\linewidth]{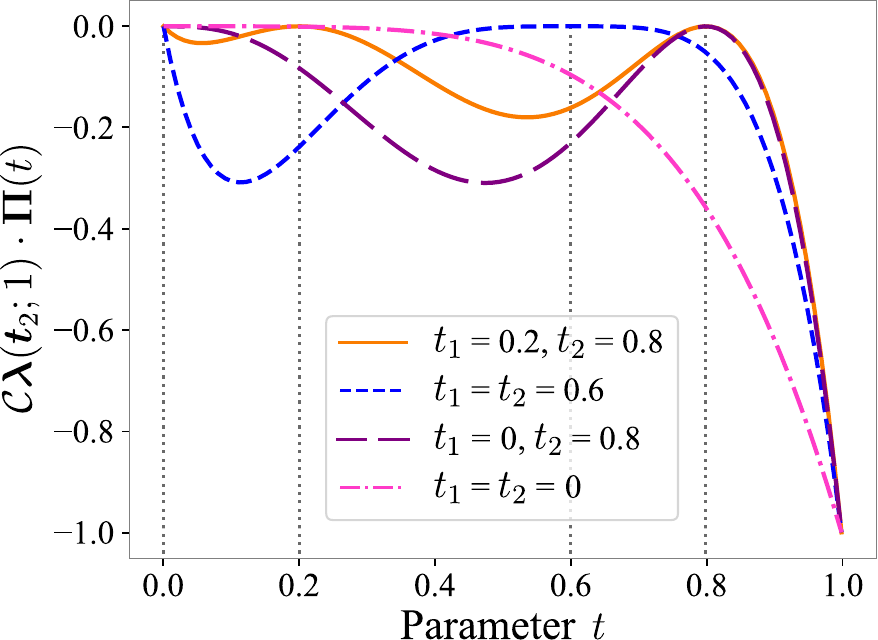}
	\caption{Function $\boldsymbol{\lambda}(\boldsymbol{t}_m; \tau)\cdot\boldsymbol{\Pi}(t)$ for $\eta_i=i/N$, $N=5$ (i.e., $m=2$), and $\tau=0$.
	The points $t = t_1$, $t = t_2$, and $t = \tau$ correspond to the global maximum.
	Depending on whether $t_1 = t_2$ or $t_2 = \tau$, the function exhibits $2^2 = 4$ different variants [see Eq.~(\ref{Eq:NumbMax-Odd})], each with a different number of maxima.
    Here, $\mathcal{C}^{-1} = \inf_{t \in [0, 1]} \boldsymbol{\lambda}(t_m;\tau) \cdot \boldsymbol{\Pi}(t)$ is a scaling factor introduced to make all plots comparable in scale.}
	\label{Fig:5DLambda}
\end{figure}

Let us start with a consideration of the scenario with an odd number of settings, $N = 2m + 1$, where $m$ is an integer.
A direct generalization of Eq.~(\ref{Eq:TightLambda3D}) is given by
	\begin{align}\label{Eq:OddPDTightLambda}
		  \boldsymbol{\lambda}(\boldsymbol{t}_m;\tau) = (-1)^{\tau+N} \mathcal{N} \left( \star \bigwedge_{k = 1}^m \Delta \boldsymbol{\Pi}(t_k,\tau) \wedge\dot{\boldsymbol{\Pi}}(t_k)  \right).
	\end{align}
Here, $\Delta \boldsymbol{\Pi}(t_k,\tau)$ is defined in Eq.~(\ref{Eq:DeltaPi}), $\tau = \pm 1$, and 
	\begin{align}\label{Eq:Vector-t_m}
		\boldsymbol{t}_m=\{t_1,\ldots,t_m\}, \quad 0 \leq t_1 \leq t_2 \leq \ldots \leq t_m \leq 1.
	\end{align}
The factor $\mathcal{N}$ denotes normalization.
The symbol $\star$ represents the Hodge star, which generalizes the cross product; see Ref.~\cite{Frankel_book} and Eq.~(\ref{Eq:Hodge}) in Appendix~\ref{Sec:Hodge}.
Applying this vector $\boldsymbol{\lambda}$, the left-hand side of the inequality (\ref{Eq:InequalityGen}) reduces to
    \begin{align}\label{Eq:LHS-Odd}
        \boldsymbol{\lambda}(\boldsymbol{t}_m; \tau) \cdot\boldsymbol{\mathcal{P}}&= \mathcal{N}
        (-1)^{\tau+1}\\
        &\times\det
       \begin{pmatrix}
       \mathcal{P}_1 & \mathcal{P}_2 & \ldots & \mathcal{P}_N  \\
       t_1-\tau & t_1^{\nu_2}-\tau & \ldots & t_1^{\nu_N}-\tau\\
       1 & \nu_2 t_1^{\nu_2 - 1} & \ldots & \nu_N t_1^{\nu_N-1}\\
       \dots & \ldots & \ldots & \ldots\\
       t_m-\tau & t_m^{\nu_2}-\tau & \ldots & t_m^{\nu_N}-\tau\\
      1 & \nu_2 t_m^{\nu_2 - 1} & \ldots & \nu_N t_m^{\nu_N-1}
       \end{pmatrix}.\nonumber
    \end{align}
The function $\boldsymbol{\lambda}(\boldsymbol{t}_m; \tau)\cdot\boldsymbol{\Pi}(t)$ attains its global maximum at all points $t = t_i$ and $t=\tau$, see Fig.~\ref{Fig:5DLambda}.
Therefore, the right-hand side of inequality (\ref{Eq:InequalityGen}) takes the form
    \begin{align}\label{Eq:SupOdd-1}
        \sup_{t \in [0, 1]} \boldsymbol{\lambda}(\boldsymbol{t}_m; \tau) &\cdot\boldsymbol{\Pi}(t)=
        \mathcal{N} (-1)^{\tau+1}\\
        &\times\det
       \begin{pmatrix}
       \tau & \tau & \ldots & \tau  \\
       t_1-\tau & t_1^{\nu_2}-\tau & \ldots & t_1^{\nu_N}-\tau\\
       1 & \nu_2 t_1^{\nu_2 - 1} & \ldots & \nu_N t_1^{\nu_N-1}\\
       \dots & \ldots & \ldots & \ldots\\
       t_m-\tau & t_m^{\nu_2}-\tau & \ldots & t_m^{\nu_N}-\tau\\
      1 & \nu_2 t_m^{\nu_2 - 1} & \ldots & \nu_N t_m^{\nu_N-1}
       \end{pmatrix}.\nonumber
    \end{align}  
In the particular case $\eta_i = i/N$, the right-hand side simplifies to
    \begin{align}\label{Eq:SupOdd-2}
        \sup_{t \in [0, 1]} \boldsymbol{\lambda}(\boldsymbol{t}_m; \tau) \cdot \boldsymbol{\Pi}(t) &=\boldsymbol{\lambda}(\boldsymbol{t}_m; \tau) \cdot \boldsymbol{\Pi}(t_i)\\
        &=\mathcal{N} \tau f(\boldsymbol{t}_m;0) f(\boldsymbol{t}_m;1) g(\boldsymbol{t}_m),\nonumber
    \end{align}
where
    \begin{align}
        & f(\boldsymbol{t}_l;a) = \prod_{k=1}^l (t_k-a)^2,\label{Eq:f} \\
        & g(\boldsymbol{t}_l) = \prod_{i < j}^l (t_i - t_j)^4.\label{Eq:g}
    \end{align}
For a detailed derivation and a proof that the corresponding inequalities form a complete set of tight inequalities, see Appendix~\ref{Sec:Proof}.
An intuitive explanation of this form of $\boldsymbol{\lambda}$ is presented at the end of this section together with the case of even $N$.

Strictly speaking, the vector $\boldsymbol{\lambda}$ in the form of Eq.~(\ref{Eq:OddPDTightLambda}) can be directly applied only when $0\leq t_1<t_2<\ldots<t_m\leq1$.
Indeed, if $t_i = t_{i+1}$ for at least one index $i$, the determinant in this equation vanishes, while the normalization factor $\mathcal{N}$ diverges.
In such cases, the indeterminacy must be resolved using l’H\^{o}pital’s rule.
The general expression for $\boldsymbol{\lambda}$, which also covers the cases with coinciding $t_i$, is given by
\begin{align}\label{Eq:LambdaOddGen}
    & \boldsymbol{\lambda}(\boldsymbol{t}_m;\tau) = (-1)^{\tau + 1} \mathcal{N} \nonumber \\
    &  \times \star \left( \bigwedge_{k \in \mathcal{U}(\boldsymbol{t}_m)} \bigwedge_{j = \delta_{t_k, \tau}}^{2 \omega(t_k) - 1 + \delta_{t_k, \tau}} \frac{\partial^j}{\partial t_k^j} \boldsymbol{\Delta \Pi}(t_k,\tau) \right),
\end{align}
where $\mathcal{U}(\boldsymbol{t}_m)$ is the set of indices which correspond to unique elements of $\boldsymbol{t}_m$, $\omega(t_k)$ is the number of times $t_k$ is included in $\boldsymbol{t}_m$.
The normalization factor $\mathcal{N}$ is defined separately for each such configuration.

The number of different configurations of $\boldsymbol{\lambda}(\boldsymbol{t}_m;\tau)$ can be determined as follows.
First, let $q = |\mathcal{U}(\boldsymbol{t}_m)|$ denote the number of unique elements in $\boldsymbol{t}_m$.
It can take values $q = 1, \ldots, m$.
Second, note that $\sum_{k \in \mathcal{U}(\boldsymbol{t}_m)} \omega(t_k) = m$.
Third, for a fixed $q$, the set $\boldsymbol{t}_m$ can be partitioned into $q$ groups of equal elements $t_i$, with group sizes given by $\omega(t_k)$.
The number of distinct partitions of $\boldsymbol{t}_m$ into $q$ groups is $\binom{m-1}{q-1}$.
Moreover, for given $q$ and $\omega(t_k)$, there exist two types of $\boldsymbol{\lambda}(\boldsymbol{t}_m;\tau)$ depending on whether any component of $\boldsymbol{t}_m$ equals $\tau$.
Thus, for a given $\tau$, there are
	\begin{align}\label{Eq:NumbMax-Odd}
		\sum_{q=1}^m 2 \binom{m-1}{q-1} = 2^m
	\end{align}
different variants of $\boldsymbol{\lambda}(\boldsymbol{t}_m;\tau)$, each having $q$ or $q+1$ global maximum points depending on whether any component of $\boldsymbol{t}_m$ equals $\tau$; see Fig.~\ref{Fig:5DLambda}.

Let us now consider the case of an even number of settings, $N = 2m$, with $m \geq 2$.
In this case, geometrical arguments similar to those presented at the end of this section lead to two types of vectors $\boldsymbol{\lambda}$,
    \begin{align}\label{Eq:EvenPDTightLambda}
		\boldsymbol{\lambda}(\boldsymbol{t}_m) = (-1)^{N-1} \mathcal{N} \left( \star \bigwedge_{k = 2}^m \Delta \boldsymbol{\Pi}(t_k,t_1) \wedge \dot{\boldsymbol{\Pi}}(t_k)\right) \wedge \dot{\boldsymbol{\Pi}}(t_1)
	\end{align}	
and	
	\begin{align}\label{Eq:EvenPDTightLambda-2}
       & \boldsymbol{\lambda}(\boldsymbol{t}_{m-1};0,1) =\nonumber\\
         & (-1)^{N} \mathcal{N}\left( \star \bigwedge_{k = 1}^{m-1} \Delta \boldsymbol{\Pi}(t_k, 0) \wedge \dot{\boldsymbol{\Pi}}(t_k)\right) \wedge \Delta \boldsymbol{\Pi}(1, 0),
    \end{align}
where $\boldsymbol{t}_m$ is defined by Eq.~(\ref{Eq:Vector-t_m})    
A proof that these vectors form a complete set of tight inequalities is given in Appendix~\ref{Sec:Proof}.

\begin{figure}[ht!]
		\centering
	    \includegraphics[width=1\linewidth]{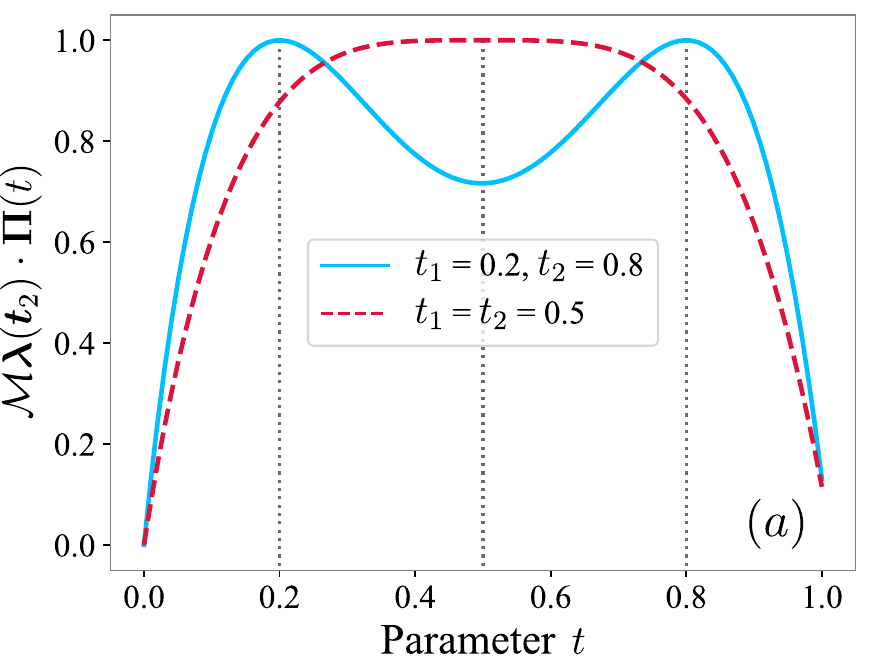}\\[2ex]
	    \includegraphics[width=1\linewidth]{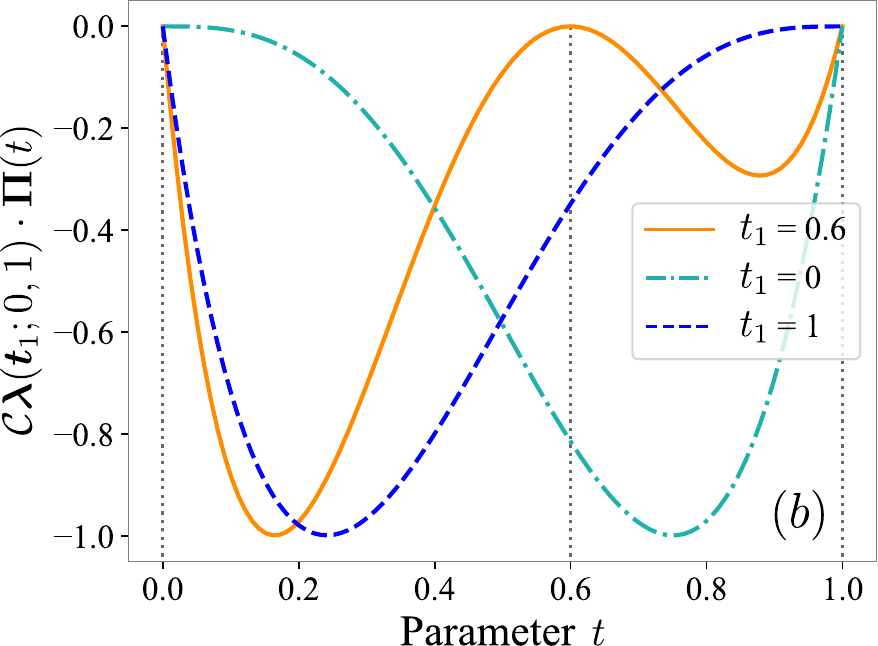}
		\caption{\label{Fig:4DLambda} Functions (a) $\boldsymbol{\lambda}(\boldsymbol{t}_m)\cdot\boldsymbol{\Pi}(t)$ and (b) $\boldsymbol{\lambda}(\boldsymbol{t}_{m-1};0,1)\cdot\boldsymbol{\Pi}(t)$ for $\eta_i = i/N$ with $N=4$ (i.e., $m=2$).
		For case (a), the points $t = t_1$ and $t = t_2$ correspond to the global maximum.
		For case (b), the points $t=0$, $t = t_1$ and $t=1$ correspond to the global maximum.
		Depending on whether $t_1 = t_2$, the function exhibits two distinct variants in case (a), and $2^3-1=7$ variants in case (b), where the cases $t_i=0$ and $t_i=1$ are also included.
        Here, $\mathcal{M}^{-1} = \sup_{t \in [0, 1]} \boldsymbol{\lambda}(t_m) \cdot \Pi(t)$ and $\mathcal{C}^{-1} = \inf_{t \in [0, 1]} \boldsymbol{\lambda}(t_{m-1};0,1) \cdot \Pi(t)$ are scaling factors introduced to make all plots comparable in scale.}
	\end{figure}

First, we analyze in detail inequality (\ref{Eq:InequalityGen}) associated with $\boldsymbol{\lambda}$ given in Eq.~(\ref{Eq:EvenPDTightLambda}).
Its left-hand side is
    \begin{align}\label{Eq:LHS-Even-1}
        \boldsymbol{\lambda}(\boldsymbol{t}_m) \cdot\boldsymbol{\mathcal{P}}= \mathcal{N}
        \det
       \begin{pmatrix}
       \mathcal{P}_1 & \mathcal{P}_2 & \ldots & \mathcal{P}_N  \\
       t_2-t_1 & t_2^{\nu_2}-t_1^{\nu_2} & \ldots & t_2^{\nu_N}-t_1^{\nu_N}\\
       1 & \nu_2 t_2^{\nu_2 - 1} & \ldots & \nu_N t_2^{\nu_N-1}\\
       \dots & \ldots & \ldots & \ldots\\
       t_m-t_1 & t_m^{\nu_2}-t_1^{\nu_2} & \ldots & t_m^{\nu_N}-t_1^{\nu_N}\\
       1 & \nu_2 t_m^{\nu_2 - 1} & \ldots & \nu_N t_m^{\nu_N-1}\\
       1 & \nu_2 t_1^{\nu_2 - 1} & \ldots & \nu_N t_1^{\nu_N-1}
       \end{pmatrix}.
    \end{align}
Since the function $\boldsymbol{\lambda}(\boldsymbol{t}_m)\cdot \boldsymbol{\Pi}(t)$ attains its global maximum at all points $t = t_i$ (see Fig.~\ref{Fig:4DLambda}a), the right-hand side is
    \begin{align}\label{Eq:SupEven-1}
         \sup_{t \in [0, 1]}& \boldsymbol{\lambda}(\boldsymbol{t}_m) \cdot\boldsymbol{\Pi}(t)= \nonumber \\
        & \mathcal{N}
        \det
       \begin{pmatrix}
       t_1 & t_1^{\nu_2} & \ldots & t_1^{\nu_N}  \\
       t_2-t_1 & t_2^{\nu_2}-t_1^{\nu_2} & \ldots & t_2^{\nu_N}-t_1^{\nu_N}\\
       1 & \nu_2 t_2^{\nu_2 - 1} & \ldots & \nu_N t_2^{\nu_N-1}\\
       \dots & \ldots & \ldots & \ldots\\
       t_m-t_1 & t_m^{\nu_2}-t_1^{\nu_2} & \ldots & t_m^{\nu_N}-t_1^{\nu_N}\\
       1 & \nu_2 t_m^{\nu_2 - 1} & \ldots & \nu_N t_m^{\nu_N-1}\\
       1 & \nu_2 t_1^{\nu_2 - 1} & \ldots & \nu_N t_1^{\nu_N-1}
       \end{pmatrix}.
    \end{align}
For the particular case $\eta_i = i/N$, the right-hand side simplifies to
    \begin{align}\label{Eq:SupEven}
        \sup_{t \in [0, 1]} \boldsymbol{\lambda}(\boldsymbol{t}_m) \cdot \boldsymbol{\Pi}(t)=\boldsymbol{\lambda}(\boldsymbol{t}_m; \tau) \cdot \boldsymbol{\Pi}(t_i) = f(\boldsymbol{t}_m;0) g(\boldsymbol{t}_m),
    \end{align}
where $f(\boldsymbol{t}_m;0)$ and $g(\boldsymbol{t}_m)$ are defined in Eqs.~(\ref{Eq:f}) and (\ref{Eq:g}), respectively.
 
Similar to the case of odd $N$, Eq.~(\ref{Eq:EvenPDTightLambda}) cannot be directly applied when $t_i = t_{i+1}$ for at least one value of $i$.
Resolving this indeterminacy using l’H\^{o}pital’s rule leads to the expression
\begin{align}\label{Eq:LambdaEvenGen-1}
    \boldsymbol{\lambda}(\boldsymbol{t}_m) = 
    (-1)^{N-1} \mathcal{N} \star &\left( \bigwedge_{k \in \mathcal{U}(\boldsymbol{t}_m)} \bigwedge_{j = \delta_{t_k, t_1}}^{2 \omega(t_k) - 1} \frac{\partial^j}{\partial t_k^j} \Big[\boldsymbol{\Pi}(t_k) \right. \nonumber \\
    & \left. - (1 - \delta_{t_k,t_1}) \boldsymbol{\Pi}(t_1)\Big] \vphantom{\bigwedge_{k \in \mathcal{U}(\boldsymbol{t}_m)}} \right),
\end{align}
where $\mathcal{U}(\boldsymbol{t}_m)$ and $\omega(t_k)$ are defined as after Eq.~(\ref{Eq:LambdaOddGen}).
The total number of distinct configurations is $2^{m-1}$.

Second, we analyze in detail the inequality associated with $\boldsymbol{\lambda}(\boldsymbol{t}_{m-1};0,1)$ given by Eq.~(\ref{Eq:EvenPDTightLambda-2}).
In this case, the left-hand side reads
    \begin{align}\label{Eq:LHS-Even}
        & \boldsymbol{\lambda}(\boldsymbol{t}_{m-1};0,1) \cdot\boldsymbol{\mathcal{P}}= \nonumber \\
        & (-1)^{\tau + N} \mathcal{N}
        \det
       \begin{pmatrix}
       \mathcal{P}_1 & \mathcal{P}_2 & \ldots & \mathcal{P}_N  \\
       t_1 & t_1^{\nu_2} & \ldots & t_1^{\nu_N}\\
        1 & \nu_2 t_1^{\nu_2 - 1} & \ldots & \nu_N t_1^{\nu_N-1}\\
       \dots & \ldots & \ldots & \ldots\\
       t_{m-1} & t_{m-1}^{\nu_2} & \ldots & t_{m-1}^{\nu_N}\\
        1 & \nu_2 t_{m-1}^{\nu_2 - 1} & \ldots & \nu_N t_{m-1}^{\nu_N-1}\\
       1 & 1 & \ldots & 1
       \end{pmatrix}.
    \end{align}
The global maximum of $\boldsymbol{\lambda}(\boldsymbol{t}_{m-1};0,1) \cdot \boldsymbol{\Pi}(t)$ is attained at the points $t = t_i$ ($i=1,\ldots,m$), $t=0$, and $t=1$, and in each case the value is zero, see Fig.~\ref{Fig:4DLambda}b.
Thus, the right-hand side is given by
    \begin{align}\label{Eq:SupEven-2}
        \sup_{t \in [0, 1]} \boldsymbol{\lambda}(\boldsymbol{t}_{m-1};0,1) \cdot \boldsymbol{\Pi}(t) = 0.
    \end{align}
The general expression, which also covers the cases of equal $t_i$ and $t_{i+1}$ and requires resolving the indeterminacy using l’H\^{o}pital’s rule, is
\begin{align}\label{Eq:LambdaEvenGen-2}
    & \boldsymbol{\lambda}(\boldsymbol{t}_{m-1};0, 1) = (-1)^{\tau + N} \mathcal{N} \star \left( \vphantom{\bigwedge_{k \in \mathcal{U}(\boldsymbol{t}_m)}} \boldsymbol{\Delta \Pi}(1, 0) \right. \nonumber \\
    & \left. \wedge \bigwedge_{k \in \mathcal{U}(\boldsymbol{t}_m)} \bigwedge_{j = \delta_{t_k, 0} + \delta_{t_k, 1}}^{2 \omega(t_k) - 1 + \delta_{t_k, 0} + \delta_{t_k, 1}} \frac{\partial^j}{\partial t_k^j} \boldsymbol{\Delta \Pi}(t_k,\tau) \right),
\end{align}
The total number of distinct configurations is $2^{m+1}-1$.

The vectors $\boldsymbol{\lambda}$ given by Eqs.~(\ref{Eq:OddPDTightLambda}), (\ref{Eq:EvenPDTightLambda}), and (\ref{Eq:EvenPDTightLambda-2}) form the set of tight inequalities defined by Eqs.~(\ref{Eq:Cond1}) and (\ref{Eq:Cond2}).
A detailed proof of this fact is provided in Appendix~\ref{Sec:Proof}.
Despite the different explicit forms of $\boldsymbol{\lambda}$ for odd and even $N$, the underlying reasoning is structurally the same.
To give an intuitive explanation, we first note that in both cases
\begin{align}
\boldsymbol{\lambda}\cdot\dot{\boldsymbol{\Pi}}(t_i) = 0
\end{align}
for $i=1,\ldots,m$.
This implies that the vector $\boldsymbol{\lambda}$ is orthogonal to all vectors $\dot{\boldsymbol{\Pi}}(t_i)$.
Consequently, the points $t_i$ satisfy the conditions for which the function $\dot{\boldsymbol{\Pi}}(t)\cdot\boldsymbol{\lambda}$ reaches its extrema, as required by Eq.~(\ref{Eq:Cond1}).
Next we observe that
\begin{align}
\boldsymbol{\lambda}\cdot\Delta\boldsymbol{\Pi}(t_i,t_k) = 0,
\end{align}
where we additionally set $t_{0,m+1} = \tau \in {0,1}$ for odd $N$.
Thus, the vector $\boldsymbol{\lambda}$ is also orthogonal to the vectors $\Delta\boldsymbol{\Pi}(t_i,t_k)$.
Using Eq.~(\ref{Eq:DeltaPi}), we conclude that $\boldsymbol{\lambda}\cdot\boldsymbol{\Pi}(t)$ takes the same value for all $t = t_i$, including $k=0,m+1$ in the odd-$N$ case.
Combining this observation with the considerations of Ref.~\cite{Kovtoniuk2024} (see also Appendix~\ref{Sec:Proof}), we conclude that these points are indeed global maxima of the functions $\boldsymbol{\lambda}\cdot\boldsymbol{\Pi}(t)$.
Their values are given explicitly by Eqs.~(\ref{Eq:SupOdd-1}) and (\ref{Eq:SupEven}).
Examples of such functions are shown in Figs.~\ref{Fig:5DLambda} and \ref{Fig:4DLambda}.

\section{Uniformly distributed efficiencies}
\label{Sec:Moments}

\subsection{Nonlinear inequalities}
\label{Sec:Moments_NI}

In this section we focus on the special case of uniformly distributed efficiencies, i.e., $\eta_i = i/N$.
In this setting, Eq.~(\ref{Eq:ConvexComb}) can be rewritten as
	\begin{align}\label{Eq:Moments-Int}
		\mathcal{P}_i = \int_0^1 dt \varrho(t) t^{i}.
	\end{align}
This implies that the no-click probabilities $\mathcal{P}_i = \mathcal{P}(0|\eta_i)$ can be simulated with classical radiation if and only if they can be treated as the moments of a probability distribution $\varrho(t)$ with $t \in [0,1]$.
Here we consider $i = 0,\ldots,N$, which ensures that $\mathcal{P}_0 = \mathcal{P}(0|\eta_0) = 1$.

Now we can apply a necessary and sufficient condition for the existence of a distribution $\varrho(t)$ whose statistical moments are given by the values $\mathcal{P}_i$, namely the Hausdorff moment problem.
According to Ref.~\cite{Krein1977}, this condition holds if and only if two quadratic forms, indexed by $k=1,2$,
\begin{align}\label{Eq:QuadraticForm}
F^{(k)}(\boldsymbol{x})=\sum_{i,j=0}^{m_{\mathrm{max}}}M^{(k)}_{i,j}x_i x_j,
\end{align}
are non-negative.
Here, $\boldsymbol{x} = \{x_1,\ldots,x_N\}$ with $x_i \in \mathbb{R}$, and $M^{(k)}_{i,j}$ together with $m_{\mathrm{max}}$ are defined as
	\begin{align}
		&M^{(1)}_{i,j}= \mathcal{P}_{i + j}, \quad m_{\mathrm{max}} = m,\\
		&M^{(2)}_{i,j}= \mathcal{P}_{i + j} - \mathcal{P}_{i + j + 2},  \quad m_{\mathrm{max}} = m - 1.
	\end{align}
for $N=2m$ and
	\begin{align}
		&M^{(1)}_{i,j}= \mathcal{P}_{i + j + 1}, \quad m_{\mathrm{max}} = m,\label{Eq:DeterOdd-1}\\
		&M^{(2)}_{i,j}= \mathcal{P}_{i + j} - \mathcal{P}_{i + j + 1}, \quad m_{\mathrm{max}} = m.\label{Eq:DeterOdd-2}
	\end{align}
for $N=2m+1$.
Since $F^{(k)}(\boldsymbol{x})$ is linear with respect to $\mathcal{P}_i$, the inequalities $F^{(k)}(\boldsymbol{x}) \geq 0$ can be considered as a special case of the inequality~(\ref{Eq:InequalityGen}), with $\boldsymbol{\lambda}$ parametrized by $\boldsymbol{x} \in \mathbb{R}^N$ and the right-hand side equal to zero.

According to Sylvester’s criterion, the quadratic form (\ref{Eq:QuadraticForm}) is non-negative if and only if all its principal minors are non-negative.
This allows us, for a given $N$, to formulate a set of nonlinear inequalities associated with minors of different orders.
Applying these inequalities requires significantly fewer computational resources compared with the optimization over $\boldsymbol{t}_m$ and $\boldsymbol{t}_{m-1}$ needed for the inequalities discussed in Sec.~\ref{Sec:N}.
The drawback of this approach is that it applies only to the case $\eta_i=i/N$.

Let us consider two simple examples.
For $N=2$ and $N=4$, only the second-order minors are involved.
They are given by
\begin{align}
\det
\begin{pmatrix}
M^{(k)}_{0, 0} & M^{(k)}_{0, 1} \\
M^{(k)}_{1, 0} & M^{(k)}_{1, 1}
\end{pmatrix} = M^{(k)}_{0, 0} M^{(k)}_{1, 1}- M^{(k)2}_{1,0}\geq 0.
\end{align}
In the case of two settings, $N=2$, we obtain two inequalities: inequality (\ref{Eq:NonlIneq2D-1}) for $\nu_2=2$ and the trivial inequality $\mathcal{P}_2\leq 1$, corresponding to $k=1$ and $k=2$, respectively.
The inequality (\ref{Eq:NonlIneq2D-2}) represents a trivial statement.
For three settings ($N=3$), we again obtain two inequalities: (\ref{Eq:IneqN3-Nonlin-1}) and (\ref{Eq:IneqN3-Nonlin-2}), corresponding to $k=1$ and $k=2$, respectively.

More involved examples with $N=4$ and $N=5$ lead to the inequalities
    \begin{align}
        \begin{pmatrix}
            M^{(k)}_{0, 0} & M^{(k)}_{0, 1} & M^{(k)}_{0, 2} \\
            M^{(k)}_{1, 0} & M^{(k)}_{1, 1} & M^{(k)}_{1, 2} \\
            M^{(k)}_{2, 0} & M^{(k)}_{2, 1} & M^{(k)}_{2, 2}
        \end{pmatrix}\geq 0.
    \end{align}
This condition implies the non-negativity of all principal minors of the $3\times 3$ matrix.
Equivalently, one may require that all three eigenvalues of the matrix are non-negative, for each $k$.

\subsection{Relation to multiplexed detection schemes}

In this section, we demonstrate the relation between our method and a photodetection technique based on multiplexed measurement schemes using several on–off detectors.
This scheme relies on balanced splitting of the signal field into $N$ modes, followed by detection of each output with an on–off detector; see Refs.~\cite{paul1996,castelletto2007,schettini2007,blanchet08,achilles03,fitch03,rehacek03}.
The measurement outcome in this scheme is the number of triggered detectors.
Assigning all detection losses to the quantum state of light, the probability distribution of observing $n$ clicks for a coherent state $\ket{\alpha}$ reads (see Ref.~\cite{sperling12a})
 	\begin{align}\label{Eq:ArrayDetectorsPOVM}
		\Pi(n|\alpha) = \binom{N}{n} e^{-(N - n)|\alpha|^2/N} \left( 1 - e^{-|\alpha|^2 / N} \right)^n.
	\end{align}
Here $n=0,\ldots,N$; however, the probability $\Pi(N|\alpha)$ is not independent due to the normalization condition and can therefore be omitted.
Accordingly, the photocounting statistics can be represented by the vector $\boldsymbol{\mathcal{P}}^{\textrm{(mp)}}$ with the components $\mathcal{P}^{\textrm{(mp)}}_n\equiv\mathcal{P}(n)$, $n=0,\ldots, N{-}1$, defined via Eq.~(\ref{Eq:PhotocountEq-P}), where the corresponding POVM is specified by Eq.~(\ref{Eq:ArrayDetectorsPOVM}).

The vector $\boldsymbol{\mathcal{P}}^{\textrm{(mp)}}$ is uniquely related to the vector $\boldsymbol{\mathcal{P}}$ describing the no-click statistics considered in this paper via
    \begin{align}
        \mathcal{P}^{\textrm{(mp)}}_n=\sum\limits_{k=1}^{N} T_{nk}\mathcal{P}_k,
    \end{align}
where the matrix elements
    \begin{align}
       T_{nk}= (-1)^{k-n+N}\binom{N}{k}\binom{k}{N-n}
    \end{align}
are obtained by expanding $\Pi(n|\alpha)$ in Eq.~(\ref{Eq:ArrayDetectorsPOVM}) in powers of $e^{-|\alpha|^2 / N}$.
Thus, the two types of statistics are equivalent in the sense that they contain the same information and share the same properties.
In particular, nonclassicality of the no-click statistics implies nonclassicality of the click statistics obtained from multiplexing detection schemes, and vice versa.
In the limit $N\rightarrow +\infty$, the statistics of the multiplexed schemes approaches that of photon-number-resolving detectors; see Ref.~\cite{sperling12a}.
This further implies that a single on–off detector with tunable efficiency, as considered in this work, can in principle access information corresponding to arbitrarily high photon-number resolution by increasing the number of efficiency settings.

The equivalence established above allows us to translate the nonlinear classicality conditions derived in Sec.~\ref{Sec:Moments_NI} directly into conditions on the click statistics $\boldsymbol{\mathcal{P}}^{\textrm{(mp)}}$ produced by multiplexing onto $N$ detection bins.
Since the matrix $T_{nk}$ in Eq.~(\ref{Eq:ArrayDetectorsPOVM}) is invertible, the no-click probabilities $\mathcal{P}_k$ can be expressed as linear combinations of the click probabilities $\mathcal{P}^{\textrm{(mp)}}_n$,
    \begin{align}\label{Eq:NoClickFromArray}
        \mathcal{P}_k=\sum_{n=0}^{N-k}\binom{N-k}{n}\binom{N}{n}^{-1}\mathcal{P}^{\textrm{(mp)}}_n,
    \end{align}
where $k=0,\ldots,N$. Substituting Eq.~(\ref{Eq:NoClickFromArray}) into the matrices $M^{(k)}_{i,j}$ recasts the quadratic forms~(\ref{Eq:QuadraticForm}) in terms of $\boldsymbol{\mathcal{P}}^{\textrm{(mp)}}$, and classicality conditions are then obtained by requiring these forms to be non-negative, which by Sylvester's criterion yields determinantal inequalities, analogously to the case of a single on-off detector with uniform efficiency.

Carrying out the substitution explicitly, for $N=2m$ the matrix elements~(\ref{Eq:QuadraticForm}) take the form
    \begin{align}\label{Eq:EvenArrayNonlinear1}
        M^{(1)}_{i,j}=\sum_{n=0}^{N-i-j}\binom{N-i-j}{n}\binom{N}{n}^{-1}\mathcal{P}^{\textrm{(mp)}}_n,
    \end{align}
and
    \begin{align}\label{Eq:EvenArrayNonlinear2}
        M^{(2)}_{i,j}=\sum_{n=1}^{N-i-j-1}\binom{N-i-j-2}{n-1}\binom{N}{n}^{-1}\mathcal{P}^{\textrm{(mp)}}_n,
    \end{align}
while for odd $N=2m+1$ they read
    \begin{align}\label{Eq:OddArrayNonlinear1}
        M^{(1)}_{i,j}=\sum_{n=0}^{N-i-j-1}\binom{N-i-j-1}{n}\binom{N}{n}^{-1}\mathcal{P}^{\textrm{(mp)}}_n,
    \end{align}
and
    \begin{align}\label{Eq:OddArrayNonlinear2}
        M^{(2)}_{i,j}=\sum_{n=1}^{N-i-j}\binom{N-i-j-1}{n-1}\binom{N}{n}^{-1}\mathcal{P}^{\textrm{(mp)}}_n.
    \end{align}
The matrix elements in Eqs.~(\ref{Eq:EvenArrayNonlinear2}) and~(\ref{Eq:OddArrayNonlinear2}) follow from simplifying the difference $\mathcal{P}_k - \mathcal{P}_{k+1}$ by means of Pascal's identity,
\begin{align}
    \binom{N-k}{n}=\binom{N-k-1}{n}+\binom{N-k-1}{n-1}.
\end{align}
The nonlinear inequalities derived in Ref.~\cite{Kovtoniuk2024} for $N=2$,
    \begin{align}\label{Eq:ArrayN2Tight}
        \left[2\mathcal{P}^{\textrm{(mp)}}_0+\mathcal{P}^{\textrm{(mp)}}_1\right]^2-4\mathcal{P}^{\textrm{(mp)}}_0\leq 0,
    \end{align}
and for $N=3$,
    \begin{align}\label{Eq:ArrayN3Lower}
        \left[\mathcal{P}^{\textrm{(mp)}}_1\right]^2\leq 3\,\mathcal{P}^{\textrm{(mp)}}_0\,\mathcal{P}^{\textrm{(mp)}}_2,
    \end{align}
    \begin{align}\label{Eq:ArrayN3Upper}
        & 3\!\left[\mathcal{P}^{\textrm{(mp)}}_1\right]^2\!\!+\!\left[\mathcal{P}^{\textrm{(mp)}}_2\right]^2\!\! \nonumber \\
        & §+3\,\mathcal{P}^{\textrm{(mp)}}_1\!\left[\mathcal{P}^{\textrm{(mp)}}_0\!+\mathcal{P}^{\textrm{(mp)}}_2\!-1\right]\!\leq\! 0,
    \end{align}
are particular cases of the requirement that all principal minors Eqs.~(\ref{Eq:EvenArrayNonlinear1})--(\ref{Eq:OddArrayNonlinear2}) are non-negative.

\section{Example: phase-squeezed coherent states}
\label{Sec:Examples}

In this section we present an example that demonstrates the applicability of our method.
Before addressing a specific state, we briefly outline the methodology for practical applications of the derived inequalities.
The simplest scenario corresponds to uniformly distributed efficiencies, $\eta_i = i/N$, considered in Sec.~\ref{Sec:Moments}.
In this case, it is sufficient to verify the positive semidefiniteness of the matrices $M^{(k)}_{i,j}$ for $k=1,2$, either by applying Sylvester’s criterion or by checking the non-negativity of their eigenvalues.

In the general case of arbitrary $\boldsymbol{\eta}=\{\eta_1,\ldots,\eta_N\}$, we need to determine the maximal violation of inequality~(\ref{Eq:InequalityGen}), which is given by
\begin{align}
V\left(\boldsymbol{\eta};\boldsymbol{\mathcal{P}}\right)=\max_{\boldsymbol{t}_l, T}\big[\boldsymbol{\lambda}(\boldsymbol{t}_l; T)\cdot\boldsymbol{\mathcal{P}}-\boldsymbol{\lambda}(\boldsymbol{t}_l; T)\cdot\boldsymbol{\Pi}(t_i)\big].
\end{align}
Here, $l=m$ or $l=m-1$, and $T$ can take the values $\tau=0$, $\tau=1$, $\varnothing$, or ${0,1}$, depending on which vector $\boldsymbol{\lambda}$ is used.
To compute this maximum, we first consider the vectors $\boldsymbol{\lambda}$ given by Eq.~(\ref{Eq:OddPDTightLambda}) for odd $N$, and by Eqs.~(\ref{Eq:EvenPDTightLambda}) and (\ref{Eq:EvenPDTightLambda-2}) for even $N$.
If the optimization yields a vector $\boldsymbol{t}_l$ such that $t_{i+1}-t_i<\varepsilon$ for at least one index $i$ (with $\varepsilon$ a fixed small value), we then proceed with the generalized vectors: Eq.~(\ref{Eq:LambdaOddGen}) for odd $N$ and Eqs.~(\ref{Eq:LambdaEvenGen-1}) and (\ref{Eq:LambdaEvenGen-2}) for even $N$.

In practical scenarios, it is reasonable to consider the case of uniformly distributed efficiencies, $\eta_i=i/N$.
However, these efficiencies are typically determined with statistical noise and are therefore known only with some uncertainty, represented by the error vector $\boldsymbol{\delta}=\{\delta_1,\ldots,\delta_N\}$, such that $\eta_i=i/N+\delta_i$.
We assume that $\boldsymbol{\delta}$ lies within a domain $\boldsymbol{\Delta}$.
To certify nonclassicality in this setting, one must verify that, for the given vector $\boldsymbol{\mathcal{P}}$, the maximal violation $V(\boldsymbol{\eta})$ remains positive throughout the entire domain $\boldsymbol{\Delta}$.
This means that if
    \begin{align}\label{Eq:IneqInac}
        \min_{\boldsymbol{\delta}\in\boldsymbol{\Delta}} V\!\left(\boldsymbol{\eta}_0+\boldsymbol{\delta};\boldsymbol{\mathcal{P}}\right)\leq0,
    \end{align}
where $\boldsymbol{\eta}_0=\{1/N,2/N,\ldots,1\}$, then there exists some $\boldsymbol{\delta}\in\boldsymbol{\Delta}$ for which the vector $\boldsymbol{\mathcal{P}}$ can be reproduced by a statistical mixture of coherent states.
Therefore, the certification of nonclassicality requires a strict violation of this inequality.

We illustrate our method using the phase-squeezed coherent state $\ket{r,\alpha_0}=\hat{D}(\alpha_0)\hat{S}(r)\ket{0}$, where $\hat{D}(\alpha_0)$ is the displacement operator, $\hat{S}(r)$ is the squeezing operator, and $\ket{0}$ is the vacuum state.
Throughout, we assume that the coherent amplitude $\alpha_0\in\mathbb{R}$ and the squeezing parameter $r>0$.
Phase-insensitive measurements, such as photocounting, are usually regarded as unsuitable for revealing nonclassicality of this state \cite{Schnabel2017}.
Indeed, the amplitude quadrature of such states is described by a probability distribution that admits a classical interpretation. 
In particular, its variance exceeds the vacuum noise level. 
Intuitively, the squared amplitude quadrature is associated with the photon number, and this intuition is commonly invoked to explain the super-Poissonian photocounting statistics for phase-squeezed coherent states.
However, Ref.~\cite{Kovtoniuk2024} demonstrated that witnessing nonclassicality of photocounting statistics for such states is possible even with imperfect photon-number resolution, as encountered in arrays of on-off detectors.
Here we further show that nonclassicality certification of these states is achievable also with a single on-off detector, by using the detection efficiency as a device setting.

The no-click probability for the squeezed coherent state $\ket{r,\alpha_0}$ is given by
    \begin{align}
        \mathcal{P}(0|\eta)&=\frac{1}{\sqrt{1+\eta(2-\eta)\sinh^2r}}\\
        &\times\exp\left[\begin{pmatrix}
            \alpha_0^\ast & -\alpha_0
        \end{pmatrix}
        \boldsymbol{\sigma}^{-1}
        \begin{pmatrix}
            -\alpha_0 \\ \alpha_0^\ast
        \end{pmatrix}
        \right],\nonumber
    \end{align}
where
    \begin{align}
        \boldsymbol{\sigma}^{-1}&=\frac{\eta}{4\left[1+\eta(2-\eta)\sinh^2r\right]}\\
        &\times\begin{pmatrix}
            \eta\cosh2r+2-\eta & -\eta \sinh 2r\\
            -\eta \sinh 2r & \eta\cosh2r+2-\eta
        \end{pmatrix}.\nonumber
    \end{align}
We construct the vector $\boldsymbol{\mathcal{P}}$ from the components $\mathcal{P}_i=\mathcal{P}(0|\eta_{\mathrm{c}}\eta_i)$, where $\eta_{\mathrm{c}}$ denotes the detection efficiency.
These results are then used in the derived inequalities to test nonclassicality.

\begin{figure}[ht!]
		\centering
	    \includegraphics[width=1\linewidth]{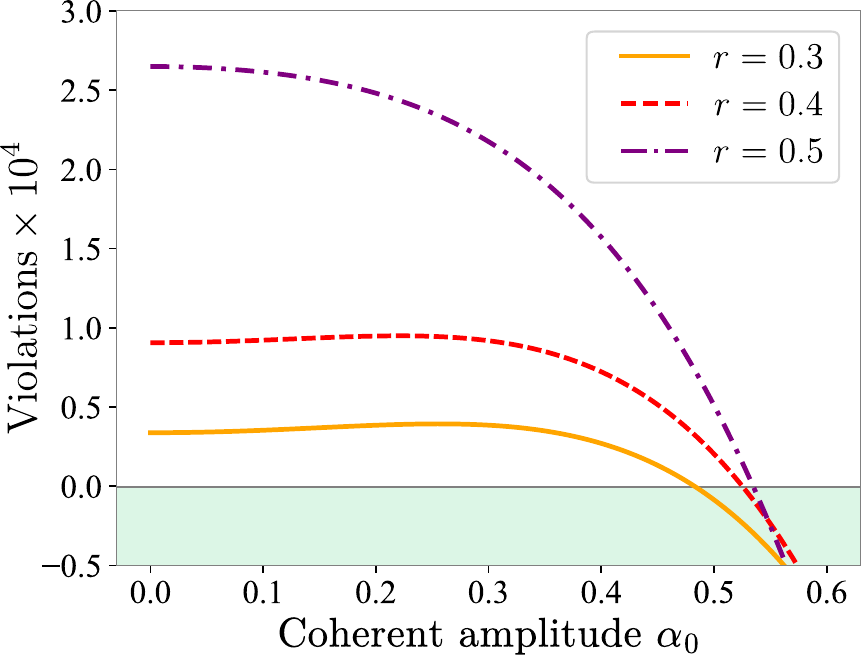}
		\caption{\label{Fig:SqCohSt-1} Violation of the nonlinear inequalities (\ref{Eq:IneqN3-Nonlin-1}) and (\ref{Eq:IneqN3-Nonlin-2}) [see Eq.~(\ref{Eq:Violation-2})] as a function of the coherent amplitude $\alpha_0$ for phase-squeezed coherent states with different squeezing parameters $r$. 
        The number of settings is $N=3$, and the detection efficiency is $\eta_{\mathrm{c}}=0.8$. 
        The shaded region corresponds to $V\leq 0$, for which no nonclassicality is detected.}
	\end{figure}

Let us consider the case $N=3$.
We begin with the simplest scenario of uniformly distributed efficiencies, $\eta_i=i/3$.
As discussed in Sec.~\ref{Sec:Moments}, the corresponding nonlinear inequalities (\ref{Eq:IneqN3-Nonlin-1}) and (\ref{Eq:IneqN3-Nonlin-2}) apply in this case.
The violations of these inequalities are quantified by
    \begin{align}\label{Eq:Violation-2} 
        V=\max\left\{\mathcal{P}^2_2{-}\mathcal{P}_1 \mathcal{P}_3; \big(\mathcal{P}_1 {-} \mathcal{P}_2\big)^2 - \big(1 - \mathcal{P}_1\big) \big(\mathcal{P}_2 - \mathcal{P}_3\big) \right\}. 
    \end{align}
The dependence of $V$ on the coherent amplitude $\alpha_0$ is shown in Fig.~\ref{Fig:SqCohSt-1}.
A clear violation is observed for small values of $\alpha_0$, which gradually disappears as $\alpha_0$ increases.
However, the magnitude of the violation is relatively small, making an accurate estimation of statistical errors essential in this regime.
This can be addressed straightforwardly by employing the inequalities in the linear form~(\ref{Eq:InequalityGen}).

\begin{figure}[ht!]
		\centering
	    \includegraphics[width=1\linewidth]{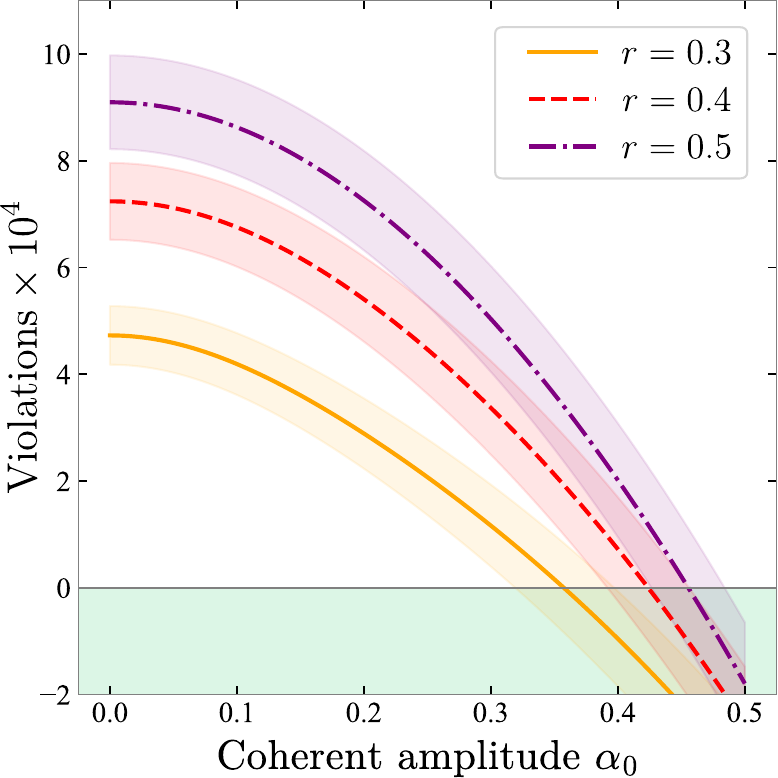}
		\caption{\label{Fig:SqCohSt-2}Violation of inequality (\ref{Eq:IneqInac}) [see Eq.~(\ref{Eq:Violation-3})] as a function of the coherent amplitude $\alpha_0$ for phase-squeezed coherent states with different squeezing parameters $r$.
        The remaining parameters are the same as in Fig.~\ref{Fig:SqCohSt-1}.
        The optimization is performed over the domain defined by $|\delta_i|\leq 0.001$ for all $i=1,\ldots,N$.
        The confidence intervals correspond to $M=10^7$ measurement trials.}
	\end{figure}

Next, we consider a scenario in which the detection efficiencies are known only within a finite uncertainty.
In this case, witnessing nonclassicality requires a violation of inequality (\ref{Eq:IneqInac}), meaning that the quantity
    \begin{align}\label{Eq:Violation-3} 
        V=\min_{\boldsymbol{\delta}\in\boldsymbol{\Delta}} V\!\left(\boldsymbol{\eta}_0+\boldsymbol{\delta};\boldsymbol{\mathcal{P}}\right) 
    \end{align}
must be non-negative.
To assess the statistical significance of such a violation, we estimate the uncertainty associated with the left-hand side of the underlying linear inequality,
$\boldsymbol{\lambda}(\boldsymbol{t}_l;T)\cdot\boldsymbol{\mathcal{P}}$.
Since this expression represents a sum of expectation values of the random variables corresponding to the components $\lambda_i$, its statistical error can be evaluated as
    \begin{align}
        \epsilon=\sqrt{\frac{1}{M}\sum_{i=1}^{N}\Var(\lambda_i)},
    \end{align}
where $M$ is the total number of experimental trials.
Here, each $\lambda_i$ is treated as a random variable that takes the value $\lambda_i$ when a no-click event occurs at the corresponding efficiency setting and $0$ otherwise.
During the optimization over the setting inaccuracies $\boldsymbol{\delta}\in\boldsymbol{\Delta}$, we select the error configuration at which the minimum in Eq.~(\ref{Eq:Violation-3}) is attained.
The resulting behavior is shown in Fig.~\ref{Fig:SqCohSt-2}.
One observes that a statistically significant violation can be achieved with a sample size of $M=10^{7}$ even under the assumption of imperfectly known transmittances.

\section{Conclusion}
\label{Sec:Conclusion}

In this paper, we have shown that, although a single on-off detector is generally regarded as unsuitable for detecting optical nonclassicality of quantum states, it can nevertheless be employed for this purpose after a simple modification.
Specifically, we propose to precede the detector with a tunable attenuator that introduces controlled additional losses into the signal.
A discrete set of $N$ attenuation values then defines a corresponding set of effective detection efficiencies, which can be interpreted as distinct detector settings.
In an experiment, one measures the no-click probabilities associated with each setting; collectively, these probabilities form a vector $\boldsymbol{\mathcal{P}}$ in an abstract $N$-dimensional Euclidean space.

The set of all vectors $\boldsymbol{\mathcal{P}}$ that can be reproduced by classical states, i.e., by statistical mixtures of coherent states, forms a convex set: the convex hull of the curve defined by the no-click probabilities of coherent states.
Because the measurement scheme considered here is not tomographically complete, vectors generated by nonclassical states may still lie within this set, implying that nonclassicality cannot always be revealed by this measurement alone.
However, if a given vector $\boldsymbol{\mathcal{P}}$ lies outside this convex set, the corresponding quantum state necessarily exhibits optical nonclassicality.

To determine whether a given vector $\boldsymbol{\mathcal{P}}$ belongs to the convex hull of classical vectors, we employ tools from convex geometry, in particular the hyperplane separation theorem.
This approach naturally leads to a continuous family of inequalities that are linear in the components of $\boldsymbol{\mathcal{P}}$.
Together, these inequalities provide a necessary and sufficient condition for membership of $\boldsymbol{\mathcal{P}}$ in the classical set.
Violation of any of these inequalities certifies the nonclassicality of the corresponding quantum state.

If uniformly distributed efficiency settings are chosen, the problem is considerably simplified.
In this case, the continuous family of linear inequalities can be replaced by a discrete set of nonlinear inequalities.
This approach is computationally efficient and straightforward to implement.
However, it is not robust against uncertainties in the efficiency values, which makes its direct application in realistic experimental scenarios challenging.

The method based on linear inequalities, however, can be naturally extended to scenarios in which the detection efficiencies are known only within finite uncertainties.
Since these inequalities are formulated for arbitrary detector settings, one can introduce an additional optimization to test whether the measured vector $\boldsymbol{\mathcal{P}}$ can be compatible with a classical description for at least one configuration of efficiencies within a predetermined uncertainty region.
If no such configuration exists, one can conclude that the proposed method reveals optical nonclassicality even under imperfect knowledge of experimental parameters, making it particularly relevant for practical experimental implementations.

We have demonstrated that our technique, while being straightforward to implement experimentally, can reveal nonclassicality in scenarios where standard methods fail.
As an illustrative example, we considered phase-squeezed coherent states.
These states constitute an emblematic case: despite being nonclassical, they exhibit super-Poissonian photocounting statistics and are therefore often regarded as unsuitable for nonclassicality tests based on phase-insensitive measurements.
Nevertheless, our approach overcomes this limitation and successfully reveals nonclassicality using only three efficiency settings.
This example highlights the strong potential of our method for experimental applications and demonstrates its usefulness in situations where conventional techniques are ineffective.

\section*{Acknowledgment}

The authors thank B. Hage and F. H\"ubner for enlightening discussions.
V.S.K. acknowledges support from the National Research Foundation of Ukraine under Project No. 2023.03/0165.
A.A.S. acknowledges support from the National Academy of Sciences of Ukraine through Project No. 0125U000031, from Simons Foundation International SFI-PD-Ukraine-00014573, PI LB, and from (VUIAS) 2025/2026 Fellowship.

\section*{Data availability}

The data that support the findings of this article are openly available~\cite{Kovtoniuk2026Data}.

% \section*{Data availability}

% The data that support the findings of this article are openly available at Ref.~\cite{}.

\appendix

\section{Hodge star}
\label{Sec:Hodge}

In this appendix we recall the definition of the Hodge star operator used in Sec.~\ref{Sec:N} for constructing the vectors $\boldsymbol{\lambda}$.
Consider the Euclidean space $\mathbb{R}^n$ and a set of $(n-1)$ vectors $\boldsymbol{v}^i$, with $i=1,\ldots,n-1$.
The Hodge star maps their exterior product to a vector in $\mathbb{R}^n$ according to
    \begin{align}\label{Eq:Hodge}
        (-1)^{n-1} \star \bigwedge_{i = 1}^{n-1} \boldsymbol{v}^i = \begin{vmatrix} \boldsymbol{e}^1 & \boldsymbol{e}^2 & \dots & \boldsymbol{e}^n \\ v^1_1 & v^1_2 & \dots & v^1_n \\ \vdots & \vdots & \ddots & \vdots \\ v^{n-1}_1 & v^{n-1}_2 & \dots & v^{n-1}_n \\ \end{vmatrix}, 
    \end{align}
where ${\boldsymbol{e}^1, \boldsymbol{e}^2, \ldots, \boldsymbol{e}^n}$ denotes an orthonormal basis of $\mathbb{R}^n$.
This construction generalizes the cross product in $\mathbb{R}^3$ to arbitrary dimension $n$.

\section{Tightness of inequalities}
\label{Sec:Proof}

    In this appendix, we prove that the inequalities associated with the vectors $\boldsymbol{\lambda}$ introduced in Sec.~\ref{Sec:N} form a complete set of tight inequalities.
    To this end, we employ methods from convex geometry. 
    In particular, our analysis relies on results from the theory of generalized moment problems and extended Chebyshev systems.
    The mathematical results used throughout this appendix can be found in Ref.~\cite{Krein1977}.
    We start with presenting general statements that apply to a broad class of measurements and experimental settings. 
    In particular, these results cover both the photocounting statistics studied in Ref.~\cite{Kovtoniuk2024} and the no-click statistics obtained from on-off detectors with tunable efficiency settings considered here.
    The explicit application of these general results to the latter case is discussed at the end of this appendix.

    Let us consider a set of $N{+}1$ functions $\{\Pi_0^\ast(t),\ldots,\Pi_N^\ast(t)\}$ that are $N$ times differentiable on the interval $t\in[a,b]$.
    This set is called an extended Chebyshev system if there exists $\sigma=\pm1$ such that
        \begin{align}
            \label{Eq:Chebyshev}
            \sigma \Delta \left( t_0, \ldots, t_N \right) > 0 \textrm{ if } a \leq t_0 \leq \ldots \leq t_N \leq b,
        \end{align}
    where
        \begin{align}\label{Eq:Delta}
            \Delta \left( t_1, \ldots, t_N \right)= \det 
           \begin{pmatrix}
            {\boldsymbol{\Pi}^\ast}^{(\omega_0)}(t_0) & \ldots & {\boldsymbol{\Pi}^\ast}^{(\omega_N)}(t_N)   
           \end{pmatrix}.
        \end{align}
    Here 
        \begin{align}
            {\boldsymbol{\Pi}^\ast}^{(\omega_i)}(t_i)=\begin{pmatrix}
            {\Pi^\ast_1}^{(\omega_i)}(t_i)&{\Pi^\ast_2}^{(\omega_i)}(t_i)&\ldots&{\Pi^\ast}^{(\omega_i)}_N(t_i)
            \end{pmatrix}^\mathrm{T},
        \end{align}
    $i=0,\ldots,N$, and the superscript $(\omega_i)$ denotes the $\omega_i$th derivative.
    The integer $\omega_i$ equals the number of occurrences of $t_i$ in the sequence $t_0,\ldots,t_{i-1}$, with the convention $\omega_0=0$.  
    In this context, a linear combination of functions from an extended Chebyshev system
        \begin{align}
            P(t) = \sum_{i = 0}^N \lambda^\ast_i \Pi_i^\ast(t) = \boldsymbol{\lambda}^\ast \cdot \boldsymbol{\Pi}^\ast(t)
        \end{align}
    is called a generalized polynomial.
    An equivalent defining property of an extended Chebyshev system is that any generalized polynomial has at most $N$ zeros on $[a,b]$, counted with multiplicities.

    Our task is to determine the conditions under which a vector $\boldsymbol{\mathcal{P}}^\ast \in \mathbb{R}^{N+1}$ belongs to the convex set generated by the vectors $\boldsymbol{\Pi}^\ast(t)$.
    In this context, a crucial role is played by vectors $\boldsymbol{\lambda}^\ast$ that correspond to generalized polynomials non-negative on the interval $t\in[a,b]$, i.e., $P(t)\ge 0$.
    The following theorem makes this connection precise.
    \begin{theorem}
        \label{Th:GenMomentsTest}
        A vector $\boldsymbol{\mathcal{P}}^\ast \in \mathbb{R}^{N+1}$ can be represented as a convex combination of the vectors $\boldsymbol{\Pi}^\ast(t)$, $t \in [a,b]$, if and only if for any non-negative generalized polynomial $P(t) = \boldsymbol{\lambda}^\ast \cdot \boldsymbol{\Pi}^\ast(t) \geq 0$, the dot product 
            \begin{align}\label{Eq:TightAlternative}
                \boldsymbol{\lambda}^\ast \cdot \boldsymbol{\mathcal{P}}^\ast \geq 0.
            \end{align}
    \end{theorem}

    The set of all vectors $\boldsymbol{\lambda}^\ast$ associated with non-negative generalized polynomials can be reduced without loss of generality to those corresponding to polynomials with exactly $N$ zeros, counting multiplicities, in the interval $t \in [a,b]$.
    We refer to such polynomials as tight polynomials.
    The following theorem provides a key result that justifies this reduction and enables an optimization of the admissible set of $\boldsymbol{\lambda}^\ast$.
    
    \begin{theorem}[Karlin]
        \label{Th:Karlin}
        Any non-negative generalized polynomial which is not tight can be uniquely decomposed into the sum of two tight generalized polynomials.
    \end{theorem}
    
    Indeed, let us suppose that $\boldsymbol{\lambda}_k^\ast \cdot \boldsymbol{\mathcal{P}}^\ast \geq 0$ holds for every $\boldsymbol{\lambda}_k^\ast$ associated with a tight generalized polynomial.
    By Karlin's theorem, any non-negative generalized polynomial $P(t) = \boldsymbol{\lambda}^\ast \cdot \boldsymbol{\Pi}^\ast(t)$ can be decomposed into the sum of two tight generalized polynomials, $P_1(t)$ and $P_2(t)$.
    This decomposition induces a corresponding decomposition of the coefficient vector, $\boldsymbol{\lambda}^\ast = \boldsymbol{\lambda}^\ast_1 + \boldsymbol{\lambda}^*_2$, where each vector $\boldsymbol{\lambda}_k^\ast$ is associated with a tight polynomial.
    Consequently,
    \begin{align}
        \boldsymbol{\lambda}^\ast \cdot \boldsymbol{\mathcal{P}}^\ast = (\boldsymbol{\lambda}^\ast_1 + \boldsymbol{\lambda}^\ast_2) \cdot \boldsymbol{\mathcal{P}}^\ast = \boldsymbol{\lambda}^\ast_1 \cdot \boldsymbol{\mathcal{P}}^\ast + \boldsymbol{\lambda}^\ast_2 \cdot \boldsymbol{\mathcal{P}}^\ast.
    \end{align}
    Since both terms on the right-hand side are non-negative by assumption, it follows that $\boldsymbol{\lambda}^\ast \cdot \boldsymbol{\mathcal{P}}^\ast \geq 0$.
    Therefore, it is sufficient to test the inequalities associated with tight generalized polynomials in order to determine whether $\boldsymbol{\mathcal{P}}^\ast$ belongs to the convex hull.

    Let us now consider a specific measurement scheme, described by POVM elements associated with the set $\{\Pi_i(t)|i=1,\ldots, N\}$.
    We define the extended set of functions by $\Pi_0^\ast(t)=1$ and $\Pi_i^\ast(t)=\Pi_i(t)$ for $i=1,\ldots,N$, and assume that this set forms an extended Chebyshev system on $[a,b]$.
    Our goal is to determine whether a given vector $\boldsymbol{\mathcal{P}}$ can be presented as a convex combination of the vectors $\boldsymbol{\Pi}(t)$.
    To this end we introduce the extended vector $\boldsymbol{\mathcal{P}}^\ast$ with components $\mathcal{P}^\ast_0 = 1$ and $\mathcal{P}^\ast_i=\mathcal{P}_i$ for $i=1,\ldots,N$.
    It is straightforward to see that the vector $\boldsymbol{\mathcal{P}}$ is a convex combination of $\boldsymbol{\Pi}(t)$ if and only if the vector $\boldsymbol{\mathcal{P}}^\ast$ is a convex combination of $\boldsymbol{\Pi}^\ast(t)$.

    In the the present setting, a generalized polynomial $\boldsymbol{\lambda}^\ast \cdot \boldsymbol{\Pi}^\ast(t)$ is non-negative, if it satisfies the inequality
        \begin{align}
            \label{Eq:Lambda0Domain}
            \lambda_0^\ast \geq \boldsymbol{\lambda} \cdot \boldsymbol{\Pi}(t).
        \end{align}
    Here we have assumed that
        \begin{align}
            \boldsymbol{\lambda}^\ast = \begin{pmatrix} \lambda_0  \\ -\boldsymbol{\lambda}\end{pmatrix}.
        \end{align}
    Additionally, inequality~(\ref{Eq:TightAlternative}) can be equivalently rewritten as
        \begin{align}
            \label{Eq:Lambda0Inequality}
            \lambda_0^\ast \geq \boldsymbol{\lambda} \cdot \boldsymbol{\mathcal{P}}.
        \end{align}
    Consider a family of non-negative generalized polynomials that share a fixed $\boldsymbol{\lambda}$ but have different $\lambda^*_0$.
    Each polynomial produces inequality (\ref{Eq:Lambda0Inequality}) with its own $\lambda^*_0$.
    Such inequalities can be replaced by a single one with the smallest admissible $\lambda^\ast_0$ since all others are implied by it
    According to Eq. (\ref{Eq:Lambda0Domain}), this minimal value is
        \begin{align}\label{Eq:TightLambda0}
            \lambda_0^\ast=\sup_{t \in [a,b]} \boldsymbol{\lambda} \cdot \boldsymbol{\Pi}(t).
        \end{align}
    Thus, we conclude that in order to verify if $\boldsymbol{\mathcal{P}^*}$ belongs to the convex hull of the vectors $\boldsymbol{\Pi}^*(t)$ one only needs to consider generalized polynomials satisfying Eq.~(\ref{Eq:TightLambda0}).
    Substituting it into the inequality (\ref{Eq:Lambda0Inequality}) directly yields the inequality (\ref{Eq:InequalityGen}).

    Now we show that tight generalized polynomials correspond to the set $\Lambda$ of tight inequalities; see Sec.~\ref{Sec:TightInequal}.
    First, by virtue of Theorem~\ref{Th:Karlin}, any vector $\boldsymbol{\lambda}^\ast$ can be decomposed into a sum of two vectors $\boldsymbol{\lambda}_k^\ast$ associated with tight generalized polynomials.
    It then follows directly that any vector $\boldsymbol{\lambda}$ defining inequalities of the form (\ref{Eq:InequalityGen}) can likewise be decomposed into a sum of corresponding vectors  $\boldsymbol{\lambda}_k$.
    Second, we need to prove that a given vector $\boldsymbol{\lambda}_k$ cannot be expressed as a linear combination of other vectors from this set.
    For this purpose, we outline a useful representation of a generalized polynomial with $N$ zeros $t_1, \ldots, t_N$, where the number of duplications of a zero indicates its multiplicity.
    Such a polynomial is determined uniquely up to a constant factor $C$ and is given by
        \begin{align}
            \label{Eq:MaxZeroCount}
            C \Delta \left( t, t_1, \ldots, t_N \right),
        \end{align}
    cf. Eq. (\ref{Eq:Delta}).
    It implies that a tight generalized polynomial cannot be decomposed into the sum of other linearly independent generalized polynomials.
    Indeed, if that were the case, they would need to have the same zeros and multiplicities as the original tight generalized polynomial.
    However, that contradicts the uniqueness property.
    This conclusion translates directly to the corresponding inequalities of the form (\ref{Eq:InequalityGen}).
    
    As the next step, we derive explicit expressions for the tight test functions.
    According to Eq. (\ref{Eq:MaxZeroCount}), a tight generalized polynomial is determined by its zeros up to a constant factor $C$.
    Since such a polynomial is non-negative on the interval, all its interior zeros must be non-nodal, i.e., their multiplicities must be even.
    We denote a tight generalized polynomial by $P(t;\boldsymbol{t}_l;T)$, where $\boldsymbol{t}_l = \left\{ t_1, \ldots, t_l \right\}$ is a set of interior zeros of size $l=(N - |T|)/2$ such that every occurrence of a zero increases its multiplicity by $2$, e.g., $\left\{0.2, 0.2, 0.9\right\}$ means that $0.2$ has a multiplicity of $4$ and $0.9$ has a multiplicity of $2$. 
    The symbol $T$ denotes the set of boundary zeros such that there are no duplications and each occurrence of a zero increases its multiplicity by $1$ and $|T|$ denotes the cardinality of this set (the number of elements).
    Depending on the specific vector $\boldsymbol{\lambda}(\boldsymbol{t}_l;T)$, under consideration, the set $T$ may contain the boundary points $a$, $b$, or be empty.

    For each parity of $N$, there are two families of tight generalized polynomials.
    If $N = 2m$, for the first family, $T = \varnothing$ and $\boldsymbol{t}_m = \left\{ t_1, \ldots, t_m \right\}$,
        \begin{align}
            \label{Eq:LowerEvenTightPolynomial}
            P(t;\boldsymbol{t}_m) = \sigma C \Delta \left( t, t_1, t_1, \ldots, t_m, t_m \right),
        \end{align}
    where $C > 0$ and $\sigma$ from Eq. (\ref{Eq:Chebyshev}) ensures that the polynomial is non-negative.
    For the second family, $T = \left\{ a, b \right\}$ and $\boldsymbol{t}_{m-1} = \left\{t_1, \ldots, t_{m-1}\right\}$,
        \begin{align}
            \label{Eq:UpperEvenTightPolynomial}
            & P(t;\boldsymbol{t}_{m-1};a,b) \nonumber \\
            & = -\sigma C \Delta \left( t, a, t_1, t_1, \ldots, t_{m-1}, t_{m-1},b \right),
        \end{align}
    where $C > 0$ and the minus sign accounts for the swap of $t$ and $a$.
    If $N = 2m + 1$, we have $\boldsymbol{t}_m = \left\{ t_1, \ldots, t_m \right\}$ for both families and $T = \left\{a\right\}$ and $T = \left\{b\right\}$ for the first and the second family, respectively,
        \begin{align}
            \label{Eq:OddTightPolynomial}
            P(t;\boldsymbol{t}_m;\tau) = (-1)^{\tau + 1} \sigma C \Delta(t, \tau, t_1, t_1, \ldots, t_m, t_m),
        \end{align}
    where $\tau = a, b$, $C > 0$.

    We now derive the explicit expression for the vectors $\boldsymbol{\lambda}$ associated with tight inequalities.
    For clarity, we restrict the derivation to the case in which the set $\boldsymbol{t}_m$ contains no duplications ($\omega_i=1$) as the methodology remains the same for the general case.
    We substitute Eq.~(\ref{Eq:Delta}) into Eq.~(\ref{Eq:LowerEvenTightPolynomial}) and perform elementary column operations---namely, subtracting the second column from the first and applying the same operation to all columns with even indices in the determinant.
    As a result, the first row acquires a single nonzero entry, which allows us to expand the determinant along this row.
    This reduces the determinant to an $N\times N$ form,
        \begin{widetext}
            \begin{align}
                P(t;\boldsymbol{t}_m) = \sigma C\det
                    = -\sigma C \det
                    \begin{pmatrix}
                        \Delta\Pi_1(t,t_1) & \dot{\Pi}_1(t_1) & \ldots & \Delta\Pi_1(t_m, t_1) & \dot{\Pi}_1(t_m) \\
                        \vdots & \vdots & \ddots & \vdots & \vdots \\
                        \Delta\Pi_N(t, t_1) & \dot{\Pi}_N(t_1) & \ldots & \Delta\Pi_N(t_m, t_m) & \dot{\Pi}_N(t_m)
                    \end{pmatrix}.
            \end{align}        
    This expression can be further rearranged as
            \begin{align}
                P(t;\boldsymbol{t}_m) = -\sigma C \left[ -\det \begin{pmatrix}\boldsymbol{\Pi}(t_1)& \dot{\boldsymbol{\Pi}}(t_1)& \ldots & \boldsymbol{\Delta\Pi}(t_m,t_1) & \dot{\boldsymbol{\Pi}}(t_m) \end{pmatrix} + \det \begin{pmatrix} \boldsymbol{\Pi}(t) & \dot{\boldsymbol{\Pi}}(t_1)& \ldots, \boldsymbol{\Delta\Pi}(t_m,t_1)&\dot{\boldsymbol{\Pi}}(t_m) \end{pmatrix} \right],
            \end{align}
    where we use vector notation for the matrix columns.         
        \end{widetext}

    Since $P(t;\boldsymbol{t}_m) = \sup_{t^\prime \in [a,b]} \boldsymbol{\lambda}(\boldsymbol{t}_m) \cdot \boldsymbol{\Pi}(t^\prime) - \boldsymbol{\lambda}(\boldsymbol{t}_m) \cdot \boldsymbol{\Pi}(t)$, where $\boldsymbol{\lambda}(\boldsymbol{t})$ is the associated tight test function, we conclude that
        \begin{align}
            & \boldsymbol{\lambda}(\boldsymbol{t}_m) \cdot \boldsymbol{\Pi}(t) \nonumber \\
            & = \sigma C \det \begin{pmatrix} \boldsymbol{\Pi}(t) & \dot{\boldsymbol{\Pi}}(t_1) & \ldots & \boldsymbol{\Delta\Pi}(t_m,t_1) & \dot{\boldsymbol{\Pi}}(t_m) \end{pmatrix} &
        \end{align}
    Recall that the determinant of a matrix formed by a vector $\mathbf{v}$ and a set of vectors $\mathbf{u}_1, \ldots, \mathbf{u}_N$ can be written as the dot product
        \begin{align}
            \det\begin{pmatrix}\mathbf{v} & \mathbf{u}_1 & \ldots & \mathbf{v}_N \end{pmatrix} = (-1)^{N-1} \mathbf{v} \cdot \star(\mathbf{u}_1 \wedge \ldots \wedge \mathbf{u}_N).
        \end{align}
    Then Eq. (\ref{Eq:TightLambdaDerivation}) can be rewritten as
        \begin{align}
            \label{Eq:TightLambdaDerivation}
            & \boldsymbol{\lambda}(\boldsymbol{t}) \cdot \boldsymbol{\Pi}(t) \nonumber \\
            & = (-1)^{N-1} \sigma C \boldsymbol{\Pi}(t) \cdot \star \left( \dot{\boldsymbol{\Pi}}(t_1) \wedge \bigwedge_{i=2}^m \boldsymbol{\Delta\Pi}(t_i,t_1) \wedge \dot{\boldsymbol{\Pi}}(t_i) \right).
        \end{align}
    Thus,
        \begin{align}
            \boldsymbol{\lambda}(\boldsymbol{t}) = (-1)^{N-1} \sigma \mathcal{N} \star \left( \dot{\boldsymbol{\Pi}}(t_1) \wedge \bigwedge_{i=2}^m \boldsymbol{\Delta\Pi}(t_i,t_1) \wedge \dot{\boldsymbol{\Pi}}(t_i) \right),
        \end{align}
    where we set $C = \mathcal{N}$.
    Similarly, one can obtain the expressions for other test functions.

    As the next step, we specify the elements $\Pi_i(t)$ as those corresponding to the no-click statistics given by Eq.~(\ref{Eq:Moments}). 
    According to Ref.~\cite{Karlin1966},
        \begin{align}
            \label{Eq:POVMSystem}
            \left\{1, t, t^{\nu_2}, \ldots, t^{\nu_N} \right\},
        \end{align}
    where $1  <  \nu_2 < \ldots < \nu_N$, is an extended Chebyshev system on $]0,1]$.
    To address the singularity at the boundary point $t=0$, we employ a limiting argument. 
    We first derive the family of tight inequalities on the truncated interval $[\epsilon, 1]$ for an arbitrary $\epsilon \in ]0,1[$.
    Observe that for any $\epsilon_1 < \epsilon_2$, the inequalities associated with $\epsilon_2$ follow directly from those associated with $\epsilon_1$. Consequently, the limiting inequalities derived as $\epsilon \to +0$ are sufficient to imply the inequalities for all $\epsilon \in ]0,1[$.

    Now, let us determine the sign $\sigma = \pm 1$ from Eq. (\ref{Eq:Chebyshev}).
    Since $\sigma$ is the same across all the domain of $t_0, \ldots, t_N$, we may restrict our attention to the case when there are no duplicates.
    As stated in Ref.~\cite{Karlin1966},
        \begin{align}
            \det\begin{pmatrix} 
                \exp(x_0 y_0) & \exp(x_0 y_1) & \dots & \exp(x_0 y_N) \\
                \exp(x_1 y_0) & \exp(x_1 y_1) & \dots & \exp(x_1 y_N) \\
                \vdots & \vdots & \ddots & \vdots \\
                \exp(x_N y_0) & \exp(x_N y_1) & \dots & \exp(x_N y_N)
            \end{pmatrix} > 0,
        \end{align}
    where $x_0 < \ldots < x_N$ and $y_0 < \ldots < y_N$.
    Substituting $x_0 = 0$, $x_i = \nu_i$, $i = 1, \ldots, N$, and $y_j = \ln t_j$, $j = 0, \ldots, N$, we obtain Eq. (\ref{Eq:Chebyshev}) for the system (\ref{Eq:POVMSystem}).
    It follows that $\sigma = 1$, which specifies the sign in Eq.~(\ref{Eq:Chebyshev}) for the considered Chebyshev system.
    
\bibliography{biblio}

%apsrev4-2.bst 2019-01-14 (MD) hand-edited version of apsrev4-1.bst
%Control: key (0)
%Control: author (8) initials jnrlst
%Control: editor formatted (1) identically to author
%Control: production of article title (0) allowed
%Control: page (0) single
%Control: year (1) truncated
%Control: production of eprint (0) enabled
\begin{thebibliography}{83}%
\makeatletter
\providecommand \@ifxundefined [1]{%
 \@ifx{#1\undefined}
}%
\providecommand \@ifnum [1]{%
 \ifnum #1\expandafter \@firstoftwo
 \else \expandafter \@secondoftwo
 \fi
}%
\providecommand \@ifx [1]{%
 \ifx #1\expandafter \@firstoftwo
 \else \expandafter \@secondoftwo
 \fi
}%
\providecommand \natexlab [1]{#1}%
\providecommand \enquote  [1]{``#1''}%
\providecommand \bibnamefont  [1]{#1}%
\providecommand \bibfnamefont [1]{#1}%
\providecommand \citenamefont [1]{#1}%
\providecommand \href@noop [0]{\@secondoftwo}%
\providecommand \href [0]{\begingroup \@sanitize@url \@href}%
\providecommand \@href[1]{\@@startlink{#1}\@@href}%
\providecommand \@@href[1]{\endgroup#1\@@endlink}%
\providecommand \@sanitize@url [0]{\catcode `\\12\catcode `\$12\catcode
  `\&12\catcode `\#12\catcode `\^12\catcode `\_12\catcode `\%12\relax}%
\providecommand \@@startlink[1]{}%
\providecommand \@@endlink[0]{}%
\providecommand \url  [0]{\begingroup\@sanitize@url \@url }%
\providecommand \@url [1]{\endgroup\@href {#1}{\urlprefix }}%
\providecommand \urlprefix  [0]{URL }%
\providecommand \Eprint [0]{\href }%
\providecommand \doibase [0]{https://doi.org/}%
\providecommand \selectlanguage [0]{\@gobble}%
\providecommand \bibinfo  [0]{\@secondoftwo}%
\providecommand \bibfield  [0]{\@secondoftwo}%
\providecommand \translation [1]{[#1]}%
\providecommand \BibitemOpen [0]{}%
\providecommand \bibitemStop [0]{}%
\providecommand \bibitemNoStop [0]{.\EOS\space}%
\providecommand \EOS [0]{\spacefactor3000\relax}%
\providecommand \BibitemShut  [1]{\csname bibitem#1\endcsname}%
\let\auto@bib@innerbib\@empty
%</preamble>
\bibitem [{\citenamefont {Titulaer}\ and\ \citenamefont
  {Glauber}(1965)}]{titulaer65}%
  \BibitemOpen
  \bibfield  {author} {\bibinfo {author} {\bibfnamefont {U.~M.}\ \bibnamefont
  {Titulaer}}\ and\ \bibinfo {author} {\bibfnamefont {R.~J.}\ \bibnamefont
  {Glauber}},\ }\bibfield  {title} {\bibinfo {title} {Correlation functions for
  coherent fields},\ }\href {https://doi.org/10.1103/PhysRev.140.B676}
  {\bibfield  {journal} {\bibinfo  {journal} {Phys. Rev.}\ }\textbf {\bibinfo
  {volume} {140}},\ \bibinfo {pages} {B676} (\bibinfo {year}
  {1965})}\BibitemShut {NoStop}%
\bibitem [{\citenamefont {Mandel}(1986)}]{mandel86}%
  \BibitemOpen
  \bibfield  {author} {\bibinfo {author} {\bibfnamefont {L.}~\bibnamefont
  {Mandel}},\ }\bibfield  {title} {\bibinfo {title} {Non-classical states of
  the electromagnetic field},\ }\href
  {https://doi.org/10.1088/0031-8949/1986/t12/005} {\bibfield  {journal}
  {\bibinfo  {journal} {Phys. Scr.}\ }\textbf {\bibinfo {volume} {T12}},\
  \bibinfo {pages} {34} (\bibinfo {year} {1986})}\BibitemShut {NoStop}%
\bibitem [{\citenamefont {Mandel}\ and\ \citenamefont
  {Wolf}(1995)}]{mandel_book}%
  \BibitemOpen
  \bibfield  {author} {\bibinfo {author} {\bibfnamefont {L.}~\bibnamefont
  {Mandel}}\ and\ \bibinfo {author} {\bibfnamefont {E.}~\bibnamefont {Wolf}},\
  }\href@noop {} {\emph {\bibinfo {title} {Optical Coherence and Quantum
  Optics}}}\ (\bibinfo  {publisher} {Cambridge University Press, Cambridge},\
  \bibinfo {year} {1995})\BibitemShut {NoStop}%
\bibitem [{\citenamefont {Vogel}\ and\ \citenamefont
  {Welsch}(2006)}]{vogel_book}%
  \BibitemOpen
  \bibfield  {author} {\bibinfo {author} {\bibfnamefont {W.}~\bibnamefont
  {Vogel}}\ and\ \bibinfo {author} {\bibfnamefont {D.-G.}\ \bibnamefont
  {Welsch}},\ }\href@noop {} {\emph {\bibinfo {title} {Quantum Optics}}}\
  (\bibinfo  {publisher} {Wiley-VCH},\ \bibinfo {address} {Weinheim},\ \bibinfo
  {year} {2006})\BibitemShut {NoStop}%
\bibitem [{\citenamefont {Agarwal}(2013)}]{agarwal_book}%
  \BibitemOpen
  \bibfield  {author} {\bibinfo {author} {\bibfnamefont {G.~S.}\ \bibnamefont
  {Agarwal}},\ }\href@noop {} {\emph {\bibinfo {title} {Quantum Optics}}}\
  (\bibinfo  {publisher} {Cambridge University Press, Cambridge},\ \bibinfo
  {year} {2013})\BibitemShut {NoStop}%
\bibitem [{\citenamefont {Schnabel}(2017)}]{Schnabel2017}%
  \BibitemOpen
  \bibfield  {author} {\bibinfo {author} {\bibfnamefont {R.}~\bibnamefont
  {Schnabel}},\ }\bibfield  {title} {\bibinfo {title} {Squeezed states of light
  and their applications in laser interferometers},\ }\href
  {https://doi.org/https://doi.org/10.1016/j.physrep.2017.04.001} {\bibfield
  {journal} {\bibinfo  {journal} {Phys. Rep.}\ }\textbf {\bibinfo {volume}
  {684}},\ \bibinfo {pages} {1} (\bibinfo {year} {2017})}\BibitemShut {NoStop}%
\bibitem [{\citenamefont {Sperling}\ and\ \citenamefont
  {Walmsley}(2018{\natexlab{a}})}]{sperling2018a}%
  \BibitemOpen
  \bibfield  {author} {\bibinfo {author} {\bibfnamefont {J.}~\bibnamefont
  {Sperling}}\ and\ \bibinfo {author} {\bibfnamefont {I.~A.}\ \bibnamefont
  {Walmsley}},\ }\bibfield  {title} {\bibinfo {title} {Quasiprobability
  representation of quantum coherence},\ }\href
  {https://doi.org/10.1103/PhysRevA.97.062327} {\bibfield  {journal} {\bibinfo
  {journal} {Phys. Rev. A}\ }\textbf {\bibinfo {volume} {97}},\ \bibinfo
  {pages} {062327} (\bibinfo {year} {2018}{\natexlab{a}})}\BibitemShut
  {NoStop}%
\bibitem [{\citenamefont {Sperling}\ and\ \citenamefont
  {Walmsley}(2018{\natexlab{b}})}]{sperling2018b}%
  \BibitemOpen
  \bibfield  {author} {\bibinfo {author} {\bibfnamefont {J.}~\bibnamefont
  {Sperling}}\ and\ \bibinfo {author} {\bibfnamefont {I.~A.}\ \bibnamefont
  {Walmsley}},\ }\bibfield  {title} {\bibinfo {title} {Quasistates and
  quasiprobabilities},\ }\href {https://doi.org/10.1103/PhysRevA.98.042122}
  {\bibfield  {journal} {\bibinfo  {journal} {Phys. Rev. A}\ }\textbf {\bibinfo
  {volume} {98}},\ \bibinfo {pages} {042122} (\bibinfo {year}
  {2018}{\natexlab{b}})}\BibitemShut {NoStop}%
\bibitem [{\citenamefont {Sperling}\ and\ \citenamefont
  {Vogel}(2020)}]{sperling2020}%
  \BibitemOpen
  \bibfield  {author} {\bibinfo {author} {\bibfnamefont {J.}~\bibnamefont
  {Sperling}}\ and\ \bibinfo {author} {\bibfnamefont {W.}~\bibnamefont
  {Vogel}},\ }\bibfield  {title} {\bibinfo {title} {Quasiprobability
  distributions for quantum-optical coherence and beyond},\ }\href
  {https://doi.org/10.1088/1402-4896/ab5501} {\bibfield  {journal} {\bibinfo
  {journal} {Phys. Scr.}\ }\textbf {\bibinfo {volume} {95}},\ \bibinfo {pages}
  {034007} (\bibinfo {year} {2020})}\BibitemShut {NoStop}%
\bibitem [{\citenamefont {Glauber}(1963)}]{glauber63c}%
  \BibitemOpen
  \bibfield  {author} {\bibinfo {author} {\bibfnamefont {R.~J.}\ \bibnamefont
  {Glauber}},\ }\bibfield  {title} {\bibinfo {title} {Coherent and incoherent
  states of the radiation field},\ }\href
  {https://doi.org/10.1103/PhysRev.131.2766} {\bibfield  {journal} {\bibinfo
  {journal} {Phys. Rev.}\ }\textbf {\bibinfo {volume} {131}},\ \bibinfo {pages}
  {2766} (\bibinfo {year} {1963})}\BibitemShut {NoStop}%
\bibitem [{\citenamefont {Sudarshan}(1963)}]{sudarshan63}%
  \BibitemOpen
  \bibfield  {author} {\bibinfo {author} {\bibfnamefont {E.~C.~G.}\
  \bibnamefont {Sudarshan}},\ }\bibfield  {title} {\bibinfo {title}
  {Equivalence of semiclassical and quantum mechanical descriptions of
  statistical light beams},\ }\href
  {https://doi.org/10.1103/PhysRevLett.10.277} {\bibfield  {journal} {\bibinfo
  {journal} {Phys. Rev. Lett.}\ }\textbf {\bibinfo {volume} {10}},\ \bibinfo
  {pages} {277} (\bibinfo {year} {1963})}\BibitemShut {NoStop}%
\bibitem [{\citenamefont {Mandel}(1979)}]{mandel79}%
  \BibitemOpen
  \bibfield  {author} {\bibinfo {author} {\bibfnamefont {L.}~\bibnamefont
  {Mandel}},\ }\bibfield  {title} {\bibinfo {title} {Sub-{P}oissonian photon
  statistics in resonance fluorescence},\ }\href
  {https://doi.org/10.1364/OL.4.000205} {\bibfield  {journal} {\bibinfo
  {journal} {Opt. Lett.}\ }\textbf {\bibinfo {volume} {4}},\ \bibinfo {pages}
  {205} (\bibinfo {year} {1979})}\BibitemShut {NoStop}%
\bibitem [{\citenamefont {Stoler}(1970)}]{Stoler1970}%
  \BibitemOpen
  \bibfield  {author} {\bibinfo {author} {\bibfnamefont {D.}~\bibnamefont
  {Stoler}},\ }\bibfield  {title} {\bibinfo {title} {Equivalence classes of
  minimum uncertainty packets},\ }\href
  {https://doi.org/10.1103/PhysRevD.1.3217} {\bibfield  {journal} {\bibinfo
  {journal} {Phys. Rev. D}\ }\textbf {\bibinfo {volume} {1}},\ \bibinfo {pages}
  {3217} (\bibinfo {year} {1970})}\BibitemShut {NoStop}%
\bibitem [{\citenamefont {Stoler}(1971)}]{Stoler1971}%
  \BibitemOpen
  \bibfield  {author} {\bibinfo {author} {\bibfnamefont {D.}~\bibnamefont
  {Stoler}},\ }\bibfield  {title} {\bibinfo {title} {Equivalence classes of
  minimum-uncertainty packets. ii},\ }\href
  {https://doi.org/10.1103/PhysRevD.4.1925} {\bibfield  {journal} {\bibinfo
  {journal} {Phys. Rev. D}\ }\textbf {\bibinfo {volume} {4}},\ \bibinfo {pages}
  {1925} (\bibinfo {year} {1971})}\BibitemShut {NoStop}%
\bibitem [{\citenamefont {Wu}\ \emph {et~al.}(1986)\citenamefont {Wu},
  \citenamefont {Kimble}, \citenamefont {Hall},\ and\ \citenamefont
  {Wu}}]{Wu1986}%
  \BibitemOpen
  \bibfield  {author} {\bibinfo {author} {\bibfnamefont {L.-A.}\ \bibnamefont
  {Wu}}, \bibinfo {author} {\bibfnamefont {H.~J.}\ \bibnamefont {Kimble}},
  \bibinfo {author} {\bibfnamefont {J.~L.}\ \bibnamefont {Hall}},\ and\
  \bibinfo {author} {\bibfnamefont {H.}~\bibnamefont {Wu}},\ }\bibfield
  {title} {\bibinfo {title} {Generation of squeezed states by parametric down
  conversion},\ }\href {https://doi.org/10.1103/PhysRevLett.57.2520} {\bibfield
   {journal} {\bibinfo  {journal} {Phys. Rev. Lett.}\ }\textbf {\bibinfo
  {volume} {57}},\ \bibinfo {pages} {2520} (\bibinfo {year}
  {1986})}\BibitemShut {NoStop}%
\bibitem [{\citenamefont {Wu}\ \emph {et~al.}(1987)\citenamefont {Wu},
  \citenamefont {Xiao},\ and\ \citenamefont {Kimble}}]{Wu1987}%
  \BibitemOpen
  \bibfield  {author} {\bibinfo {author} {\bibfnamefont {L.-A.}\ \bibnamefont
  {Wu}}, \bibinfo {author} {\bibfnamefont {M.}~\bibnamefont {Xiao}},\ and\
  \bibinfo {author} {\bibfnamefont {H.~J.}\ \bibnamefont {Kimble}},\ }\bibfield
   {title} {\bibinfo {title} {Squeezed states of light from an optical
  parametric oscillator},\ }\href {https://doi.org/10.1364/JOSAB.4.001465}
  {\bibfield  {journal} {\bibinfo  {journal} {J. Opt. Soc. Am. B}\ }\textbf
  {\bibinfo {volume} {4}},\ \bibinfo {pages} {1465} (\bibinfo {year}
  {1987})}\BibitemShut {NoStop}%
\bibitem [{\citenamefont {Vahlbruch}\ \emph {et~al.}(2016)\citenamefont
  {Vahlbruch}, \citenamefont {Mehmet}, \citenamefont {Danzmann},\ and\
  \citenamefont {Schnabel}}]{Vahlbruch}%
  \BibitemOpen
  \bibfield  {author} {\bibinfo {author} {\bibfnamefont {H.}~\bibnamefont
  {Vahlbruch}}, \bibinfo {author} {\bibfnamefont {M.}~\bibnamefont {Mehmet}},
  \bibinfo {author} {\bibfnamefont {K.}~\bibnamefont {Danzmann}},\ and\
  \bibinfo {author} {\bibfnamefont {R.}~\bibnamefont {Schnabel}},\ }\bibfield
  {title} {\bibinfo {title} {Detection of 15 {dB} squeezed states of light and
  their application for the absolute calibration of photoelectric quantum
  efficiency},\ }\href {https://doi.org/10.1103/PhysRevLett.117.110801}
  {\bibfield  {journal} {\bibinfo  {journal} {Phys. Rev. Lett.}\ }\textbf
  {\bibinfo {volume} {117}},\ \bibinfo {pages} {110801} (\bibinfo {year}
  {2016})}\BibitemShut {NoStop}%
\bibitem [{\citenamefont {Agarwal}\ and\ \citenamefont
  {Tara}(1992)}]{agarwal92}%
  \BibitemOpen
  \bibfield  {author} {\bibinfo {author} {\bibfnamefont {G.~S.}\ \bibnamefont
  {Agarwal}}\ and\ \bibinfo {author} {\bibfnamefont {K.}~\bibnamefont {Tara}},\
  }\bibfield  {title} {\bibinfo {title} {Nonclassical character of states
  exhibiting no squeezing or sub-{P}oissonian statistics},\ }\href
  {https://doi.org/10.1103/PhysRevA.46.485} {\bibfield  {journal} {\bibinfo
  {journal} {Phys. Rev. A}\ }\textbf {\bibinfo {volume} {46}},\ \bibinfo
  {pages} {485} (\bibinfo {year} {1992})}\BibitemShut {NoStop}%
\bibitem [{\citenamefont {Reid}\ and\ \citenamefont {Walls}(1986)}]{reid1986}%
  \BibitemOpen
  \bibfield  {author} {\bibinfo {author} {\bibfnamefont {M.~D.}\ \bibnamefont
  {Reid}}\ and\ \bibinfo {author} {\bibfnamefont {D.~F.}\ \bibnamefont
  {Walls}},\ }\bibfield  {title} {\bibinfo {title} {Violations of classical
  inequalities in quantum optics},\ }\href
  {https://doi.org/10.1103/PhysRevA.34.1260} {\bibfield  {journal} {\bibinfo
  {journal} {Phys. Rev. A}\ }\textbf {\bibinfo {volume} {34}},\ \bibinfo
  {pages} {1260} (\bibinfo {year} {1986})}\BibitemShut {NoStop}%
\bibitem [{\citenamefont {Hillery}(1987)}]{Hillery1987}%
  \BibitemOpen
  \bibfield  {author} {\bibinfo {author} {\bibfnamefont {M.}~\bibnamefont
  {Hillery}},\ }\bibfield  {title} {\bibinfo {title} {Nonclassical distance in
  quantum optics},\ }\href {https://doi.org/10.1103/PhysRevA.35.725} {\bibfield
   {journal} {\bibinfo  {journal} {Phys. Rev. A}\ }\textbf {\bibinfo {volume}
  {35}},\ \bibinfo {pages} {725} (\bibinfo {year} {1987})}\BibitemShut
  {NoStop}%
\bibitem [{\citenamefont {Lee}(1991)}]{Lee1991}%
  \BibitemOpen
  \bibfield  {author} {\bibinfo {author} {\bibfnamefont {C.~T.}\ \bibnamefont
  {Lee}},\ }\bibfield  {title} {\bibinfo {title} {Measure of the
  nonclassicality of nonclassical states},\ }\href
  {https://doi.org/10.1103/PhysRevA.44.R2775} {\bibfield  {journal} {\bibinfo
  {journal} {Phys. Rev. A}\ }\textbf {\bibinfo {volume} {44}},\ \bibinfo
  {pages} {R2775} (\bibinfo {year} {1991})}\BibitemShut {NoStop}%
\bibitem [{\citenamefont {Agarwal}(1993)}]{agarwal93}%
  \BibitemOpen
  \bibfield  {author} {\bibinfo {author} {\bibfnamefont {G.}~\bibnamefont
  {Agarwal}},\ }\bibfield  {title} {\bibinfo {title} {Nonclassical
  characteristics of the marginals for the radiation field},\ }\href
  {https://doi.org/http://dx.doi.org/10.1016/0030-4018(93)90059-E} {\bibfield
  {journal} {\bibinfo  {journal} {Opt. Commun.}\ }\textbf {\bibinfo {volume}
  {95}},\ \bibinfo {pages} {109 } (\bibinfo {year} {1993})}\BibitemShut
  {NoStop}%
\bibitem [{\citenamefont {Klyshko}(1996)}]{klyshko1996}%
  \BibitemOpen
  \bibfield  {author} {\bibinfo {author} {\bibfnamefont {D.~N.}\ \bibnamefont
  {Klyshko}},\ }\bibfield  {title} {\bibinfo {title} {Observable signs of
  nonclassical light},\ }\href
  {https://doi.org/https://doi.org/10.1016/0375-9601(96)00091-6} {\bibfield
  {journal} {\bibinfo  {journal} {Phys. Lett. A}\ }\textbf {\bibinfo {volume}
  {213}},\ \bibinfo {pages} {7} (\bibinfo {year} {1996})}\BibitemShut {NoStop}%
\bibitem [{\citenamefont {Vogel}(2000)}]{vogel00}%
  \BibitemOpen
  \bibfield  {author} {\bibinfo {author} {\bibfnamefont {W.}~\bibnamefont
  {Vogel}},\ }\bibfield  {title} {\bibinfo {title} {Nonclassical states: An
  observable criterion},\ }\href {https://doi.org/10.1103/PhysRevLett.84.1849}
  {\bibfield  {journal} {\bibinfo  {journal} {Phys. Rev. Lett.}\ }\textbf
  {\bibinfo {volume} {84}},\ \bibinfo {pages} {1849} (\bibinfo {year}
  {2000})}\BibitemShut {NoStop}%
\bibitem [{\citenamefont {Richter}\ and\ \citenamefont
  {Vogel}(2002)}]{richter02}%
  \BibitemOpen
  \bibfield  {author} {\bibinfo {author} {\bibfnamefont {T.}~\bibnamefont
  {Richter}}\ and\ \bibinfo {author} {\bibfnamefont {W.}~\bibnamefont
  {Vogel}},\ }\bibfield  {title} {\bibinfo {title} {Nonclassicality of quantum
  states: A hierarchy of observable conditions},\ }\href
  {https://doi.org/10.1103/PhysRevLett.89.283601} {\bibfield  {journal}
  {\bibinfo  {journal} {Phys. Rev. Lett.}\ }\textbf {\bibinfo {volume} {89}},\
  \bibinfo {pages} {283601} (\bibinfo {year} {2002})}\BibitemShut {NoStop}%
\bibitem [{\citenamefont {Asb\'oth}\ \emph {et~al.}(2005)\citenamefont
  {Asb\'oth}, \citenamefont {Calsamiglia},\ and\ \citenamefont
  {Ritsch}}]{Asboth2005}%
  \BibitemOpen
  \bibfield  {author} {\bibinfo {author} {\bibfnamefont {J.~K.}\ \bibnamefont
  {Asb\'oth}}, \bibinfo {author} {\bibfnamefont {J.}~\bibnamefont
  {Calsamiglia}},\ and\ \bibinfo {author} {\bibfnamefont {H.}~\bibnamefont
  {Ritsch}},\ }\bibfield  {title} {\bibinfo {title} {Computable measure of
  nonclassicality for light},\ }\href
  {https://doi.org/10.1103/PhysRevLett.94.173602} {\bibfield  {journal}
  {\bibinfo  {journal} {Phys. Rev. Lett.}\ }\textbf {\bibinfo {volume} {94}},\
  \bibinfo {pages} {173602} (\bibinfo {year} {2005})}\BibitemShut {NoStop}%
\bibitem [{\citenamefont {Kiesel}\ \emph {et~al.}(2008)\citenamefont {Kiesel},
  \citenamefont {Vogel}, \citenamefont {Parigi}, \citenamefont {Zavatta},\ and\
  \citenamefont {Bellini}}]{kiesel08}%
  \BibitemOpen
  \bibfield  {author} {\bibinfo {author} {\bibfnamefont {T.}~\bibnamefont
  {Kiesel}}, \bibinfo {author} {\bibfnamefont {W.}~\bibnamefont {Vogel}},
  \bibinfo {author} {\bibfnamefont {V.}~\bibnamefont {Parigi}}, \bibinfo
  {author} {\bibfnamefont {A.}~\bibnamefont {Zavatta}},\ and\ \bibinfo {author}
  {\bibfnamefont {M.}~\bibnamefont {Bellini}},\ }\bibfield  {title} {\bibinfo
  {title} {Experimental determination of a nonclassical {Glauber-Sudarshan $P$}
  function},\ }\href {https://doi.org/10.1103/PhysRevA.78.021804} {\bibfield
  {journal} {\bibinfo  {journal} {Phys. Rev. A}\ }\textbf {\bibinfo {volume}
  {78}},\ \bibinfo {pages} {021804(R)} (\bibinfo {year} {2008})}\BibitemShut
  {NoStop}%
\bibitem [{\citenamefont {Rivas}\ and\ \citenamefont {Luis}(2009)}]{rivas2009}%
  \BibitemOpen
  \bibfield  {author} {\bibinfo {author} {\bibfnamefont {A.}~\bibnamefont
  {Rivas}}\ and\ \bibinfo {author} {\bibfnamefont {A.}~\bibnamefont {Luis}},\
  }\bibfield  {title} {\bibinfo {title} {Nonclassicality of states and
  measurements by breaking classical bounds on statistics},\ }\href
  {https://doi.org/10.1103/PhysRevA.79.042105} {\bibfield  {journal} {\bibinfo
  {journal} {Phys. Rev. A}\ }\textbf {\bibinfo {volume} {79}},\ \bibinfo
  {pages} {042105} (\bibinfo {year} {2009})}\BibitemShut {NoStop}%
\bibitem [{\citenamefont {Kiesel}\ and\ \citenamefont
  {Vogel}(2010)}]{kiesel10}%
  \BibitemOpen
  \bibfield  {author} {\bibinfo {author} {\bibfnamefont {T.}~\bibnamefont
  {Kiesel}}\ and\ \bibinfo {author} {\bibfnamefont {W.}~\bibnamefont {Vogel}},\
  }\bibfield  {title} {\bibinfo {title} {Nonclassicality filters and
  quasiprobabilities},\ }\href {https://doi.org/10.1103/PhysRevA.82.032107}
  {\bibfield  {journal} {\bibinfo  {journal} {Phys. Rev. A}\ }\textbf {\bibinfo
  {volume} {82}},\ \bibinfo {pages} {032107} (\bibinfo {year}
  {2010})}\BibitemShut {NoStop}%
\bibitem [{\citenamefont {Kiesel}\ \emph {et~al.}(2011)\citenamefont {Kiesel},
  \citenamefont {Vogel}, \citenamefont {Bellini},\ and\ \citenamefont
  {Zavatta}}]{kiesel11b}%
  \BibitemOpen
  \bibfield  {author} {\bibinfo {author} {\bibfnamefont {T.}~\bibnamefont
  {Kiesel}}, \bibinfo {author} {\bibfnamefont {W.}~\bibnamefont {Vogel}},
  \bibinfo {author} {\bibfnamefont {M.}~\bibnamefont {Bellini}},\ and\ \bibinfo
  {author} {\bibfnamefont {A.}~\bibnamefont {Zavatta}},\ }\bibfield  {title}
  {\bibinfo {title} {Nonclassicality quasiprobability of single-photon-added
  thermal states},\ }\href {https://doi.org/10.1103/PhysRevA.83.032116}
  {\bibfield  {journal} {\bibinfo  {journal} {Phys. Rev. A}\ }\textbf {\bibinfo
  {volume} {83}},\ \bibinfo {pages} {032116} (\bibinfo {year}
  {2011})}\BibitemShut {NoStop}%
\bibitem [{\citenamefont {Sperling}\ \emph
  {et~al.}(2012{\natexlab{a}})\citenamefont {Sperling}, \citenamefont {Vogel},\
  and\ \citenamefont {Agarwal}}]{sperling12c}%
  \BibitemOpen
  \bibfield  {author} {\bibinfo {author} {\bibfnamefont {J.}~\bibnamefont
  {Sperling}}, \bibinfo {author} {\bibfnamefont {W.}~\bibnamefont {Vogel}},\
  and\ \bibinfo {author} {\bibfnamefont {G.~S.}\ \bibnamefont {Agarwal}},\
  }\bibfield  {title} {\bibinfo {title} {Sub-binomial light},\ }\href
  {https://doi.org/10.1103/PhysRevLett.109.093601} {\bibfield  {journal}
  {\bibinfo  {journal} {Phys. Rev. Lett.}\ }\textbf {\bibinfo {volume} {109}},\
  \bibinfo {pages} {093601} (\bibinfo {year} {2012}{\natexlab{a}})}\BibitemShut
  {NoStop}%
\bibitem [{\citenamefont {Bartley}\ \emph {et~al.}(2013)\citenamefont
  {Bartley}, \citenamefont {Donati}, \citenamefont {Jin}, \citenamefont
  {Datta}, \citenamefont {Barbieri},\ and\ \citenamefont
  {Walmsley}}]{bartley13}%
  \BibitemOpen
  \bibfield  {author} {\bibinfo {author} {\bibfnamefont {T.~J.}\ \bibnamefont
  {Bartley}}, \bibinfo {author} {\bibfnamefont {G.}~\bibnamefont {Donati}},
  \bibinfo {author} {\bibfnamefont {X.-M.}\ \bibnamefont {Jin}}, \bibinfo
  {author} {\bibfnamefont {A.}~\bibnamefont {Datta}}, \bibinfo {author}
  {\bibfnamefont {M.}~\bibnamefont {Barbieri}},\ and\ \bibinfo {author}
  {\bibfnamefont {I.~A.}\ \bibnamefont {Walmsley}},\ }\bibfield  {title}
  {\bibinfo {title} {Direct observation of sub-binomial light},\ }\href
  {https://doi.org/10.1103/PhysRevLett.110.173602} {\bibfield  {journal}
  {\bibinfo  {journal} {Phys. Rev. Lett.}\ }\textbf {\bibinfo {volume} {110}},\
  \bibinfo {pages} {173602} (\bibinfo {year} {2013})}\BibitemShut {NoStop}%
\bibitem [{\citenamefont {Sperling}\ \emph {et~al.}(2013)\citenamefont
  {Sperling}, \citenamefont {Vogel},\ and\ \citenamefont
  {Agarwal}}]{sperling13b}%
  \BibitemOpen
  \bibfield  {author} {\bibinfo {author} {\bibfnamefont {J.}~\bibnamefont
  {Sperling}}, \bibinfo {author} {\bibfnamefont {W.}~\bibnamefont {Vogel}},\
  and\ \bibinfo {author} {\bibfnamefont {G.~S.}\ \bibnamefont {Agarwal}},\
  }\bibfield  {title} {\bibinfo {title} {Correlation measurements with on-off
  detectors},\ }\href {https://doi.org/10.1103/PhysRevA.88.043821} {\bibfield
  {journal} {\bibinfo  {journal} {Phys. Rev. A}\ }\textbf {\bibinfo {volume}
  {88}},\ \bibinfo {pages} {043821} (\bibinfo {year} {2013})}\BibitemShut
  {NoStop}%
\bibitem [{\citenamefont {Agudelo}\ \emph {et~al.}(2013)\citenamefont
  {Agudelo}, \citenamefont {Sperling},\ and\ \citenamefont
  {Vogel}}]{Agudelo2013}%
  \BibitemOpen
  \bibfield  {author} {\bibinfo {author} {\bibfnamefont {E.}~\bibnamefont
  {Agudelo}}, \bibinfo {author} {\bibfnamefont {J.}~\bibnamefont {Sperling}},\
  and\ \bibinfo {author} {\bibfnamefont {W.}~\bibnamefont {Vogel}},\ }\bibfield
   {title} {\bibinfo {title} {Quasiprobabilities for multipartite quantum
  correlations of light},\ }\href {https://doi.org/10.1103/PhysRevA.87.033811}
  {\bibfield  {journal} {\bibinfo  {journal} {Phys. Rev. A}\ }\textbf {\bibinfo
  {volume} {87}},\ \bibinfo {pages} {033811} (\bibinfo {year}
  {2013})}\BibitemShut {NoStop}%
\bibitem [{\citenamefont {Park}\ \emph {et~al.}(2015)\citenamefont {Park},
  \citenamefont {Zhang}, \citenamefont {Lee}, \citenamefont {Ji}, \citenamefont
  {Um}, \citenamefont {Lv}, \citenamefont {Kim},\ and\ \citenamefont
  {Nha}}]{park2015a}%
  \BibitemOpen
  \bibfield  {author} {\bibinfo {author} {\bibfnamefont {J.}~\bibnamefont
  {Park}}, \bibinfo {author} {\bibfnamefont {J.}~\bibnamefont {Zhang}},
  \bibinfo {author} {\bibfnamefont {J.}~\bibnamefont {Lee}}, \bibinfo {author}
  {\bibfnamefont {S.-W.}\ \bibnamefont {Ji}}, \bibinfo {author} {\bibfnamefont
  {M.}~\bibnamefont {Um}}, \bibinfo {author} {\bibfnamefont {D.}~\bibnamefont
  {Lv}}, \bibinfo {author} {\bibfnamefont {K.}~\bibnamefont {Kim}},\ and\
  \bibinfo {author} {\bibfnamefont {H.}~\bibnamefont {Nha}},\ }\bibfield
  {title} {\bibinfo {title} {Testing nonclassicality and non-{G}aussianity in
  phase space},\ }\href {https://doi.org/10.1103/PhysRevLett.114.190402}
  {\bibfield  {journal} {\bibinfo  {journal} {Phys. Rev. Lett.}\ }\textbf
  {\bibinfo {volume} {114}},\ \bibinfo {pages} {190402} (\bibinfo {year}
  {2015})}\BibitemShut {NoStop}%
\bibitem [{\citenamefont {Park}\ and\ \citenamefont {Nha}(2015)}]{park2015b}%
  \BibitemOpen
  \bibfield  {author} {\bibinfo {author} {\bibfnamefont {J.}~\bibnamefont
  {Park}}\ and\ \bibinfo {author} {\bibfnamefont {H.}~\bibnamefont {Nha}},\
  }\bibfield  {title} {\bibinfo {title} {Demonstrating nonclassicality and
  non-{G}aussianity of single-mode fields: {Bell}-type tests using generalized
  phase-space distributions},\ }\href
  {https://doi.org/10.1103/PhysRevA.92.062134} {\bibfield  {journal} {\bibinfo
  {journal} {Phys. Rev. A}\ }\textbf {\bibinfo {volume} {92}},\ \bibinfo
  {pages} {062134} (\bibinfo {year} {2015})}\BibitemShut {NoStop}%
\bibitem [{\citenamefont {Luis}\ \emph {et~al.}(2015)\citenamefont {Luis},
  \citenamefont {Sperling},\ and\ \citenamefont {Vogel}}]{luis15}%
  \BibitemOpen
  \bibfield  {author} {\bibinfo {author} {\bibfnamefont {A.}~\bibnamefont
  {Luis}}, \bibinfo {author} {\bibfnamefont {J.}~\bibnamefont {Sperling}},\
  and\ \bibinfo {author} {\bibfnamefont {W.}~\bibnamefont {Vogel}},\ }\bibfield
   {title} {\bibinfo {title} {Nonclassicality phase-space functions: More
  insight with fewer detectors},\ }\href
  {https://doi.org/10.1103/PhysRevLett.114.103602} {\bibfield  {journal}
  {\bibinfo  {journal} {Phys. Rev. Lett.}\ }\textbf {\bibinfo {volume} {114}},\
  \bibinfo {pages} {103602} (\bibinfo {year} {2015})}\BibitemShut {NoStop}%
\bibitem [{\citenamefont {Miranowicz}\ \emph
  {et~al.}(2015{\natexlab{a}})\citenamefont {Miranowicz}, \citenamefont
  {Bartkiewicz}, \citenamefont {Pathak}, \citenamefont
  {Pe\ifmmode~\check{r}\else \v{r}\fi{}ina}, \citenamefont {Chen},\ and\
  \citenamefont {Nori}}]{Miranowicz2015a}%
  \BibitemOpen
  \bibfield  {author} {\bibinfo {author} {\bibfnamefont {A.}~\bibnamefont
  {Miranowicz}}, \bibinfo {author} {\bibfnamefont {K.}~\bibnamefont
  {Bartkiewicz}}, \bibinfo {author} {\bibfnamefont {A.}~\bibnamefont {Pathak}},
  \bibinfo {author} {\bibfnamefont {J.}~\bibnamefont {Pe\ifmmode~\check{r}\else
  \v{r}\fi{}ina}}, \bibinfo {author} {\bibfnamefont {Y.-N.}\ \bibnamefont
  {Chen}},\ and\ \bibinfo {author} {\bibfnamefont {F.}~\bibnamefont {Nori}},\
  }\bibfield  {title} {\bibinfo {title} {Statistical mixtures of states can be
  more quantum than their superpositions: Comparison of nonclassicality
  measures for single-qubit states},\ }\href
  {https://doi.org/10.1103/PhysRevA.91.042309} {\bibfield  {journal} {\bibinfo
  {journal} {Phys. Rev. A}\ }\textbf {\bibinfo {volume} {91}},\ \bibinfo
  {pages} {042309} (\bibinfo {year} {2015}{\natexlab{a}})}\BibitemShut
  {NoStop}%
\bibitem [{\citenamefont {Miranowicz}\ \emph
  {et~al.}(2015{\natexlab{b}})\citenamefont {Miranowicz}, \citenamefont
  {Bartkiewicz}, \citenamefont {Lambert}, \citenamefont {Chen},\ and\
  \citenamefont {Nori}}]{Miranowicz2015b}%
  \BibitemOpen
  \bibfield  {author} {\bibinfo {author} {\bibfnamefont {A.}~\bibnamefont
  {Miranowicz}}, \bibinfo {author} {\bibfnamefont {K.}~\bibnamefont
  {Bartkiewicz}}, \bibinfo {author} {\bibfnamefont {N.}~\bibnamefont
  {Lambert}}, \bibinfo {author} {\bibfnamefont {Y.-N.}\ \bibnamefont {Chen}},\
  and\ \bibinfo {author} {\bibfnamefont {F.}~\bibnamefont {Nori}},\ }\bibfield
  {title} {\bibinfo {title} {Increasing relative nonclassicality quantified by
  standard entanglement potentials by dissipation and unbalanced beam
  splitting},\ }\href {https://doi.org/10.1103/PhysRevA.92.062314} {\bibfield
  {journal} {\bibinfo  {journal} {Phys. Rev. A}\ }\textbf {\bibinfo {volume}
  {92}},\ \bibinfo {pages} {062314} (\bibinfo {year}
  {2015}{\natexlab{b}})}\BibitemShut {NoStop}%
\bibitem [{\citenamefont {Yadin}\ \emph {et~al.}(2018)\citenamefont {Yadin},
  \citenamefont {Binder}, \citenamefont {Thompson}, \citenamefont
  {Narasimhachar}, \citenamefont {Gu},\ and\ \citenamefont {Kim}}]{Yadin2019}%
  \BibitemOpen
  \bibfield  {author} {\bibinfo {author} {\bibfnamefont {B.}~\bibnamefont
  {Yadin}}, \bibinfo {author} {\bibfnamefont {F.~C.}\ \bibnamefont {Binder}},
  \bibinfo {author} {\bibfnamefont {J.}~\bibnamefont {Thompson}}, \bibinfo
  {author} {\bibfnamefont {V.}~\bibnamefont {Narasimhachar}}, \bibinfo {author}
  {\bibfnamefont {M.}~\bibnamefont {Gu}},\ and\ \bibinfo {author}
  {\bibfnamefont {M.~S.}\ \bibnamefont {Kim}},\ }\bibfield  {title} {\bibinfo
  {title} {Operational resource theory of continuous-variable
  nonclassicality},\ }\href {https://doi.org/10.1103/PhysRevX.8.041038}
  {\bibfield  {journal} {\bibinfo  {journal} {Phys. Rev. X}\ }\textbf {\bibinfo
  {volume} {8}},\ \bibinfo {pages} {041038} (\bibinfo {year}
  {2018})}\BibitemShut {NoStop}%
\bibitem [{\citenamefont {Luo}\ and\ \citenamefont {Zhang}(2019)}]{Luo2019}%
  \BibitemOpen
  \bibfield  {author} {\bibinfo {author} {\bibfnamefont {S.}~\bibnamefont
  {Luo}}\ and\ \bibinfo {author} {\bibfnamefont {Y.}~\bibnamefont {Zhang}},\
  }\bibfield  {title} {\bibinfo {title} {Quantifying nonclassicality via
  {W}igner-{Y}anase skew information},\ }\href
  {https://doi.org/10.1103/PhysRevA.100.032116} {\bibfield  {journal} {\bibinfo
   {journal} {Phys. Rev. A}\ }\textbf {\bibinfo {volume} {100}},\ \bibinfo
  {pages} {032116} (\bibinfo {year} {2019})}\BibitemShut {NoStop}%
\bibitem [{\citenamefont {Pe\v{r}ina}\ \emph {et~al.}(2020)\citenamefont
  {Pe\v{r}ina}, \citenamefont {Mich\'{a}lek},\ and\ \citenamefont
  {Haderka}}]{Perina2020}%
  \BibitemOpen
  \bibfield  {author} {\bibinfo {author} {\bibfnamefont {J.}~\bibnamefont
  {Pe\v{r}ina}}, \bibinfo {author} {\bibfnamefont {V.}~\bibnamefont
  {Mich\'{a}lek}},\ and\ \bibinfo {author} {\bibfnamefont {O.}~\bibnamefont
  {Haderka}},\ }\bibfield  {title} {\bibinfo {title} {Non-classicality of
  optical fields as observed in photocount and photon-number distributions},\
  }\href {https://doi.org/10.1364/OE.405548} {\bibfield  {journal} {\bibinfo
  {journal} {Opt. Express}\ }\textbf {\bibinfo {volume} {28}},\ \bibinfo
  {pages} {32620} (\bibinfo {year} {2020})}\BibitemShut {NoStop}%
\bibitem [{\citenamefont {Bohmann}\ and\ \citenamefont
  {Agudelo}(2020)}]{bohmann2020}%
  \BibitemOpen
  \bibfield  {author} {\bibinfo {author} {\bibfnamefont {M.}~\bibnamefont
  {Bohmann}}\ and\ \bibinfo {author} {\bibfnamefont {E.}~\bibnamefont
  {Agudelo}},\ }\bibfield  {title} {\bibinfo {title} {Phase-space inequalities
  beyond negativities},\ }\href
  {https://doi.org/10.1103/PhysRevLett.124.133601} {\bibfield  {journal}
  {\bibinfo  {journal} {Phys. Rev. Lett.}\ }\textbf {\bibinfo {volume} {124}},\
  \bibinfo {pages} {133601} (\bibinfo {year} {2020})}\BibitemShut {NoStop}%
\bibitem [{\citenamefont {Bohmann}\ \emph {et~al.}(2020)\citenamefont
  {Bohmann}, \citenamefont {Agudelo},\ and\ \citenamefont
  {Sperling}}]{bohmann2020b}%
  \BibitemOpen
  \bibfield  {author} {\bibinfo {author} {\bibfnamefont {M.}~\bibnamefont
  {Bohmann}}, \bibinfo {author} {\bibfnamefont {E.}~\bibnamefont {Agudelo}},\
  and\ \bibinfo {author} {\bibfnamefont {J.}~\bibnamefont {Sperling}},\
  }\bibfield  {title} {\bibinfo {title} {Probing nonclassicality with matrices
  of phase-space distributions},\ }\href
  {https://doi.org/10.22331/q-2020-10-15-343} {\bibfield  {journal} {\bibinfo
  {journal} {{Quantum}}\ }\textbf {\bibinfo {volume} {4}},\ \bibinfo {pages}
  {343} (\bibinfo {year} {2020})}\BibitemShut {NoStop}%
\bibitem [{\citenamefont {Semenov}\ and\ \citenamefont
  {Klimov}(2021)}]{Semenov2021}%
  \BibitemOpen
  \bibfield  {author} {\bibinfo {author} {\bibfnamefont {A.~A.}\ \bibnamefont
  {Semenov}}\ and\ \bibinfo {author} {\bibfnamefont {A.~B.}\ \bibnamefont
  {Klimov}},\ }\bibfield  {title} {\bibinfo {title} {Dual form of the
  phase-space classical simulation problem in quantum optics},\ }\href
  {https://doi.org/10.1088/1367-2630/ac40cc} {\bibfield  {journal} {\bibinfo
  {journal} {New J. Phys.}\ }\textbf {\bibinfo {volume} {23}},\ \bibinfo
  {pages} {123046} (\bibinfo {year} {2021})}\BibitemShut {NoStop}%
\bibitem [{\citenamefont {Innocenti}\ \emph {et~al.}(2022)\citenamefont
  {Innocenti}, \citenamefont {Lachman},\ and\ \citenamefont
  {Filip}}]{Innocenti2022}%
  \BibitemOpen
  \bibfield  {author} {\bibinfo {author} {\bibfnamefont {L.}~\bibnamefont
  {Innocenti}}, \bibinfo {author} {\bibfnamefont {L.}~\bibnamefont {Lachman}},\
  and\ \bibinfo {author} {\bibfnamefont {R.}~\bibnamefont {Filip}},\ }\bibfield
   {title} {\bibinfo {title} {Nonclassicality detection from few {Fock}-state
  probabilities},\ }\href {https://doi.org/10.1038/s41534-022-00538-y}
  {\bibfield  {journal} {\bibinfo  {journal} {npj Quantum Inf.}\ }\textbf
  {\bibinfo {volume} {8}},\ \bibinfo {pages} {30} (\bibinfo {year}
  {2022})}\BibitemShut {NoStop}%
\bibitem [{\citenamefont {Innocenti}\ \emph {et~al.}(2023)\citenamefont
  {Innocenti}, \citenamefont {Lachman},\ and\ \citenamefont
  {Filip}}]{Innocenti2023}%
  \BibitemOpen
  \bibfield  {author} {\bibinfo {author} {\bibfnamefont {L.}~\bibnamefont
  {Innocenti}}, \bibinfo {author} {\bibfnamefont {L.}~\bibnamefont {Lachman}},\
  and\ \bibinfo {author} {\bibfnamefont {R.}~\bibnamefont {Filip}},\ }\bibfield
   {title} {\bibinfo {title} {Coherence-based operational nonclassicality
  criteria},\ }\href {https://doi.org/10.1103/PhysRevLett.131.160201}
  {\bibfield  {journal} {\bibinfo  {journal} {Phys. Rev. Lett.}\ }\textbf
  {\bibinfo {volume} {131}},\ \bibinfo {pages} {160201} (\bibinfo {year}
  {2023})}\BibitemShut {NoStop}%
\bibitem [{\citenamefont {Fiur\'a\ifmmode~\check{s}\else
  \v{s}\fi{}ek}(2024)}]{Fiurasek2024}%
  \BibitemOpen
  \bibfield  {author} {\bibinfo {author} {\bibfnamefont {J.}~\bibnamefont
  {Fiur\'a\ifmmode~\check{s}\else \v{s}\fi{}ek}},\ }\bibfield  {title}
  {\bibinfo {title} {Tight nonclassicality criteria for unbalanced {H}anbury
  {B}rown--{T}wiss measurement scheme with click detectors},\ }\href
  {https://doi.org/10.1103/PhysRevA.109.033713} {\bibfield  {journal} {\bibinfo
   {journal} {Phys. Rev. A}\ }\textbf {\bibinfo {volume} {109}},\ \bibinfo
  {pages} {033713} (\bibinfo {year} {2024})}\BibitemShut {NoStop}%
\bibitem [{\citenamefont {Kovtoniuk}\ \emph {et~al.}(2024)\citenamefont
  {Kovtoniuk}, \citenamefont {Stolyarov}, \citenamefont {Kliushnichenko},\ and\
  \citenamefont {Semenov}}]{Kovtoniuk2024}%
  \BibitemOpen
  \bibfield  {author} {\bibinfo {author} {\bibfnamefont {V.~S.}\ \bibnamefont
  {Kovtoniuk}}, \bibinfo {author} {\bibfnamefont {E.~V.}\ \bibnamefont
  {Stolyarov}}, \bibinfo {author} {\bibfnamefont {O.~V.}\ \bibnamefont
  {Kliushnichenko}},\ and\ \bibinfo {author} {\bibfnamefont {A.~A.}\
  \bibnamefont {Semenov}},\ }\bibfield  {title} {\bibinfo {title} {Tight
  inequalities for nonclassicality of measurement statistics},\ }\href
  {https://doi.org/10.1103/PhysRevA.109.053710} {\bibfield  {journal} {\bibinfo
   {journal} {Phys. Rev. A}\ }\textbf {\bibinfo {volume} {109}},\ \bibinfo
  {pages} {053710} (\bibinfo {year} {2024})}\BibitemShut {NoStop}%
\bibitem [{\citenamefont {Manikandan}\ and\ \citenamefont
  {Wilczek}(2025)}]{Manikandan2025}%
  \BibitemOpen
  \bibfield  {author} {\bibinfo {author} {\bibfnamefont {S.~K.}\ \bibnamefont
  {Manikandan}}\ and\ \bibinfo {author} {\bibfnamefont {F.}~\bibnamefont
  {Wilczek}},\ }\bibfield  {title} {\bibinfo {title} {Testing the
  coherent-state description of radiation fields},\ }\href
  {https://doi.org/10.1103/PhysRevA.111.033705} {\bibfield  {journal} {\bibinfo
   {journal} {Phys. Rev. A}\ }\textbf {\bibinfo {volume} {111}},\ \bibinfo
  {pages} {033705} (\bibinfo {year} {2025})}\BibitemShut {NoStop}%
\bibitem [{\citenamefont {Horodecki}\ and\ \citenamefont
  {Jannathan}(2013)}]{Horodecki2013}%
  \BibitemOpen
  \bibfield  {author} {\bibinfo {author} {\bibfnamefont {M.}~\bibnamefont
  {Horodecki}}\ and\ \bibinfo {author} {\bibfnamefont {O.}~\bibnamefont
  {Jannathan}},\ }\bibfield  {title} {\bibinfo {title} {(quantumness in the
  context of) resource theories},\ }\href
  {https://doi.org/10.1142/S0217979213450197} {\bibfield  {journal} {\bibinfo
  {journal} {Int. J. Mod. Phys. B}\ }\textbf {\bibinfo {volume} {27}},\
  \bibinfo {pages} {1345019} (\bibinfo {year} {2013})}\BibitemShut {NoStop}%
\bibitem [{\citenamefont {Chitambar}\ and\ \citenamefont
  {Gour}(2019)}]{Chitambar2019}%
  \BibitemOpen
  \bibfield  {author} {\bibinfo {author} {\bibfnamefont {E.}~\bibnamefont
  {Chitambar}}\ and\ \bibinfo {author} {\bibfnamefont {G.}~\bibnamefont
  {Gour}},\ }\bibfield  {title} {\bibinfo {title} {Quantum resource theories},\
  }\href {https://doi.org/10.1103/RevModPhys.91.025001} {\bibfield  {journal}
  {\bibinfo  {journal} {Rev. Mod. Phys.}\ }\textbf {\bibinfo {volume} {91}},\
  \bibinfo {pages} {025001} (\bibinfo {year} {2019})}\BibitemShut {NoStop}%
\bibitem [{\citenamefont {Streltsov}\ \emph {et~al.}(2017)\citenamefont
  {Streltsov}, \citenamefont {Adesso},\ and\ \citenamefont
  {Plenio}}]{Streltsov2017}%
  \BibitemOpen
  \bibfield  {author} {\bibinfo {author} {\bibfnamefont {A.}~\bibnamefont
  {Streltsov}}, \bibinfo {author} {\bibfnamefont {G.}~\bibnamefont {Adesso}},\
  and\ \bibinfo {author} {\bibfnamefont {M.~B.}\ \bibnamefont {Plenio}},\
  }\bibfield  {title} {\bibinfo {title} {Colloquium: Quantum coherence as a
  resource},\ }\href {https://doi.org/10.1103/RevModPhys.89.041003} {\bibfield
  {journal} {\bibinfo  {journal} {Rev. Mod. Phys.}\ }\textbf {\bibinfo {volume}
  {89}},\ \bibinfo {pages} {041003} (\bibinfo {year} {2017})}\BibitemShut
  {NoStop}%
\bibitem [{\citenamefont {Ge}\ \emph {et~al.}(2020)\citenamefont {Ge},
  \citenamefont {Jacobs}, \citenamefont {Asiri}, \citenamefont {Foss-Feig},\
  and\ \citenamefont {Zubairy}}]{Ge2020a}%
  \BibitemOpen
  \bibfield  {author} {\bibinfo {author} {\bibfnamefont {W.}~\bibnamefont
  {Ge}}, \bibinfo {author} {\bibfnamefont {K.}~\bibnamefont {Jacobs}}, \bibinfo
  {author} {\bibfnamefont {S.}~\bibnamefont {Asiri}}, \bibinfo {author}
  {\bibfnamefont {M.}~\bibnamefont {Foss-Feig}},\ and\ \bibinfo {author}
  {\bibfnamefont {M.~S.}\ \bibnamefont {Zubairy}},\ }\bibfield  {title}
  {\bibinfo {title} {Operational resource theory of nonclassicality via quantum
  metrology},\ }\href {https://doi.org/10.1103/PhysRevResearch.2.023400}
  {\bibfield  {journal} {\bibinfo  {journal} {Phys. Rev. Res.}\ }\textbf
  {\bibinfo {volume} {2}},\ \bibinfo {pages} {023400} (\bibinfo {year}
  {2020})}\BibitemShut {NoStop}%
\bibitem [{\citenamefont {Ge}\ and\ \citenamefont {Zubairy}(2020)}]{Ge2020b}%
  \BibitemOpen
  \bibfield  {author} {\bibinfo {author} {\bibfnamefont {W.}~\bibnamefont
  {Ge}}\ and\ \bibinfo {author} {\bibfnamefont {M.~S.}\ \bibnamefont
  {Zubairy}},\ }\bibfield  {title} {\bibinfo {title} {Evaluating single-mode
  nonclassicality},\ }\href {https://doi.org/10.1103/PhysRevA.102.043703}
  {\bibfield  {journal} {\bibinfo  {journal} {Phys. Rev. A}\ }\textbf {\bibinfo
  {volume} {102}},\ \bibinfo {pages} {043703} (\bibinfo {year}
  {2020})}\BibitemShut {NoStop}%
\bibitem [{\citenamefont {Rahimi-Keshari}\ \emph {et~al.}(2016)\citenamefont
  {Rahimi-Keshari}, \citenamefont {Ralph},\ and\ \citenamefont
  {Caves}}]{rahimi-keshari16}%
  \BibitemOpen
  \bibfield  {author} {\bibinfo {author} {\bibfnamefont {S.}~\bibnamefont
  {Rahimi-Keshari}}, \bibinfo {author} {\bibfnamefont {T.~C.}\ \bibnamefont
  {Ralph}},\ and\ \bibinfo {author} {\bibfnamefont {C.~M.}\ \bibnamefont
  {Caves}},\ }\bibfield  {title} {\bibinfo {title} {Sufficient conditions for
  efficient classical simulation of quantum optics},\ }\href
  {https://doi.org/10.1103/PhysRevX.6.021039} {\bibfield  {journal} {\bibinfo
  {journal} {Phys. Rev. X}\ }\textbf {\bibinfo {volume} {6}},\ \bibinfo {pages}
  {021039} (\bibinfo {year} {2016})}\BibitemShut {NoStop}%
\bibitem [{\citenamefont {Chabaud}\ \emph {et~al.}(2024)\citenamefont
  {Chabaud}, \citenamefont {Ghobadi}, \citenamefont {Beigi},\ and\
  \citenamefont {Rahimi-Keshari}}]{Chabaud2024}%
  \BibitemOpen
  \bibfield  {author} {\bibinfo {author} {\bibfnamefont {U.}~\bibnamefont
  {Chabaud}}, \bibinfo {author} {\bibfnamefont {R.}~\bibnamefont {Ghobadi}},
  \bibinfo {author} {\bibfnamefont {S.}~\bibnamefont {Beigi}},\ and\ \bibinfo
  {author} {\bibfnamefont {S.}~\bibnamefont {Rahimi-Keshari}},\ }\bibfield
  {title} {\bibinfo {title} {Phase-space negativity as a computational resource
  for quantum kernel methods},\ }\href
  {https://doi.org/10.22331/q-2024-11-07-1519} {\bibfield  {journal} {\bibinfo
  {journal} {{Quantum}}\ }\textbf {\bibinfo {volume} {8}},\ \bibinfo {pages}
  {1519} (\bibinfo {year} {2024})}\BibitemShut {NoStop}%
\bibitem [{\citenamefont {Kelley}\ and\ \citenamefont
  {Kleiner}(1964)}]{kelley64}%
  \BibitemOpen
  \bibfield  {author} {\bibinfo {author} {\bibfnamefont {P.~L.}\ \bibnamefont
  {Kelley}}\ and\ \bibinfo {author} {\bibfnamefont {W.~H.}\ \bibnamefont
  {Kleiner}},\ }\bibfield  {title} {\bibinfo {title} {Theory of electromagnetic
  field measurement and photoelectron counting},\ }\href
  {https://doi.org/10.1103/PhysRev.136.A316} {\bibfield  {journal} {\bibinfo
  {journal} {Phys. Rev.}\ }\textbf {\bibinfo {volume} {136}},\ \bibinfo {pages}
  {A316} (\bibinfo {year} {1964})}\BibitemShut {NoStop}%
\bibitem [{\citenamefont {Sperling}\ \emph
  {et~al.}(2012{\natexlab{b}})\citenamefont {Sperling}, \citenamefont {Vogel},\
  and\ \citenamefont {Agarwal}}]{sperling12a}%
  \BibitemOpen
  \bibfield  {author} {\bibinfo {author} {\bibfnamefont {J.}~\bibnamefont
  {Sperling}}, \bibinfo {author} {\bibfnamefont {W.}~\bibnamefont {Vogel}},\
  and\ \bibinfo {author} {\bibfnamefont {G.~S.}\ \bibnamefont {Agarwal}},\
  }\bibfield  {title} {\bibinfo {title} {True photocounting statistics of
  multiple on-off detectors},\ }\href
  {https://doi.org/10.1103/PhysRevA.85.023820} {\bibfield  {journal} {\bibinfo
  {journal} {Phys. Rev. A}\ }\textbf {\bibinfo {volume} {85}},\ \bibinfo
  {pages} {023820} (\bibinfo {year} {2012}{\natexlab{b}})}\BibitemShut
  {NoStop}%
\bibitem [{\citenamefont {Semenov}\ \emph {et~al.}(2024)\citenamefont
  {Semenov}, \citenamefont {Samelin}, \citenamefont {Boldt}, \citenamefont
  {Sch\"unemann}, \citenamefont {Reiher}, \citenamefont {Vogel},\ and\
  \citenamefont {Hage}}]{Semenov2024}%
  \BibitemOpen
  \bibfield  {author} {\bibinfo {author} {\bibfnamefont {A.~A.}\ \bibnamefont
  {Semenov}}, \bibinfo {author} {\bibfnamefont {J.}~\bibnamefont {Samelin}},
  \bibinfo {author} {\bibfnamefont {C.}~\bibnamefont {Boldt}}, \bibinfo
  {author} {\bibfnamefont {M.}~\bibnamefont {Sch\"unemann}}, \bibinfo {author}
  {\bibfnamefont {C.}~\bibnamefont {Reiher}}, \bibinfo {author} {\bibfnamefont
  {W.}~\bibnamefont {Vogel}},\ and\ \bibinfo {author} {\bibfnamefont
  {B.}~\bibnamefont {Hage}},\ }\bibfield  {title} {\bibinfo {title}
  {Photocounting measurements with dead time and afterpulses in the
  continuous-wave regime},\ }\href
  {https://doi.org/10.1103/PhysRevA.109.013701} {\bibfield  {journal} {\bibinfo
   {journal} {Phys. Rev. A}\ }\textbf {\bibinfo {volume} {109}},\ \bibinfo
  {pages} {013701} (\bibinfo {year} {2024})}\BibitemShut {NoStop}%
\bibitem [{\citenamefont {Uzunova}\ and\ \citenamefont
  {Semenov}(2022)}]{Uzunova2022}%
  \BibitemOpen
  \bibfield  {author} {\bibinfo {author} {\bibfnamefont {V.~A.}\ \bibnamefont
  {Uzunova}}\ and\ \bibinfo {author} {\bibfnamefont {A.~A.}\ \bibnamefont
  {Semenov}},\ }\bibfield  {title} {\bibinfo {title} {Photocounting statistics
  of superconducting nanowire single-photon detectors},\ }\href
  {https://doi.org/10.1103/PhysRevA.105.063716} {\bibfield  {journal} {\bibinfo
   {journal} {Phys. Rev. A}\ }\textbf {\bibinfo {volume} {105}},\ \bibinfo
  {pages} {063716} (\bibinfo {year} {2022})}\BibitemShut {NoStop}%
\bibitem [{\citenamefont {Stolyarov}\ \emph {et~al.}(2023)\citenamefont
  {Stolyarov}, \citenamefont {Kliushnichenko}, \citenamefont {Kovtoniuk},\ and\
  \citenamefont {Semenov}}]{Stolyarov2023}%
  \BibitemOpen
  \bibfield  {author} {\bibinfo {author} {\bibfnamefont {E.~V.}\ \bibnamefont
  {Stolyarov}}, \bibinfo {author} {\bibfnamefont {O.~V.}\ \bibnamefont
  {Kliushnichenko}}, \bibinfo {author} {\bibfnamefont {V.~S.}\ \bibnamefont
  {Kovtoniuk}},\ and\ \bibinfo {author} {\bibfnamefont {A.~A.}\ \bibnamefont
  {Semenov}},\ }\bibfield  {title} {\bibinfo {title} {Photon-number resolution
  with microwave {J}osephson photomultipliers},\ }\href
  {https://doi.org/10.1103/PhysRevA.108.063710} {\bibfield  {journal} {\bibinfo
   {journal} {Phys. Rev. A}\ }\textbf {\bibinfo {volume} {108}},\ \bibinfo
  {pages} {063710} (\bibinfo {year} {2023})}\BibitemShut {NoStop}%
\bibitem [{\citenamefont {Cahill}\ and\ \citenamefont
  {Glauber}(1969{\natexlab{a}})}]{cahill69}%
  \BibitemOpen
  \bibfield  {author} {\bibinfo {author} {\bibfnamefont {K.~E.}\ \bibnamefont
  {Cahill}}\ and\ \bibinfo {author} {\bibfnamefont {R.~J.}\ \bibnamefont
  {Glauber}},\ }\bibfield  {title} {\bibinfo {title} {Density operators and
  quasiprobability distributions},\ }\href
  {https://doi.org/10.1103/PhysRev.177.1882} {\bibfield  {journal} {\bibinfo
  {journal} {Phys. Rev.}\ }\textbf {\bibinfo {volume} {177}},\ \bibinfo {pages}
  {1882} (\bibinfo {year} {1969}{\natexlab{a}})}\BibitemShut {NoStop}%
\bibitem [{\citenamefont {Cahill}\ and\ \citenamefont
  {Glauber}(1969{\natexlab{b}})}]{cahill69a}%
  \BibitemOpen
  \bibfield  {author} {\bibinfo {author} {\bibfnamefont {K.~E.}\ \bibnamefont
  {Cahill}}\ and\ \bibinfo {author} {\bibfnamefont {R.~J.}\ \bibnamefont
  {Glauber}},\ }\bibfield  {title} {\bibinfo {title} {Ordered expansions in
  boson amplitude operators},\ }\href
  {https://doi.org/10.1103/PhysRev.177.1857} {\bibfield  {journal} {\bibinfo
  {journal} {Phys. Rev.}\ }\textbf {\bibinfo {volume} {177}},\ \bibinfo {pages}
  {1857} (\bibinfo {year} {1969}{\natexlab{b}})}\BibitemShut {NoStop}%
\bibitem [{\citenamefont {Spring}\ \emph {et~al.}(2013)\citenamefont {Spring},
  \citenamefont {J.~Metcalf}, \citenamefont {Humphreys}, \citenamefont
  {Kolthammer}, \citenamefont {Jin}, \citenamefont {Barbieri}, \citenamefont
  {Datta}, \citenamefont {Thomas-Peter}, \citenamefont {Langford},
  \citenamefont {Kundys}, \citenamefont {Gates}, \citenamefont {Smith},
  \citenamefont {Smith},\ and\ \citenamefont {Walmsley}}]{Spring2013}%
  \BibitemOpen
  \bibfield  {author} {\bibinfo {author} {\bibfnamefont {J.~B.}\ \bibnamefont
  {Spring}}, \bibinfo {author} {\bibfnamefont {B.}~\bibnamefont {J.~Metcalf}},
  \bibinfo {author} {\bibfnamefont {P.~C.}\ \bibnamefont {Humphreys}}, \bibinfo
  {author} {\bibfnamefont {W.~S.}\ \bibnamefont {Kolthammer}}, \bibinfo
  {author} {\bibfnamefont {X.-M.}\ \bibnamefont {Jin}}, \bibinfo {author}
  {\bibfnamefont {M.}~\bibnamefont {Barbieri}}, \bibinfo {author}
  {\bibfnamefont {A.}~\bibnamefont {Datta}}, \bibinfo {author} {\bibfnamefont
  {N.}~\bibnamefont {Thomas-Peter}}, \bibinfo {author} {\bibfnamefont {N.~K.}\
  \bibnamefont {Langford}}, \bibinfo {author} {\bibfnamefont {D.}~\bibnamefont
  {Kundys}}, \bibinfo {author} {\bibfnamefont {J.~C.}\ \bibnamefont {Gates}},
  \bibinfo {author} {\bibfnamefont {B.~J.}\ \bibnamefont {Smith}}, \bibinfo
  {author} {\bibfnamefont {P.~G.~R.}\ \bibnamefont {Smith}},\ and\ \bibinfo
  {author} {\bibfnamefont {I.~A.}\ \bibnamefont {Walmsley}},\ }\bibfield
  {title} {\bibinfo {title} {Boson sampling on a photonic chip},\ }\href
  {https://doi.org/10.1126/science.1231692} {\bibfield  {journal} {\bibinfo
  {journal} {Science}\ }\textbf {\bibinfo {volume} {339}},\ \bibinfo {pages}
  {798} (\bibinfo {year} {2013})}\BibitemShut {NoStop}%
\bibitem [{\citenamefont {Moreva}\ \emph {et~al.}(2017)\citenamefont {Moreva},
  \citenamefont {Traina}, \citenamefont {Forneris}, \citenamefont {Degiovanni},
  \citenamefont {Ditalia~Tchernij}, \citenamefont {Picollo}, \citenamefont
  {Brida}, \citenamefont {Olivero},\ and\ \citenamefont
  {Genovese}}]{Moreva2017}%
  \BibitemOpen
  \bibfield  {author} {\bibinfo {author} {\bibfnamefont {E.}~\bibnamefont
  {Moreva}}, \bibinfo {author} {\bibfnamefont {P.}~\bibnamefont {Traina}},
  \bibinfo {author} {\bibfnamefont {J.}~\bibnamefont {Forneris}}, \bibinfo
  {author} {\bibfnamefont {I.~P.}\ \bibnamefont {Degiovanni}}, \bibinfo
  {author} {\bibfnamefont {S.}~\bibnamefont {Ditalia~Tchernij}}, \bibinfo
  {author} {\bibfnamefont {F.}~\bibnamefont {Picollo}}, \bibinfo {author}
  {\bibfnamefont {G.}~\bibnamefont {Brida}}, \bibinfo {author} {\bibfnamefont
  {P.}~\bibnamefont {Olivero}},\ and\ \bibinfo {author} {\bibfnamefont
  {M.}~\bibnamefont {Genovese}},\ }\bibfield  {title} {\bibinfo {title} {Direct
  experimental observation of nonclassicality in ensembles of single-photon
  emitters},\ }\href {https://doi.org/10.1103/PhysRevB.96.195209} {\bibfield
  {journal} {\bibinfo  {journal} {Phys. Rev. B}\ }\textbf {\bibinfo {volume}
  {96}},\ \bibinfo {pages} {195209} (\bibinfo {year} {2017})}\BibitemShut
  {NoStop}%
\bibitem [{\citenamefont {Ob\ifmmode~\check{s}\else \v{s}\fi{}il}\ \emph
  {et~al.}(2018)\citenamefont {Ob\ifmmode~\check{s}\else \v{s}\fi{}il},
  \citenamefont {Lachman}, \citenamefont {Pham}, \citenamefont
  {Le\ifmmode~\check{s}\else \v{s}\fi{}und\'ak}, \citenamefont {Hucl},
  \citenamefont {\ifmmode \check{C}\else
  \v{C}\fi{}\'{\i}\ifmmode~\check{z}\else \v{z}\fi{}ek}, \citenamefont
  {Hrabina}, \citenamefont {\ifmmode~\check{C}\else \v{C}\fi{}\'{\i}p},
  \citenamefont {Slodi\ifmmode~\check{c}\else \v{c}\fi{}ka},\ and\
  \citenamefont {Filip}}]{Obsil2018}%
  \BibitemOpen
  \bibfield  {author} {\bibinfo {author} {\bibfnamefont {P.}~\bibnamefont
  {Ob\ifmmode~\check{s}\else \v{s}\fi{}il}}, \bibinfo {author} {\bibfnamefont
  {L.}~\bibnamefont {Lachman}}, \bibinfo {author} {\bibfnamefont
  {T.}~\bibnamefont {Pham}}, \bibinfo {author} {\bibfnamefont {A.}~\bibnamefont
  {Le\ifmmode~\check{s}\else \v{s}\fi{}und\'ak}}, \bibinfo {author}
  {\bibfnamefont {V.}~\bibnamefont {Hucl}}, \bibinfo {author} {\bibfnamefont
  {M.}~\bibnamefont {\ifmmode \check{C}\else
  \v{C}\fi{}\'{\i}\ifmmode~\check{z}\else \v{z}\fi{}ek}}, \bibinfo {author}
  {\bibfnamefont {J.}~\bibnamefont {Hrabina}}, \bibinfo {author} {\bibfnamefont
  {O.}~\bibnamefont {\ifmmode~\check{C}\else \v{C}\fi{}\'{\i}p}}, \bibinfo
  {author} {\bibfnamefont {L.}~\bibnamefont {Slodi\ifmmode~\check{c}\else
  \v{c}\fi{}ka}},\ and\ \bibinfo {author} {\bibfnamefont {R.}~\bibnamefont
  {Filip}},\ }\bibfield  {title} {\bibinfo {title} {Nonclassical light from
  large ensembles of trapped ions},\ }\href
  {https://doi.org/10.1103/PhysRevLett.120.253602} {\bibfield  {journal}
  {\bibinfo  {journal} {Phys. Rev. Lett.}\ }\textbf {\bibinfo {volume} {120}},\
  \bibinfo {pages} {253602} (\bibinfo {year} {2018})}\BibitemShut {NoStop}%
\bibitem [{\citenamefont {Bohmann}\ \emph {et~al.}(2019)\citenamefont
  {Bohmann}, \citenamefont {Qi}, \citenamefont {Vogel},\ and\ \citenamefont
  {Chekhova}}]{Bohmann2019}%
  \BibitemOpen
  \bibfield  {author} {\bibinfo {author} {\bibfnamefont {M.}~\bibnamefont
  {Bohmann}}, \bibinfo {author} {\bibfnamefont {L.}~\bibnamefont {Qi}},
  \bibinfo {author} {\bibfnamefont {W.}~\bibnamefont {Vogel}},\ and\ \bibinfo
  {author} {\bibfnamefont {M.}~\bibnamefont {Chekhova}},\ }\bibfield  {title}
  {\bibinfo {title} {Detection-device-independent verification of nonclassical
  light},\ }\href {https://doi.org/10.1103/PhysRevResearch.1.033178} {\bibfield
   {journal} {\bibinfo  {journal} {Phys. Rev. Res.}\ }\textbf {\bibinfo
  {volume} {1}},\ \bibinfo {pages} {033178} (\bibinfo {year}
  {2019})}\BibitemShut {NoStop}%
\bibitem [{\citenamefont {Lachman}\ \emph {et~al.}(2024)\citenamefont
  {Lachman}, \citenamefont {Radko}, \citenamefont {Bergamin}, \citenamefont
  {Andersen}, \citenamefont {Huck},\ and\ \citenamefont {Filip}}]{Lachman2024}%
  \BibitemOpen
  \bibfield  {author} {\bibinfo {author} {\bibfnamefont {L.~c.~v.}\
  \bibnamefont {Lachman}}, \bibinfo {author} {\bibfnamefont {I.~P.}\
  \bibnamefont {Radko}}, \bibinfo {author} {\bibfnamefont {M.}~\bibnamefont
  {Bergamin}}, \bibinfo {author} {\bibfnamefont {U.~L.}\ \bibnamefont
  {Andersen}}, \bibinfo {author} {\bibfnamefont {A.}~\bibnamefont {Huck}},\
  and\ \bibinfo {author} {\bibfnamefont {R.}~\bibnamefont {Filip}},\ }\bibfield
   {title} {\bibinfo {title} {Experimental certification of level dynamics in
  single-photon emitters},\ }\href
  {https://doi.org/10.1103/PhysRevResearch.6.033254} {\bibfield  {journal}
  {\bibinfo  {journal} {Phys. Rev. Res.}\ }\textbf {\bibinfo {volume} {6}},\
  \bibinfo {pages} {033254} (\bibinfo {year} {2024})}\BibitemShut {NoStop}%
\bibitem [{\citenamefont {Boyd}\ and\ \citenamefont
  {Vandenberghe}(2004)}]{boyd_book}%
  \BibitemOpen
  \bibfield  {author} {\bibinfo {author} {\bibfnamefont {S.}~\bibnamefont
  {Boyd}}\ and\ \bibinfo {author} {\bibfnamefont {L.}~\bibnamefont
  {Vandenberghe}},\ }\href@noop {} {\emph {\bibinfo {title} {Convex
  Optimization}}}\ (\bibinfo  {publisher} {Cambridge University Press},\
  \bibinfo {address} {Cambridge, England},\ \bibinfo {year} {2004})\BibitemShut
  {NoStop}%
\bibitem [{\citenamefont {Kovtoniuk}\ \emph {et~al.}(2022)\citenamefont
  {Kovtoniuk}, \citenamefont {Yeremenko}, \citenamefont {Ryl}, \citenamefont
  {Vogel},\ and\ \citenamefont {Semenov}}]{Kovtoniuk2022}%
  \BibitemOpen
  \bibfield  {author} {\bibinfo {author} {\bibfnamefont {V.~S.}\ \bibnamefont
  {Kovtoniuk}}, \bibinfo {author} {\bibfnamefont {I.~S.}\ \bibnamefont
  {Yeremenko}}, \bibinfo {author} {\bibfnamefont {S.}~\bibnamefont {Ryl}},
  \bibinfo {author} {\bibfnamefont {W.}~\bibnamefont {Vogel}},\ and\ \bibinfo
  {author} {\bibfnamefont {A.~A.}\ \bibnamefont {Semenov}},\ }\bibfield
  {title} {\bibinfo {title} {Nonclassical correlations of radiation in relation
  to {Bell} nonlocality},\ }\href {https://doi.org/10.1103/PhysRevA.105.063722}
  {\bibfield  {journal} {\bibinfo  {journal} {Phys. Rev. A}\ }\textbf {\bibinfo
  {volume} {105}},\ \bibinfo {pages} {063722} (\bibinfo {year}
  {2022})}\BibitemShut {NoStop}%
\bibitem [{\citenamefont {Karlin}\ and\ \citenamefont
  {Studden}(1966)}]{Karlin1966}%
  \BibitemOpen
  \bibfield  {author} {\bibinfo {author} {\bibfnamefont {S.}~\bibnamefont
  {Karlin}}\ and\ \bibinfo {author} {\bibfnamefont {W.}~\bibnamefont
  {Studden}},\ }\href@noop {} {\emph {\bibinfo {title} {Tchebycheff Systems:
  With Applications in Analysis and Statistics}}},\ Pure and Applied
  Mathematics: Interscience\ (\bibinfo  {publisher} {Interscience Publishers},\
  \bibinfo {address} {New York},\ \bibinfo {year} {1966})\BibitemShut {NoStop}%
\bibitem [{\citenamefont {Krein}\ and\ \citenamefont
  {Nudel'man}(1977)}]{Krein1977}%
  \BibitemOpen
  \bibfield  {author} {\bibinfo {author} {\bibfnamefont {M.~G.}\ \bibnamefont
  {Krein}}\ and\ \bibinfo {author} {\bibfnamefont {A.~A.}\ \bibnamefont
  {Nudel'man}},\ }\href@noop {} {\emph {\bibinfo {title} {The Markov Moment
  Problem and Extremal Problems}}},\ Translations of mathematical monographs,
  vol. 50\ (\bibinfo  {publisher} {American Mathematical Society},\ \bibinfo
  {address} {Rhode Island},\ \bibinfo {year} {1977})\BibitemShut {NoStop}%
\bibitem [{\citenamefont {de~Dios~Pont}\ \emph {et~al.}(2023)\citenamefont
  {de~Dios~Pont}, \citenamefont {Ivanisvili},\ and\ \citenamefont
  {Madrid}}]{Pont2023}%
  \BibitemOpen
  \bibfield  {author} {\bibinfo {author} {\bibfnamefont {J.}~\bibnamefont
  {de~Dios~Pont}}, \bibinfo {author} {\bibfnamefont {P.}~\bibnamefont
  {Ivanisvili}},\ and\ \bibinfo {author} {\bibfnamefont {J.}~\bibnamefont
  {Madrid}},\ }\href@noop {} {\bibinfo {title} {A new proof of the description
  of the convex hull of space curves with totally positive torsion}} (\bibinfo
  {year} {2023}),\ \Eprint {https://arxiv.org/abs/2201.12932} {arXiv:2201.12932
  [math.PR]} \BibitemShut {NoStop}%
\bibitem [{\citenamefont {Frankel}(2012)}]{Frankel_book}%
  \BibitemOpen
  \bibfield  {author} {\bibinfo {author} {\bibfnamefont {T.}~\bibnamefont
  {Frankel}},\ }\href@noop {} {\emph {\bibinfo {title} {The geometry of
  physics: {A}n introduction}}},\ \bibinfo {edition} {3rd}\ ed.\ (\bibinfo
  {publisher} {Cambridge University Press},\ \bibinfo {address} {Cambridge,
  England},\ \bibinfo {year} {2012})\BibitemShut {NoStop}%
\bibitem [{\citenamefont {Paul}\ \emph {et~al.}(1996)\citenamefont {Paul},
  \citenamefont {T\"orm\"a}, \citenamefont {Kiss},\ and\ \citenamefont
  {Jex}}]{paul1996}%
  \BibitemOpen
  \bibfield  {author} {\bibinfo {author} {\bibfnamefont {H.}~\bibnamefont
  {Paul}}, \bibinfo {author} {\bibfnamefont {P.}~\bibnamefont {T\"orm\"a}},
  \bibinfo {author} {\bibfnamefont {T.}~\bibnamefont {Kiss}},\ and\ \bibinfo
  {author} {\bibfnamefont {I.}~\bibnamefont {Jex}},\ }\bibfield  {title}
  {\bibinfo {title} {Photon chopping: New way to measure the quantum state of
  light},\ }\href {https://doi.org/10.1103/PhysRevLett.76.2464} {\bibfield
  {journal} {\bibinfo  {journal} {Phys. Rev. Lett.}\ }\textbf {\bibinfo
  {volume} {76}},\ \bibinfo {pages} {2464} (\bibinfo {year}
  {1996})}\BibitemShut {NoStop}%
\bibitem [{\citenamefont {Castelletto}\ \emph {et~al.}(2007)\citenamefont
  {Castelletto}, \citenamefont {Degiovanni}, \citenamefont {Schettini},\ and\
  \citenamefont {Migdall}}]{castelletto2007}%
  \BibitemOpen
  \bibfield  {author} {\bibinfo {author} {\bibfnamefont {S.~A.}\ \bibnamefont
  {Castelletto}}, \bibinfo {author} {\bibfnamefont {I.~P.}\ \bibnamefont
  {Degiovanni}}, \bibinfo {author} {\bibfnamefont {V.}~\bibnamefont
  {Schettini}},\ and\ \bibinfo {author} {\bibfnamefont {A.~L.}\ \bibnamefont
  {Migdall}},\ }\bibfield  {title} {\bibinfo {title} {Reduced deadtime and
  higher rate photon-counting detection using a multiplexed detector array},\
  }\href {https://doi.org/10.1080/09500340600779579} {\bibfield  {journal}
  {\bibinfo  {journal} {J. Mod. Opt.}\ }\textbf {\bibinfo {volume} {54}},\
  \bibinfo {pages} {337} (\bibinfo {year} {2007})}\BibitemShut {NoStop}%
\bibitem [{\citenamefont {Schettini}\ \emph {et~al.}(2007)\citenamefont
  {Schettini}, \citenamefont {Polyakov}, \citenamefont {Degiovanni},
  \citenamefont {Brida}, \citenamefont {Castelletto},\ and\ \citenamefont
  {Migdall}}]{schettini2007}%
  \BibitemOpen
  \bibfield  {author} {\bibinfo {author} {\bibfnamefont {V.}~\bibnamefont
  {Schettini}}, \bibinfo {author} {\bibfnamefont {S.~V.}\ \bibnamefont
  {Polyakov}}, \bibinfo {author} {\bibfnamefont {I.~P.}\ \bibnamefont
  {Degiovanni}}, \bibinfo {author} {\bibfnamefont {G.}~\bibnamefont {Brida}},
  \bibinfo {author} {\bibfnamefont {S.}~\bibnamefont {Castelletto}},\ and\
  \bibinfo {author} {\bibfnamefont {A.~L.}\ \bibnamefont {Migdall}},\
  }\bibfield  {title} {\bibinfo {title} {Implementing a multiplexed system of
  detectors for higher photon counting rates},\ }\href
  {https://doi.org/10.1109/JSTQE.2007.902846} {\bibfield  {journal} {\bibinfo
  {journal} {IEEE J. Sel. Top. Quantum Electron.}\ }\textbf {\bibinfo {volume}
  {13}},\ \bibinfo {pages} {978} (\bibinfo {year} {2007})}\BibitemShut
  {NoStop}%
\bibitem [{\citenamefont {Blanchet}\ \emph {et~al.}(2008)\citenamefont
  {Blanchet}, \citenamefont {Devaux}, \citenamefont {Furfaro},\ and\
  \citenamefont {Lantz}}]{blanchet08}%
  \BibitemOpen
  \bibfield  {author} {\bibinfo {author} {\bibfnamefont {J.-L.}\ \bibnamefont
  {Blanchet}}, \bibinfo {author} {\bibfnamefont {F.}~\bibnamefont {Devaux}},
  \bibinfo {author} {\bibfnamefont {L.}~\bibnamefont {Furfaro}},\ and\ \bibinfo
  {author} {\bibfnamefont {E.}~\bibnamefont {Lantz}},\ }\bibfield  {title}
  {\bibinfo {title} {Measurement of sub-shot-noise correlations of spatial
  fluctuations in the photon-counting regime},\ }\href
  {https://doi.org/10.1103/PhysRevLett.101.233604} {\bibfield  {journal}
  {\bibinfo  {journal} {Phys. Rev. Lett.}\ }\textbf {\bibinfo {volume} {101}},\
  \bibinfo {pages} {233604} (\bibinfo {year} {2008})}\BibitemShut {NoStop}%
\bibitem [{\citenamefont {Achilles}\ \emph {et~al.}(2003)\citenamefont
  {Achilles}, \citenamefont {Silberhorn}, \citenamefont {\'{S}liwa},
  \citenamefont {Banaszek},\ and\ \citenamefont {Walmsley}}]{achilles03}%
  \BibitemOpen
  \bibfield  {author} {\bibinfo {author} {\bibfnamefont {D.}~\bibnamefont
  {Achilles}}, \bibinfo {author} {\bibfnamefont {C.}~\bibnamefont
  {Silberhorn}}, \bibinfo {author} {\bibfnamefont {C.}~\bibnamefont
  {\'{S}liwa}}, \bibinfo {author} {\bibfnamefont {K.}~\bibnamefont
  {Banaszek}},\ and\ \bibinfo {author} {\bibfnamefont {I.~A.}\ \bibnamefont
  {Walmsley}},\ }\bibfield  {title} {\bibinfo {title} {Fiber-assisted detection
  with photon number resolution},\ }\href
  {https://doi.org/10.1364/OL.28.002387} {\bibfield  {journal} {\bibinfo
  {journal} {Opt. Lett.}\ }\textbf {\bibinfo {volume} {28}},\ \bibinfo {pages}
  {2387} (\bibinfo {year} {2003})}\BibitemShut {NoStop}%
\bibitem [{\citenamefont {Fitch}\ \emph {et~al.}(2003)\citenamefont {Fitch},
  \citenamefont {Jacobs}, \citenamefont {Pittman},\ and\ \citenamefont
  {Franson}}]{fitch03}%
  \BibitemOpen
  \bibfield  {author} {\bibinfo {author} {\bibfnamefont {M.~J.}\ \bibnamefont
  {Fitch}}, \bibinfo {author} {\bibfnamefont {B.~C.}\ \bibnamefont {Jacobs}},
  \bibinfo {author} {\bibfnamefont {T.~B.}\ \bibnamefont {Pittman}},\ and\
  \bibinfo {author} {\bibfnamefont {J.~D.}\ \bibnamefont {Franson}},\
  }\bibfield  {title} {\bibinfo {title} {Photon-number resolution using
  time-multiplexed single-photon detectors},\ }\href
  {https://doi.org/10.1103/PhysRevA.68.043814} {\bibfield  {journal} {\bibinfo
  {journal} {Phys. Rev. A}\ }\textbf {\bibinfo {volume} {68}},\ \bibinfo
  {pages} {043814} (\bibinfo {year} {2003})}\BibitemShut {NoStop}%
\bibitem [{\citenamefont {\ifmmode \check{R}\else
  \v{R}\fi{}eh\'a\ifmmode~\check{c}\else \v{c}\fi{}ek}\ \emph
  {et~al.}(2003)\citenamefont {\ifmmode \check{R}\else
  \v{R}\fi{}eh\'a\ifmmode~\check{c}\else \v{c}\fi{}ek}, \citenamefont {Hradil},
  \citenamefont {Haderka}, \citenamefont {Pe\ifmmode~\check{r}\else
  \v{r}\fi{}ina},\ and\ \citenamefont {Hamar}}]{rehacek03}%
  \BibitemOpen
  \bibfield  {author} {\bibinfo {author} {\bibfnamefont {J.}~\bibnamefont
  {\ifmmode \check{R}\else \v{R}\fi{}eh\'a\ifmmode~\check{c}\else
  \v{c}\fi{}ek}}, \bibinfo {author} {\bibfnamefont {Z.}~\bibnamefont {Hradil}},
  \bibinfo {author} {\bibfnamefont {O.}~\bibnamefont {Haderka}}, \bibinfo
  {author} {\bibfnamefont {J.}~\bibnamefont {Pe\ifmmode~\check{r}\else
  \v{r}\fi{}ina}},\ and\ \bibinfo {author} {\bibfnamefont {M.}~\bibnamefont
  {Hamar}},\ }\bibfield  {title} {\bibinfo {title} {Multiple-photon resolving
  fiber-loop detector},\ }\href {https://doi.org/10.1103/PhysRevA.67.061801}
  {\bibfield  {journal} {\bibinfo  {journal} {Phys. Rev. A}\ }\textbf {\bibinfo
  {volume} {67}},\ \bibinfo {pages} {061801(R)} (\bibinfo {year}
  {2003})}\BibitemShut {NoStop}%
\bibitem [{\citenamefont {Kovtoniuk}\ \emph {et~al.}(2026)\citenamefont
  {Kovtoniuk}, \citenamefont {Bohmann},\ and\ \citenamefont
  {Semenov}}]{Kovtoniuk2026Data}%
  \BibitemOpen
  \bibfield  {author} {\bibinfo {author} {\bibfnamefont {V.~S.}\ \bibnamefont
  {Kovtoniuk}}, \bibinfo {author} {\bibfnamefont {M.}~\bibnamefont {Bohmann}},\
  and\ \bibinfo {author} {\bibfnamefont {A.~A.}\ \bibnamefont {Semenov}},\
  }\href {https://doi.org/10.5281/zenodo.20524274} {\bibinfo {title} {Codebase
  for reproducing figures of ``{N}onclassical photocounting statistics with a
  single on-off detector'', zenodo, version 1.0, doi:10.5281/zenodo.20524274}}
  (\bibinfo {year} {2026})\BibitemShut {NoStop}%
\end{thebibliography}%

\end{document}